\documentstyle[preprint,aps]{revtex}

\begin{document}

\draft

\title{Gravitational Radiation from Rotational Instabilites in
Compact Stellar Cores with Stiff Equations of State}

\author{Janet L. Houser and Joan M. Centrella}

\address
{Department of Physics and Atmospheric Science,
 Drexel University,
Philadelphia, PA 19104}

\maketitle

\begin{abstract}
We carry out 3-D numerical simulations of the dynamical 
instability in rapidly rotating stars initially modeled
as polytropes with $n = 1.5, 1.0,$ and $0.5$.  The
calculations are done with a SPH code using
Newtonian gravity, and the gravitational radiation is
calculated in the quadrupole limit.
All models develop the global $m=2$ bar mode, with mass
and angular momentum being shed from the ends of the
bar in two trailing spiral arms. 
The models then undergo successive episodes of
core recontraction and spiral arm ejection, 
with the number of these episodes increasing
as $n$ decreases: this results
in longer-lived gravitational wave signals
for stiffer models.
This instability may operate in a stellar core
that has expended its nuclear fuel and is prevented
from further collapse due to centrifugal forces.
The actual values of the gravitational radiation
amplitudes and frequencies depend sensitively on the
radius of the star $R_{\rm eq}$ at which the
instability develops.

\end{abstract}

\pacs{PACS numbers: 04.30.Db,
 04.80.Nn, 95.55.Ym, 97.60.-s}

\section{INTRODUCTION}

The direct detection of gravitational radiation from astrophysical 
sources is one of the greatest challenges of our day.  With 
interferometers such as LIGO \cite{LIGO92}, VIRGO \cite{VIRGO}, and
GEO \cite{GEO} under construction, and a new generation of spherical
resonant mass detectors under study 
\cite{resonant,TIGA}, the detailed modeling
of these sources takes a high priority.

One promising source of gravitational waves is the development
of rotational instabilities in dense stellar cores or compact objects
\cite{schutz86}. Consider, for example, 
a rapidly rotating stellar core that has expended its
nuclear fuel and is unable to collapse to neutron star size due to
centrifugal forces. If such an object underwent a rotational 
instability, it could possibly shed 
enough angular momentum to allow collapse to a supernova 
\cite{thorne96a,thorne96b}.  
Alternatively, a neutron star that is 
spun up by accretion
of mass from a binary companion may reach fast 
enough rotation rates to go unstable 
\cite{wagoner,schutz89}. 

Global rotational instabilities in fluids arise from 
nonradial ``toroidal'' 
modes $e^{\pm im\phi}$, where $m=2$ is known as the ``bar mode''.
It is convenient to parametrize them by 
\begin{equation}
	\beta = T_{\rm rot}/|W|,
\label{beta} \end{equation}
where $T_{\rm rot}$ is the rotational
 kinetic energy and $W$ is the 
gravitational potential energy 
\cite{tassoul,ST,DT85}. 
 We focus on the bar instability
since it is expected to be the fastest growing mode.  This
 instability
can occur under two different physical mechanisms.  The 
{\em dynamical} bar
instability is driven by Newtonian hydrodynamics and gravity.
It operates for fairly large values 
of the stability parameter $\beta > \beta_{\rm d}$ 
and develops on 
a timescale of approximately one bar rotation period. 
In contrast, the {\em secular}
instability arises from dissipative processes such as gravitational
radiation reaction and viscosity. It occurs for
$\beta_{\rm s} < \beta < \beta_{\rm d}$ 
and develops on a timescale of several
rotation periods or longer \cite{schutz89}. 
For the constant density,
incompressible, uniformly rotating Maclaurin spheroids
we have $\beta_{\rm s} \approx 0.14$ and
 $\beta_{\rm d} \approx 0.27$.  
In the case of differentially rotating
fluids with a polytropic equation of state
\begin{equation}
	P  = K \rho^{\Gamma} = K \rho^{1+ 1/n},
\label{poly}  
\end{equation}
where $n$ is the polytropic index and $K$ is a constant
that depends on the entropy, early studies indicated that 
the secular and dynamical bar 
instabilities should occur at about these same values of
$\beta$ \cite{ST,DT85,managan85,IFD}. More recent work 
\cite{ITDPY} shows that both
the angular momentum distribution and, to a lesser degree, the
polytropic index affect the value of 
$\beta_{\rm s}$ at which the $m=2$
secular instability sets in.  
For the dynamical bar instability
Pickett, et al. \cite{PDD} demonstrate 
that, for $n=1.5$ polytropes, the $m=2$ 
dynamical stability limit
$\beta_{\rm d} \approx 0.27$ is valid for 
centrally condensed initial angular momentum
distributions that are similar to those
of Maclaurin spheroids.  However, for angular momentum distributions
with somewhat extended disk-like regions, both one- and
two-armed spiral instablities appear at considerably lower
values of $\beta$.

The work presented in this paper is part of a research program 
aimed at calculating the gravitational radiation produced when a
rapidly rotating stellar core undergoes the {\em dynamical}
bar instability.  These studies are carried out using 3-D
numerical simulations.  The gravitational field is purely
Newtonian, and the gravitational radiation produced is calculated
using the quadrupole approximation; the back reaction of the 
radiation on the system is not included.  The rapidly rotating
cores are initially modeled as polytropes with 
$\beta \approx 0.3$, which is just above the dynamical stability
limit.  In order to sustain such high rotational kinetic energy
they must be rotating differentially
\cite{tassoul,LRS-APJS}. Such objects could form, for example,
when the cores of massive stars collapse on a dynamical time scale.

Much of the 
previous work in this area has concentrated on polytropes with
$n = 1.5$.  This case has been investigated by Tohline and
collaborators \cite{TDM,DGTB,WT87} and more recently by 
Pickett, et al. \cite{PDD} in the context of star formation.
These studies primarily use an Eulerian code that 
imposes the polytropic equation of
state~(\ref{poly}) throughout the evolution instead of solving
an energy equation; thus, the entropy generation by shocks during
the later stages of evolution is not taken into account.  The
work of Houser, Centrella, and Smith \cite{PRL} was the first 
to model the fluid using an energy
equation and to calculate the resulting gravitational radiation;
these calculations were carried out using the
Lagrangian smooth particle hydrodynamics (SPH) method.
(New \cite{new_phd} has recently performed similar calculations
using an improved version of Tohline's Eulerian code.)
Smith, Houser, and Centrella \cite{comparison}
 also updated the earlier work of Ref. \cite{DGTB} by carrying
out a detailed comparison study of this model with two different
3-D codes, one using Eulerian techniques and the other based on 
SPH.

In this paper we extend our calculations to objects having stiff
polytropic equations of state using SPH
\cite{jlh_phd}.  We 
use $\beta \approx 0.3$ and consider the cases
$n = 1.5, 1.0$, and $0.5$, which correspond to $\Gamma = 5/3, 2$, 
and $3$.  Previously, Williams and Tohline \cite{WT87}
studied the initial development of the bar instability in
similar models with $n = 0.8, 1.0, 1.3, 1.5$, and $1.8$;
longer evolutions were carried out in Ref. \cite{WT88} for the
cases $n=0.8$ and $n=1.8$.  Their work was done with an Eulerian
code in cylindrical coordinates $(\varpi,\phi,z,)$ that used
the diffusive donor cell advection method and imposed the
polytropic equation of state~(\ref{poly}) throughout the
runs.  In addition, they modeled only the region 
$0 \le \phi <  \pi$ in the angular coordinate, so that only
even toroidal modes were represented.  Our simulations do not
suffer from these restrictions, and include the calculation of
the gravitational radiation.

This paper is organized as follows.
Section~\ref{sim-tech} contains
a brief description of the techniques used in our
simulations.  The construction of initial axisymmetric models
with $\beta \approx 0.3$ is 
discussed in Sec.~\ref{initial} and the dynamical evolution of
the models is presented in Sec.~\ref{evol}.  Analysis of the
instabilities using
Fourier components is given in Sec.~\ref{modes} and the
gravitational radiation produced by the models
is presented in
Sec.~\ref{GW}.  The paper concludes with a 
summary and discussion of our
results in Sec.~\ref{discuss}.

\section{Simulation Techniques}
\label{sim-tech}

Detailed descriptions of the basic techniques used to
 carry out these simulations have
been presented previously 
in Refs. \cite{comparison} and \cite{CM}.
  We therefore
give a only brief description of these methods in this section,
and refer the reader to the literature for further information.

SPH is a Lagrangian method
in which the fluid is modeled as a collection of fluid elements
of finite extent described by a smoothing kernel
\cite{SPH}.
 We have used the
 implementation of SPH by Hernquist \& Katz \cite{HK} known as
 TREESPH, which allows variable smoothing lengths and individual
particle timesteps. For the runs discussed in this paper,
we smooth over $N_{\cal S} = 64$ neighbors for kernel
interpolation.
Shocks are handled using 
an artificial viscosity modified by the curl of the velocity
field, with the user-specified 
coefficients of the 
linear and quadratic terms taking the values 
$\alpha = 0.25$ and $\beta_{\rm AV} = 1.0$;
see Refs. \cite{HK,comparison,CM} for more details.
 The gravitational 
forces in this code are purely Newtonian, and are calculated 
using
a hierarchical tree method optimized for vector 
computers\cite{tree}.   This
leads to a significant gain in efficiency and allows the use of
larger numbers of particles than would be possible with methods
 that
simply sum over all possible pairs of particles.

We calculate the gravitational radiation produced
by the instabilities using the
quadrupole approximation, which is valid for nearly Newtonian 
sources
\cite{MTW}.  The reduced (i.e., traceless) quadrupole moment
of the source is given by
\begin{equation}
{{I\mkern-6.8mu\raise0.3ex\hbox{-}}}_{ij} = \int\rho \,(x_i x_j -
{\textstyle{\frac{1}{3}}}
\delta_{ij} r^2) \:d^3 r,
\label{Iij}  
\end{equation}
where $i,j=1,2,3$ are spatial indices and $r=(x^2 + y^2 +
z^2)^{1/2}$ is the distance to the source.
For an observer situated on the axis at $\theta=0, \phi=0$ of a 
spherical coordinate system with its origin located at the center
of mass of the source, the amplitude of the gravitational waves
for the two
polarization states takes the simple form
\begin{eqnarray}
h_{+}&=&\frac{G}{c^4}\frac{1}{r}( {\skew6\ddot{
{I\mkern-6.8mu\raise0.3ex\hbox{-}}}}_{xx}- {\skew6\ddot{
{I\mkern-6.8mu\raise0.3ex\hbox{-}}}}_{yy}),
\label{hplus-axis}\\
h_{\times} &=&\frac{G}{c^4}\frac{2}{r} {\skew6\ddot{
{I\mkern-6.8mu\raise0.3ex\hbox{-}}}}_{xy} ,\label{hcross-axis}
\end{eqnarray}
where an overdot indicates a time derivative $d/dt$.
The gravitational wave luminosity is given by
\begin{equation}
L = \frac{dE}{dt} = {\frac15}\frac{G}{c^5} \left\langle
{I\mkern-6.8mu\raise0.3ex\hbox{-}}^{(3)}_{ij}
{I\mkern-6.8mu\raise0.3ex\hbox{-}}^{(3)}_{ij} \right \rangle ,
\label{lum} 
\end{equation}
and the rate at which
 angular momentum is lost through gravitational radiation is
\begin{equation}
\frac{dJ_i}{dt} = {\frac{2}{5}}\frac{G}{c^5} \epsilon_{ijk}
 \left\langle {I\mkern-6.8mu\raise0.3ex\hbox{-}}^{(2)}_{jm}
{I\mkern-6.8mu\raise0.3ex\hbox{-}}^{(3)}_{km}  \right \rangle.
\label{dJidt} \end{equation}
Here there is an implied sum on repeated indices, 
the superscript $(3)$
indicates the third time derivative, and the 
 angle brackets
indicate an average over several wave periods. 
 Since such averaging
is not well-defined for these burst sources, we 
display instead
the unaveraged quantities $(G/5c^5)\textstyle{
{I\mkern-6.8mu
\raise0.3ex\hbox{-}}^{(3)}_{ij}
{I\mkern-6.8mu\raise0.3ex\hbox{-}}^{(3)}_{ij}}$ and 
$(2G/5c^5){\textstyle
\epsilon_{ijk}{I\mkern-6.8mu\raise0.3ex\hbox{-}}^{(2)}_{jm}
{I\mkern-6.8mu\raise0.3ex\hbox{-}}^{(3)}_{km}}$ below.  The energy
emitted as gravitational radiation is
\begin{equation}
\Delta E = \int L\; dt
\label{delta-E} \end{equation}
 and
the angular momentum carried away by the waves is
\begin{equation}
\Delta J_i = \int (dJ_i/dt)dt.
\label{delta-Ji} \end{equation}
Finally,
the gravitational wave energy spectrum $dE/df$, which
 gives the 
energy emitted as gravitational radiation per unit
 frequency interval, takes the form \cite{thorne87} 
\begin{equation}
\frac{dE}{df} =  \frac{c^3}{G}
 \frac{\pi}{2} (4 \pi r^2) f^2 \langle
\langle |\tilde h_+ (f)|^2 + |\tilde h_{\times}(f)|^2 \rangle
\rangle ,
\label{dEdf}  \end{equation}
where $\tilde h(f)$ is the Fourier transform of $h(t)$. 
The double angle brackets in Eq.~(\ref{dEdf}) denote an average
 over all source angles.

We calculate the
 reduced quadrupole moment 
${{I\mkern-6.8mu\raise0.3ex\hbox{-}}}_{ij}$
and its derivatives using the methods developed
in Ref. \cite{CM}.  In particular,
particle positions, velocities, and accelerations already 
present in 
the code are used to obtain
${\skew6\dot{I\mkern-6.8mu\raise0.3ex\hbox{-}}}_{ij}$ and
${\skew6\ddot{I\mkern-6.8mu\raise0.3ex\hbox{-}}}_{ij}$,
yielding expressions similar to those in Ref. \cite{FE}.
This results in waveforms that 
are very smooth functions of time and 
require no filtering or smoothing to remove numerical noise.
However, the luminosity $L$ and angular momentum lost by
gravitational radiation $dJ_i/dt$
do contain the time derivative of the 
particle acceleration; this is
taken numerically and therefore introduces 
some noise.  To remove this noise, we
smooth the luminosity data
using simple averaging over a fixed time interval
of $0.1 t_{\rm D}$
centered on each point. Here, $t_{\rm D}$ is the dynamical time 
for a spherical star 
of mass $M$ and (equatorial) radius $R_{\rm eq}$ and
is defined by
\begin{equation}
t_{\rm D}= \left ( \frac{R_{\rm eq}^3}{GM} \right ) ^{1/2} .
\label{tD}
\end{equation}
 In general this procedure produces very smooth 
luminosity profiles
 \cite{CM} and makes a 
negligible change in the integrated luminosity
$\Delta E$,
which gives the energy emitted as gravitational radiation.
The profiles of $dJ_i/dt$ are not smoothed.

\section{Initial Models}
\label{initial}

The initial conditions for our simulations consist of rotating
axisymmetric equilibrium fluid models having $\beta \approx 0.3$
and polytropic index $n$.  The dynamical bar instability then
grows from nonaxisymmetric perturbations due to particle
discreteness in the models.  We use a two-step procedure to generate
the initial models.  First, a self-consistent field (SCF) method
is used to produce an equilibrium model on a grid.  Then, a particle
fit to the SCF model is performed to generate initial data for
TREESPH.
In this section, we describe the
construction of these equilibrium models.

\subsection{Self-Consistent Field Method}

The first step is to use the SCF method (\cite{SC};
see also \cite{OM,BO,hachisu86})
 to generate axisymmetric equilibrium models.  The SCF
procedure derives from an integral formulation of
 the equations of 
hydrodynamic equilibrium which automatically incorporates the
boundary conditions.  We use cylindrical coordinates 
$(\varpi,z)$ and a uniformly-zoned grid 
of $N_{\varpi}$ radial and $N_z$ axial zones.
An initial ``guess'' density distribution 
$\rho(\varpi,z)$ is given, and the 
gravitational potential is calculated using a Legendre polynomial
expansion to solve Poisson's equation \cite{hachisu86}. 
 A rotation law in which angular momentum is constant on 
cylinders is specified.  This takes
the general form $j(m) = j(m(\varpi))$,
where $j(m)$ is the specific angular momentum 
and $m(\varpi)$ here denotes the dimensionless
mass fraction interior to the cylinder of radius $\varpi$ 
\cite{OM}.  Following the convention of earlier work 
(e.g. Refs. \cite{TDM,DGTB,WT87,WT88,BO} )
 we use the rotation law for the uniformly
rotating, constant density Maclaurin spheroids, 
\begin{equation}
 j(m) = {\frac{5}{2}}{\frac{J}{M}}
\left [ 1 - (1-m)^{2/3} \right ],
\label{rotn-law}  \end{equation}
where $J$ is the total angular momentum and $M$ is the total 
mass.  Applying this rotation law to polytropes, which do 
not have constant density, produces differentially rotating
models.  The rotation law~(\ref{rotn-law})
is used to calculate a rotational
potential, which is then used with the gravitational potential to
compute an improved density distribution.  This process is 
then repeated, iterating until convergence is achieved.

Rotation causes the resulting 
models to be flattened, so that
$R_{\rm p} < R_{\rm eq}$, where $R_{\rm p}$ is
the polar radius and $R_{\rm eq}$ is the equatorial radius.
The freely specifiable quantities in this method are the 
dimensionless form of the 
rotation law $h(m) = (M/J)j(m)$, $n$, $R_{\rm eq}$,
and the axis ratio $R_{\rm p}/R_{\rm eq}$.
Upon convergence to a solution of the equations of 
hydrodynamic equilibrium,
this procedure gives the density $\rho(\varpi,z)$, 
the angular velocity
$\Omega(\varpi)$, $M$, $J$, and $\beta$.
To get a dimensional model, we
specify the entropy, which is given by the
constant $K$ in the polytropic equation of 
state~(\ref{poly}), and the maximum density.

One measure of the accuracy of the initial equilibrium models 
comes from the virial theorem.  
For a fluid system this gives \cite{tassoul}
\begin{equation}
2T + W + 3\Pi = 0,
\label{virial_thm}
\end{equation}
where $T$ is the total kinetic energy and 
$\Pi = \int P dV$ is the volume integral of the
pressure.  Using this we define the virial relation VR by
\cite{hachisu86}
\begin{equation}
VR = \left | \frac{2T + W + 3\Pi}{W} \right |.
\label{virial}
\end{equation}

Using the SCF method, we generated initial models for the
cases $n = 1.5, 1.0$, and $0.5$; we refer to these as
SCF1, SCF2, and SCF3, respectively.  
The parameters of these
initial models are given in Table~\ref{table_SCF}.
Notice that a finer grid was used for the $n=1.5$ model SCF1;
this was done because it is more centrally condensed
than the other two cases.

\subsection{Generation of Particle Models}
\label{gen_particle}

Once the SCF equilibrium model has been produced on a cylindrical
grid, it must be transformed into a form readable by TREESPH.
This is done by performing a particle fit to the density
profile $\rho(\varpi,z)$ given by the SCF method.   
The simplest technique for doing this
 randomly distributes particles within the
probability distribution $\rho(\varpi,z)$ using a ``rejection'' 
method \cite{NumRec,CM}; this technique was used to generate
the initial conditions for the runs performed in Ref. \cite{PRL}. 
Due to the underlying
Poisson distribution of particles, 
these initial models contain
relatively large positive and negative density fluctuations.
The resulting internal noise causes spurious entropy 
generation and partially masks the signal of the bar mode
instability. 
Because of these problems, we developed
methods to produce ``colder''
initial particle models with less internal noise
\cite{comparison,jlh_phd}.

To generate these cold models, we
use equipotential surfaces to determine 
the function $M(\Phi)$, which is
the mass interior to an equipotential.  This information
then allows us to obtain
the desired physical properties of the SCF model by reordering the
particles from a physically simpler model.  
This simpler model is created
by placing particles within  the 
known stellar boundary to
obtain a uniform density particle distribution.  
Then, using a chosen set
of equipotential surfaces, the mass
interior  to these surfaces $M(\Phi)$
is computed in both the SCF and 
uniform density models.
The actual number of surfaces used is taken to equal
the number of zones in the $\varpi$ direction used to generate
the SCF model, $N_{pot}=N_{\varpi}$;
c.f. Table~\ref{table_SCF}.
By a direct comparison between the resulting SCF and uniform 
density mass
functions, a systematic contraction, or repositioning, of 
particle positions
from their original locations in the uniform density model
can be performed.  This results in a particle
model which realistically reproduces
the SCF density profile, and does not suffer from the density
fluctuations found in the random particle method
\cite{jlh_phd}.

To implement this procedure, 
we use the rotational ($\Phi_{rot}$) and
gravitational ($\Phi_{grav}$) potentials, which are
 natural by-products
of the SCF method, to define the total surface 
equipotential $\Phi_{surf}$.  A uniform 3-D Cartesian grid centered
on the star is then created inside a cube having length 
$2R_{\rm eq}$.  Particles are placed at each of the grid nodes,
and a particle is accepted into the model if it lies inside the
boundary of the star, producing a 3-D uniform
density particle representation with the exact 
physical shape of the SCF stellar model.
This uniform distribution must
now be transformed into the more centrally-condensed polytropic
model given by the SCF method.  
In practice we do this by systematically
``contracting'' the particle positions 
in the original
uniform density model
along radial vectors
(i.e.,  moving them toward the center).
Comparison of the mass functions 
$M(\Phi)$ for the two models tells how to move the
particles along radial rays to achieve the SCF density
distribution.
Since the SCF model is more centrally condensed than the 
uniform density distribution, the particles are moved
toward the stellar center to their new positions.
This repositioning of particles then reproduces the 
SCF density distribution. 
When the contraction process is completed, 
each particle is assigned
an angular velocity $\Omega$ to reproduce 
the rotation law given in Eq.~(\ref{rotn-law}).

Using this contraction method, we generated initial models
with $\beta \approx 0.3$ and $n = 1.5$ using 
particle numbers in the range
$N \sim 2000 - 32,000$. These SPH models form 
 the basis of the
comparison study with Eulerian methods reported in
Ref. \cite{comparison}.  Overall, these models do not suffer
from the large density fluctuations present in the randomly
generated models and, when evolved using TREESPH, better
reproduce the basic features of the bar mode instability.

To generate the initial models for the runs presented here,
we incorporated several improvements to this method.  
First of all, 
we added an iterative procedure to the contraction process.
The initial repositioning of
particles is identical to that presented above.  
However, once the
uniform density model has been contracted, 
the mass interior to the
equipotential surfaces is recalculated for the particle model.  
The radial
contraction is again applied to all particles, and the 
process is iterated
until the difference between the initial and final positions
of all particles 
is less than a given tolerance, here chosen to be
typically $\sim 0.5 N_{\varpi}^{-1}$.
We also modified 
the initial placement of the particles within the uniform
density model.  
A box with extent  $\pm R_{eq}$ in the 
$x$ and $y$ directions and $\pm R_p$ in the $z$ direction
 was used instead of a cubical box.
Squeezing the particles along the $z$ direction 
to fit within the range of $R_{\rm p}$ 
should initially position them
 closer to their final equilibrium 
positions, thus
making the contraction method more efficient.
Also, to eliminate systematic errors due to contraction, 
the particle 
planes of constant $z$ were displaced in the $\varpi$ direction
by $\pm 1/4$ of the interparticle
spacing for even and odd $z$-planes, respectively. 
Overall, this iterated contraction method produces particle
models that better reproduce the SCF density distribution. 

Using this iterated contraction procedure, we constructed 
three models with $\beta \approx 0.3$ 
using
 $N \sim 16000$ for
different values of the polytropic index $n$.  
We refer to these as 
Run 1 ($n=1.5$), Run 2 ($n=1.0$), and Run 3 ($n=0.5$); see 
Table~\ref{table_params}.
Fig.~\ref{6.20} shows the normalized equatorial plane
density for these initial models. Notice that the density values
calculated at the particle positions 
(shown as dots in the figure) match the
SCF profiles (shown as solid lines)
with very little scatter.  The 
angular velocity profiles for all particles are shown 
in Fig.~\ref{6.21} and also reproduce the SCF values with
very little scatter.

Rigorously, this repositioning of particles should be 
carried out by following normal, rather than radial, 
vectors.  We have used radial contraction here for simplicity
and computational speed.
  For spherical systems, the normal and radial
vectors coincide. However, as the equilibrium model becomes
increasingly oblate due to rotation, contracting the
particles along their radial vectors becomes less accurate.
In practice, we find the radial contraction method
models the equatorial plane density well, as shown in
Fig.~\ref{6.20}.  
However, under-densities are observed in the regions around the
rotation axis for increasing 
values of $|z|$.  When the models are evolved forward in
time, this causes a slight redistribution of mass and
angular momentum in the inner regions.
Based on comparisons of the $n=1.5$ singly contracted
models with Eulerian runs in Ref. \cite{comparison},
we do not believe that this small adjustment of the initial
models significantly affects the evolution of the bar mode 
instability.  The improved iterated contraction method
reduces this under-density somewhat for the $n=1.5$ case.
 As $n$ decreases, the polytrope becomes less centrally 
condensed and hence is closer to the uniform density model
prior to contraction.  As a result the under-densities
decrease for Run 2, and almost disappear for Run 3.

\section{Dynamical Evolution}
\label{evol}

The initial particle models for Runs 1, 2, and 3
generated using the iterated contraction
method were evolved in time
using TREESPH.  The case $n=1.5$ has already been 
the subject of our detailed comparison study using
Eulerian and SPH methods reported in Ref. \cite{comparison}.
Run 1 comprises the evolution of an improved initial
model with $n=1.5$ and shows the same general behavior seen
in that previous study; it is included here as a 
benchmark for comparison with the stiffer equations
of state.

The dynamical evolution
of these models is displayed visually in 
Figs.~\ref{7.5} -~\ref{8.4}.  Plots showing the particle
positions projected onto the equatorial plane 
are shown for Run 1 with $n=1.5$ in Fig.~\ref{7.5}, Run 2
with $n=1.0$ in Fig.~\ref{8.1} and Run 3 with $n=0.5$ in
Fig.~\ref{8.3}.  Corresponding contour plots covering the
same spatial area in the equatorial plane are shown for
Run 1 in Fig.~\ref{7.6}, Run 2 in Fig.~\ref{8.2}, and
Run 3 in Fig.~\ref{8.4}.  Time is measured in units of the
dynamical time $t_{\rm D}$ for a spherical star with radius
$R_{\rm eq}$ as defined in Eq.~(\ref{tD}); recall that 
these rapidly rotating models have significant 
rotational flattening.  All models are 
rotating in the counterclockwise direction.
Runs 1 and 2 were stopped at $t_{\rm f} = 35 t_{\rm D}$
and $t_{\rm f} = 50 t_{\rm D}$, respectively, by which times
these models had essentially stopped evolving.  Run 3 was
still evolving at $t_{\rm f} = 60 t_{\rm D}$
 when it was stopped because of the need to save
computer time.
Table~\ref{table_params} displays several important parameters
of these models.

The three models exhibit
certain basic features in common;
c.f. \cite{DT85,TDM,DGTB,PRL,comparison,WT88}.
Nonaxisymmetric structure 
grows spontaneously out of deviations from
axisymmetry, caused in these models 
by particle discreteness.
A global bar-shaped
structure develops in which the amplitude of the $m=2$ mode 
grows
exponentially in time; see Sec.~\ref{modes}.  
During the bar mode's growth, 
trailing spiral 
arms develop as mass is 
shed from both ends of the bar.  The bar and 
spiral arms exert
gravitational torques that cause angular momentum to 
be transported outward
from the core and lost from the ends of the bar.
The spiral arms expand supersonically, causing shock
heating and dissipation.
Careful examination of the contour plots shows that the cores
recontract toward an axisymmetric state after the initial growth
of the bar and ejection of spiral arms.  These systems
undergo two or more such episodes, depending on $n$; we will
describe this in more detail below.
Table~\ref{data-hydro} shows that the amounts of
mass and angular momentum in the cores at the ends of the
Runs 1 and 2 are very similar, as are the final
values of the stability parameter $\beta$.  Here we define
the core to contain material within cylindrical
 radius $\varpi = R_{\rm eq}$, 
where $R_{\rm eq}$ is the initial
equatorial radius. Since Run 3 was still evolving when 
the simulation was ended, we do not know what its final
values will be. Overall, a significant amount of
angular momentum is removed from the core by a relatively
small amount of mass.

Figure~\ref{8.8} displays the mass and angular momentum 
distributions for the three runs in the initial
[frames (a) and (b)] and final [frames (c) and
(d)] states.  Here, $m(\varpi)/M$ and
$J(\varpi)/J_0$  are the normalized 
mass and angular momentum within 
cylindrical radius $\varpi$, respectively. 
$M$ is the total mass and
$J_0$ is the initial total angular momentum.
Notice that at the final times
the curves for both the mass and angular
momentum distributions intersect near $\varpi \sim
R_{\rm eq}$.  This is consistent with the fact that
the models all shed roughly the same 
total amount of mass
and angular momentum, and have final cores with 
radii $\varpi \sim R_{\rm eq}$.

Figs.~\ref{7.5}(l) and~\ref{8.1}(u) show that 
the final states of both Runs 1
and 2 exhibit a flattened ``double halo'' structure,
consisting of a denser, inner region surrounded by a 
more diffuse,
extended outer distribution of matter.  
(We do not know if Run 3
will have the same double halo structure at the
conclusion of its evolution since the model was not
run long enough to reach its final state.)
This double halo may result
from differences in 
the angular momentum carried by the mass when it is shed.
During the first spiral arm ejection phase, 
the system has a higher
value of $\beta$ and the mass is shed from 
the ends of the bar with a
greater angular momentum.
This mass moves into the vacuum 
carrying a fraction of the initial
angular momentum of the system, and eventually 
distributes itself
about the remaining core to produce the outer halo.  
The inner halo is formed when the system
undergoes the second spiral arm ejection
episode at a lower 
value of $\beta$, and the mass is shed with
lower angular momentum into a region that already
has mass from the first episode.

For example,
Fig.~\ref{7.7} shows the mass $\Delta m/M$ and
angular momentum $\Delta J/J_0$ distributions for 
Run 1 at $t=13.4 t_D$ (solid line) and
$t=18.5 t_D$ (dashed line); these times correspond
to the first and second episodes of
mass shedding through spiral arms and are 
shown in 
the contour frames (e) and (i) in Fig.~\ref{7.6}.
Here $\Delta m$ is the amount of mass within a
cylindrical shell of thickness $d\varpi$ at
radius $\varpi$, and similarly for $\Delta J$.
Fig.~\ref{7.7}(a) and (b) show the first set
of spiral arms represented as a 
localized concentration of mass directly outside
the stellar core in the region $1 \lesssim
\varpi/R_{\rm eq} \lesssim
2$ at $t=13.4 t_D$.  As the model evolves, 
this mass expands into the surrounding
vacuum.  The amount of mass shed during the second
episode is much less than during the first;
c.f.\ Table~\ref{table-mass-shed}.  This can be
easily seen when we examine Fig.~\ref{7.7}(b), 
which zooms in on the ejected mass.
Fig.~\ref{7.7}(c) and (d) show 
that the material ejected from the primary
instability carries a higher amount of angular
 momentum than the mass
ejected after the second spiral arm
ejection phase. 

Comparison of Figs.~\ref{7.5} -~\ref{8.4} shows that 
several important model properties depend on the stiffness
of the equation of state. For example, 
the spatial deformation of the initially axisymmetric  
model into an
elongated bar-shaped figure
increases as $n$ decreases and the fluid description 
approaches that of an incompressible
fluid \cite{WT87}.
The widths of the spiral
arms and bar also depend on the polytropic index, both 
decreasing as $n$ decreases.  And, as already mentioned,
the system undergoes
more spiral arm ejection phases
as the equation of state stiffens.

Recall that for stiffer polytropes, the
density profiles become less centrally condensed, 
as shown in Fig.~\ref{6.20}.  
This greater amount of mass near the stellar boundary 
causes the material at the edge to be more tightly
bound.
Also, since the same rotation law Eq.~(\ref{rotn-law})
is applied to all runs, 
the less compressible models exhibit a smaller degree
of differential rotation. This can be seen by examining 
Figs.~\ref{6.21} and~\ref{final_omega}, which show,
respectively, the
angular velocities of the models at the initial and final
times.
The results of these effects can be clearly
seen in frame (c) of 
Figs.~\ref{7.5},~\ref{8.1}, and~\ref {8.3},
which are the particle position plots 
for Runs 1, 2 and 3, respectively.  Comparison of these
frames shows the mass at the stellar 
boundary becoming
less diffuse as the polytropic index decreases.
Overall, these effects 
contribute to the development of the less tightly 
wound spiral arm pattern found in models 
with lower polytropic indices \cite{WT87}.

The behavior of the stability parameter for these runs
is shown in Figure~\ref{8.6}.  The solid lines give
$\beta = T_{\rm rot}/|W|$ and the dashed lines show
$T/|W|$, where $T_{\rm rot}$ is the rotational kinetic energy
and $T$ is the total kinetic energy.
Comparison with Figs.~\ref{7.5} -~\ref{8.4} shows that $\beta$
decreases sharply from its initial value 
as the bar instability develops, dropping to a local 
minimum as the bar reaches its maximum elongation.
For stiffer polytropes, the temporal
location of the initial growth of the bar occurs later
in the evolution, with the minimum value of $\beta$ 
occurring at $\sim 13
t_D$, $17 t_D$, and $ 19 t_D$ for Runs 1, 2, and 3, 
respectively. 
When the results of Williams and Tohline\cite{WT88} are
converted to time measured in units of the dynamical time
$t_{\rm D}$, they also show that the instability reaches
a nonlinear amplitude later as $n$ is decreased.
Notice also that 
the value of the first local minimum of $\beta$, 
which corresponds
to the end of the first
spiral arm ejection phase, decreases with $n$; 
c.f. \cite{WT88}.  This behavior reflects the
fact that as $n$ decreases, the
maximum elongation of the central
bar-like region, and hence its moment of inertia, increases.
Assuming angular momentum conservation in the core, 
this causes the
minimum kinetic energy to decrease with $n$.

At the end of the first spiral arm ejection phase, 
the core recontracts and $\beta$
increases again.
As the models evolve forward in time, they undergo successive
periods of spiral arm ejection and core recontraction.
The number of these episodes increases as the
equation of state stiffens, with Run 1 showing 2 spiral
arm ejection phases and Run 2 showing 4. Run 3 undergoes
5 such episodes before the run was stopped; 
we expect it would
exhibit several more if allowed to run to later times. 
Table~\ref{table-mass-shed} shows the 
cumulative amount of mass and
angular momentum shed from the core ($\varpi \le R_{\rm eq}$)
after each spiral arm ejection phase. 
Notice that the cumulative amount of angular momentum
lost by the core after each spiral arm 
ejection episode decreases
 as the equation of state
stiffens.
Therefore, assuming that the cores conserve angular momentum
between periods of spiral arm ejection,
the stiffer cores recontract to a
higher angular velocity (and a larger $\beta$),
and deform to a greater elongation (and a smaller $\beta$)
than more compressible ones, as shown in
Fig.~\ref{8.6}.  Also, since the cores lose angular
momentum with each spiral arm ejection, 
each successive episode produces 
a smaller maximum and a larger minimum value of 
$\beta$.

Now consider
 Run 3, which undergoes 5 periods of core recontraction
corresponding to the local maxima of $\beta$ 
seen in Fig.~\ref{8.6}. 
Its contracted core, displayed in the contour plots in 
frames (g), (k), (o), (r), and (w) of Fig.~\ref{8.4}, 
shows a ``parallelogram-like''
structure.  This feature 
becomes stronger as $n$ is decreased, with
an intercomparison of frame (g) of 
Figs.~\ref{7.6}, ~\ref{8.2}, and~\ref{8.4}
showing the
emergence of this feature as we 
progress through the polytropic sequence
toward stiffer equations of state.
An explanation of this feature is as follows.
When core recontraction begins,
the ends of the bar move
toward the central regions.
A more compressible model is 
better able to
increase its central density in response to this
material forced toward the center.
 However, as $n$ decreases, the material in the center
cannot be easily compressed, thus forcing the fluid 
to move in a direction
perpendicular to the contracting bar.
This produces the observed parallelogram-like
structure, or ``anti-bar''.
We shall see in Sec.~\ref{modes} below that the
$m=4$ mode is also present in these simulations;
this may provide a degree of freedom that allows 
the formation of this feature.

Another important difference observed as
$n$ is changed concerns the long term 
behavior of the models.  The spiral arms in
Runs 1 and 2 eventually merge as the systems evolve, resulting
in a late time state consisting of a 
{\em nearly} axisymmetric
central remnant of 
extent $\sim R_{\rm eq}$ surrounded by a flattened double halo.
After a comparable period of time, the core of Run 3 was still
quite elongated and the spiral arms were just beginning to
merge. 
One explanation for this longer-lived elongation 
in the $n=0.5$ case is as follows.
Consider an equilibrium sequence of uniformly rotating
axisymmetric polytropes parametrized by
$\beta$.
As $\beta$ increases along such a sequence, a point is
eventually reached at which mass is lost at the equator.
Uniformly rotating polytropes with $n \ge 0.808$
($\Gamma \le 2.24$) reach this mass-shedding limit
before the point
at which ellipsoidal configurations can exist
 \cite{tassoul,LRS-APJS}.
Although Fig.~\ref{final_omega} shows that 
the central remnants in these runs 
are differentially rotating, we
believe that a similar mechanism may be operating here
(see also \cite{ZCM1,ZCM2}), causing
the cores of Runs 1 and 2 
to be nearly axisymmetric at the end of the run.  

One major difference observed between our work and that of
Tohline, Durisen
 and collaborators is the final outcome of the
simulations.
In all of our runs with $n=1.5$ and $n=1.0$, we find that
the systems evolve to a final state consisting of a
{\em nearly} axisymmetric central remnant surrounded by
an extended disk-like halo \cite{comparison,PRL}. 
This behavior was observed in both the $n=1.5$
Eulerian and SPH runs investigated in the comparison study
reported in Ref. \cite{comparison} as well as the
work reported here.  In contrast, all the long Eulerian
evolutions reported by these other researchers
(refs. \cite{PDD,TDM,DGTB,WT88}) 
resulted in a bar-like central core surrounded
by a ring of material.  (Interestingly, in the very low resolution
SPH run reported in Ref. \cite{DGTB}, the low-density material
did form an extended disk.)  Currently,
we do not understand the reason for
these differences in the final state.  One possible explanation
is that these other researchers
 do not evolve an energy equation,
and hence cannot model the shocks which occur in the outer, 
low density regions \cite{PDD}. 
 Clearly, this is an important
issue and efforts are underway to resolve
these questions.

\section{Analysis of Fourier Components}
\label{modes}

We quantify the development of the dynamical instability 
by studying the behavior of various Fourier components
in the density using
cylindrical coordinates $(\varpi,\phi,z)$.  The
density in a ring of fixed $\varpi$ and $z$ is analyzed 
using the complex Fourier series
\begin{equation}
        \rho(\varpi,\phi,z)=\sum_{m=-\infty}^{+\infty}
        C_m(\varpi,z) e^{im\phi} \,.
\end{equation}
The  azimuthal Fourier decomposition of the density 
distribution
for various components $m$ is expressed in 
terms of the amplitudes
$C_m$, defined by
\begin{equation}
        C_m(\varpi,z) = \frac{1}{2 \pi} \int_0^{2 \pi} \rho(\varpi,
        \phi,z) e^{-im\phi} d\phi.
\end{equation}
\cite{Powers,TDM}. 
The relative normalized amplitude is then defined
by
\begin{equation}
        |A_m| = |C_m|/|C_0|,
\end{equation}
where $C_0(\varpi,z) = \bar{\rho}(\varpi,z)$ is the 
mean density in the
ring under examination.  The integration is performed 
over the azimuthal
coordinate ($0 \leq \phi < 2\pi$) while $\varpi$
and $z$ remain fixed.  In
this way, the analysis can be carried out in 
``density rings'' for different values of 
$\varpi$ and $z$.
 
To apply this procedure to the SPH simulations, 
the particle model is
interpolated onto a cylindrical grid at pre-chosen 
time intervals
(typically every $0.01 t_D$) using kernel estimation
\cite{HK}. 
The grid used here
consists of $66$, $34$, and $16$ zones 
in the $\varpi$,
$\phi$, and $z$ directions, respectively.
Analysis of the density in rings at different values
of $\varpi$ in the equatorial plane ($z=0$) gives 
quantitative information about the
development of the global $m=2$ mode 
visually seen
in Figures~\ref{7.5} - \ref{8.4},
as well as other Fourier components that may be present.
 
Examining how the normalized amplitude changes in time
yields the growth rate $d \ln |A_m|/dt$
of the various Fourier components in our
models.  In practice, this
 is obtained by fitting a straight line 
through the data points in the
time interval during which the function $\ln |A_m|$
is linearly growing (thus giving an exponential
growth rate for $|A_m|$). The endpoints
of this time interval
are chosen ``by eye''.  
A clearly defined linear region typically 
lasts for a relatively short
time interval, and the value of the 
slope is sensitive to the endpoints defining 
the interval. 
 
Also, by examining the complex phase 
$\phi_m$ of a Fourier component, where
\begin{equation}
        \phi_m (\varpi,z) = \tan^{-1} 
\left[ \frac{\mbox{Im}(-C_m)}
        {\mbox{Re}(C_m)} \right] \,,
        \label{phase_m}
\end{equation}
we can describe global non-axisymmetric 
structure propagating in the
azimuthal direction. The development out of the 
initial noise of such a
global mode with a well-defined angular eigenfrequency 
allows us to write
\begin{equation}
        \phi_m = \sigma_m t,
\end{equation}
where $\phi_m$ is the phase angle of the disturbance, 
and $\sigma_m$ is the
eigenfrequency.  The relation between
the pattern speed 
$\Omega_{{\rm pat},m}$ of the $m^{th}$ 
structure and the phase
angle $\phi_m$ is
then \cite{WT87}
\begin{equation}
    \Omega_{{\rm pat},m}(\varpi,z) \equiv \frac{1}{m} 
\frac{d \phi_m}{d t} =
      \frac{\sigma_m}{m} \,.
        \label{pattern-speed}
\end{equation}
Notice that for the $m=2$ bar mode, the 
eigenfrequency $\sigma_2$
is twice the rotational speed of the bar, and the 
bar rotation  period is
$T_{\rm bar} = 2\pi m/\sigma_m = 4 \pi/ \sigma_2$.  
 
The eigenfrequency is thus 
obtained by a simple
calculation once the period is known.  
The period $T_m$ of the $m^{th}$
disturbance is determined
directly from the period of the 
cosine of the phase angle $\phi_m$ versus time.  
We use the function $\cos
\phi_m$ rather than $\phi_m$ itself 
due to the multi-valued nature
of the inverse trigonometric function 
$\arctan$ (Eq.~(\ref{phase_m})),
which would require us to artificially insert multiples
of $\pi$ in order to keep the function continuous.  
In practice, the
half-period is obtained by locating successive differences in
time between pairs of neighboring extrema of $\cos \phi_m$. 
Once the
half-period is known, the  eigenfrequency $\sigma_m$ 
can be obtained from
Eq.~(\ref{pattern-speed}).  
Overall, we find that the $m=2$ instability grows
on a time scale of approximately one bar rotation period,
as expected. 

The linearized tensor viral analysis (TVE) can also be used 
to calculate the bar
mode amplitude $|A_m|$ and phase angle $\phi_m$.  
Although this method is
exact only for small oscillations of uniform density, 
incompressible
ellipsoids \cite{chandra69}, it has proven a useful 
point of
comparison for numerical simulations when adapted to 
the study of rotating
compressible fluids \cite{tassoul,TDM,WT87}.  
Table~\ref{tve-n}
shows the TVE growth rates $d\ln |A_2|/dt$ and 
eigenfrequencies $\sigma_2$
for the $m=2$ mode for the cases
$n=1.5$ and $n=1.0$ with $\beta = 0.31$
reported by Williams and
Tohline in Ref. \cite{WT87}, 
where we have converted from their units.
Notice that as $n$ decreases, both the growth
rate and eigenfrequency also decrease.
We were unable to find TVE results in the
literature for 
$n=0.5$.

Fig.~\ref{modes-n=1.5} shows the amplitudes of the Fourier
components $m=1,2,3,$ and $4$
for Run 1
in the density ring $\varpi = 0.36 R_{\rm eq}$ in the 
equatorial plane $z=0$
as functions of time.  As expected, the
development of the $m=2$ disturbance dominates the initial
evolution, with the other components growing at later times.
Both the $m=2$ and $m=4$ components show an initial period
of exponential growth.  Since this takes place at various
cylindrical radii $\varpi$ throughout the model, we identify
these disturbances as global modes.  The initial peak in the
$m=2$ amplitude corresponds to 
the maximum elongation of the bar
and the minimum value of $\beta$. The 
 detailed structure after the
initial growth of the bar mode varies somewhat with $\varpi$,
as the density in various parts of the star fluctuates due to 
the complex motions involved in the contraction and 
re-expansion of the core.

The growth rate calculated for the $m=2$ and 
$m=4$ modes is rather sensitive
to the specific time interval over which the linear fit to 
$\ln|A_m|$ is performed.  For the $m=2$ mode
in Run 1, we typically
get $d\ln|A_2|/dt \sim 0.6 t_{\rm D}^{-1}$, and
$d\ln|A_4|/dt \sim 0.8 - 1 t_{\rm D}^{-1}$.  The calculation
of the eigenfrequencies is more robust, yielding
$\sigma_2 \sim 1.9 t_{\rm D}^{-1}$ and
$\sigma_4 \sim 3.8 t_{\rm D}^{-1}$.  Both modes reach their
peak amplitudes at about the same time, then drop to local
minima and grow again.  Since the pattern speeds of the 
these modes are nearly the same, 
$\Omega_{{\rm pat},2} \sim \Omega_{{\rm pat},4} \sim 0.95$,
this suggests that the $m=4$ mode is a harmonic of the
bar mode and not an independent mode \cite{WT87}.

We also see that the $m=1$ and $m=3$ Fourier components grow
somewhat, although not in the global and coherent fashion
exhibited by the $m=2$ and $m=4$ modes.  Recent work in the
area of star formation \cite{PDD,Bonnell} has highlighted
the importance of the $m=1$ case.

Fig.~\ref{modes-n=1.0} displays the amplitudes of the Fourier
components $m=1, 2, 3,$ and $4$
for Run 2 at the same value of $\varpi$
 used above.  Again, the $m=2$ and
$m=4$ components emerge as exponentially growing global modes,
with the bar mode dominating the early stages of the 
evolution.  The growth rates in this case are even more
sensitive to the time interval chosen for the linear fit than
was the case for Run 1.  We  find 
$d\ln|A_2|/dt \sim 0.5 - 0.8 t_{\rm D}^{-1}$ and
$d\ln|A_4|/dt \sim 0.9 - 1.3 t_{\rm D}^{-1}$. Again, the
eigenfrequencies are less dependent on the time interval
chosen and take the values $\sigma_2 \sim 1.5 t_{\rm D}^{-1}$
and $\sigma_4 \sim 2.9 t_{\rm D}^{-1}$.  The pattern speeds
are thus 
$\Omega_{{\rm pat},2} \sim \Omega_{{\rm pat},4} 
\sim 0.7 t_{\rm D}^{-1}$, implying
again that
the $m=4$ mode is a harmonic of the bar mode.

Finally, the amplitudes of the Fourier components 
$m=1, 2, 3,$ and $4$ for Run 3
are shown in Fig.~\ref{modes-n=0.5}. 
 The $m=1$ and $m=3$ Fourier 
components are stronger in this case, although they do not
appear to develop into global modes.  The $m=2$ 
and $m=4$ disturbances do develop into global modes and
appear to be more strongly coupled than before.  For example,
the initial exponential growth rate of the bar mode (in the
time interval $11.5 t_{\rm D}^{-1} \lesssim t \lesssim
15 t_{\rm D}^{-1}$) is 
$d\ln|A_2|/dt \sim 0.4 t_{\rm D}^{-1}$.  Then, the bar mode
growth rate increases sharply to 
$d\ln|A_2|/dt \sim 2.2 t_{\rm D}^{-1}$; this may due to 
coupling with the $m=4$ mode, which initially grows at the
rate $d\ln|A_4|/dt \sim 2.2 t_{\rm D}^{-1}$.  The eigenfrequencies
are $\sigma_2 \sim 1.2 t_{\rm D}^{-1}$ and
$\sigma_4 \sim 2.3 t_{\rm D}^{-1}$, so that
$\Omega_{{\rm pat},2} \sim \Omega_{{\rm pat},4}
\sim 0.6 t_{\rm D}^{-1}$.

Overall, the amount of structure seen in the Fourier components
increases as $n$ decreases.  This reflects the fact that the 
stiffer fluids show more internal fluctuations as the cores
expand and recontract.  The eigenfrequencies $\sigma_2$
do show the decrease with $n$ predicted by the TVE
analysis as given in Table~\ref{tve-n}.  
This is due to the fact that the stiffer equations of
state produce longer bars, which rotate more slowly.
However, it is
more difficult to assess the trends in the growth rates of the
bar mode. 
If we consider the
initial exponential 
growth period of the bar mode in Run 3, then
it does grow at a slower rate than in Run 1 until 
the $m=4$ mode starts to grow.  
 As the fluids become stiffer and the number of
spiral arm ejection episodes increases, the matter in the
cores oscillates more.  The coupling between the  
$m=2$ and $m=4$ also grows stronger; this may be linked to
the development of the anti-bar discussed in 
Sec.~\ref{evol}.  In particular, the elongation of the
anti-bar in Run 3 seen in Fig.~\ref{8.4} (g) at 
$t = 25.2 t_{\rm D}$ occurs at roughly 
the same time as the
second maximum in $\ln |A_4|$ shown in 
Fig.~\ref{modes-n=0.5}.  

\section{Gravitational Radiation}
\label{GW}

The time-changing quadrupole moment caused by the
development of the bar instability generates
gravitational waves.  The initial
development of the bar mode produces a
burst of radiation, followed
by a weaker signal due to the subsequent
expansions and recontractions of the core.
Overall, the gravitational wave 
signal lasts for a longer
time as the equation of state stiffens
and the systems undergo more episodes of spiral
arm ejection.
Some interesting properties of the gravitational
radiation produced by these models are given in
Table~\ref{data-GW}.

The gravitational waveform $rh_+$ for an observer
on the axis at $\theta=0, \phi=0$ of a 
spherical coordinate system centered on the source
is shown in Fig.~\ref{8.14} for these runs.  
Comparison of the waveforms with the contour plots in
Figs.~\ref{7.6}, ~\ref{8.2}, and~\ref{8.4} shows
that indeed the onset of the burst coincides with the
development of the primary instability.  
Notice that the maximum amplitude of the waveform
does not vary significantly with the equation of
state; see Table~\ref{data-GW}. As the 
core recontracts back to a more axisymmetric state,
the amplitude of the waveform decreases.  The
successive periods of spiral arm ejection and
core recontraction produce additional bursts
of gravitational waves; these have decreasing
amplitudes because the maximum elongation of the
core drops with each episode.  Runs 1 and 2
show the weak signal of a slightly non-axisymmetric
remnant in their final states, whereas
Run 3 shows the much stronger signal of its
still-evolving, more elongated core.  Also, as the
equation of state stiffens the frequency
of the waves decreases, since
the more elongated bars produced for smaller $n$
rotate more slowly.

The gravitational wave luminosity $L$ for these runs
is displayed in Fig.~\ref{8.15}.  Notice that the
peak amplitude of the luminosity decreases as $n$
decreases.  For a non-axisymmetric
object rotating rigidly about the
$z$ axis, the luminosity takes the form
\begin{equation}
	L = \frac{dE}{dt} =
-\frac{32}{5} \frac{G}{c^5}(I_x - I_y)\Omega^6,
\label{lum-ST}
\end{equation}
where $I_x$ and $I_y$ are the moments of inertia about the
$x$ and $y$ axes, respectively, and $\Omega$ is the
rotational angular velocity \cite{ST}.  
Since the term in $\Omega^6$ dominates, the peak
luminosity should
decrease as $n$ decreases and the 
central bar-like structures rotate more
slowly, as shown in Fig.~\ref{8.15}.  

It is interesting to examine the structure of the
luminosity profiles. The luminosity 
of Run 1 shows peaks at
$t \sim 13 t_{\rm D}$ and $t \sim 19 t_{\rm D}$;
these correspond to the
primary and secondary spiral arm ejection episodes.
Run 2 shows two closely spaced peaks at
$t \sim 15 t_{\rm D}$ and $t \sim 19 t_{\rm D}$,
followed by other peaks at
$t \sim 25 t_{\rm D}$ and $t \sim 33 t_{\rm D}$.
The first two peaks are associated with the
initial period of spiral arm ejection; 
c.f.\ Fig.~\ref{8.6} (b). The remaining two peaks
correspond to subsequent episodes.  The successively
smaller amplitudes reflect the fact
 that the angular velocity
of the core decreases as angular momentum is shed 
on each subsequent episode.
Finally, Fig.~\ref{8.15} (c) shows that Run 3
has two closely spaced luminosity peaks at 
$t \sim 17 t_{\rm D}$ and $t \sim 21 t_{\rm D}$,
which are again associated with the initial
burst.  The local miniumum in the luminosity at
$t \sim 25 t_{\rm D}$ occurs at the time of
core recontraction, as can be seen by comparing
with Figs.~\ref{8.4} (g) and~\ref{8.6} (c). At later
times it is more difficult to discern individual bursts in
the luminosity function, a trend that is also seen in the
waveform shown in Fig.~\ref{8.14} (c).

The energy emitted as gravitational waves $\Delta E/Mc^2$
is shown in Fig.~\ref{8.16}.  In the case of Runs 1 and 2,
$\Delta E/Mc^2$ grows due to the initial and secondary
bursts, and levels off when the cores reach their nearly
axisymmetric final states.  For Run 3, this quantity
grows almost linearly with time and has not yet
leveled off by the end of the simulation, indicating that
the core is still quite nonaxisymmetric.

Fig.~\ref{8.17a} shows the rate at which 
angular momentum is carried by the waves, $dJ_z/dt$.
As was the case with the luminosity, 
 we see structure in this quantity that corresponds
to periods of spiral arm ejection and core
recontraction.  The angular momentum $\Delta J_z$
carried by the gravitational waves is displayed
in Fig.~\ref{8.17b} and shows features similar to
those found in $\Delta E/M$.

Finally, the gravitational wave energy spectrum
$dE/df$ is displayed in Fig.~\ref{8.18} and shows
that the peak frequency 
of the gravitational radiation $f_{\rm grav}$ 
decreases
as the equation of state stiffens. 
Table~\ref{table-freq} shows that,
as expected, 
$2f_{\rm bar} \sim f_{\rm grav}$,
where $f_{\rm bar} = (1/2)\sigma_2/2\pi$
is the rotational frequency of the bar.
Notice, however, that the
rotational frequencies $2f_{\rm bar}$ are slightly
lower than $f_{\rm grav}$.  
This is due to the fact that, while the 
the eigenfrequency $\sigma_2$ is calculated 
only during the
initial development of the bar instability,
$f_{\rm grav}$ is computed for the entire 
evolution of the model and thus includes the higher
rotational velocities obtained 
when the cores
recontract.

\section{Summary and Discussion}
\label{discuss}

We have carried out numerical simulations of the dynamical
instability in rapidly rotating stars initially modeled as
polytropes with $n = 1.5, 1.0,$ and $0.5$. These calculations
have been done using a 3-D SPH code with $N\sim 16,000$ 
particles.  The code has a purely Newtonian gravitational
field, and the gravitational radiation is calculated in the
quadrupole approximation.  The back reaction of the
gravitational radiation is not included.

All models exhibit the growth of the global $m=2$ 
bar mode, with mass and angular momentum being shed
from the ends of the bar to form two trailing spiral
arms.  In general, as $n$ decreases the central bar
becomes narrower and more elongated.  Once the central
core has reached its maximum elongation, it begins to 
recontract toward a more axisymmetric state. This primary
instability is followed by successive episodes of 
spiral arm ejection and core recontraction, with the 
number of these episodes increasing for stiffer equations
of state.  At the end of the simulations, the models with
$n=1.5$ and $n=1.0$ have settled into a state with a nearly
axisymmetric core of radius $\sim R_{\rm eq}$, where 
$R_{\rm eq}$ is the initial equatorial radius,
surrounded by a flattened disk-like halo that contains
$\sim 10\%$ of the total mass and $\sim 30\%$ of the
total angular momentum.
Since these models have $\beta_{\rm s} < \beta 
< \beta_{\rm d}$, they are expected to continue evolving
under the secular instability \cite{LS-nascent}.
 The model with $n=0.5$ had a 
fairly elongated core and was still evolving when that
run was terminated.

The development of the instability produces a burst of
gravitational radiation.  The maximum amplitude of the 
waveform $r|h|$ does not vary significantly with the
polytropic index, whereas the frequency of the waves 
decreases somewhat as $n$ decreases.  This lowering of
the frequency with $n$ reflects the fact that the stiffer
polytropes produce more elongated bars, which rotate more
slowly; it also results in a decrease in the peak
gravitational wave luminosity with $n$.   Since the
stiffer models undergo more episodes of spiral arm
ejection and core recontraction, they produce longer-lived
gravitational wave signals from the dynamical instability,  
with the total amount of
energy and angular momentum emitted in the form of
gravitational radiation increasing as $n$ decreases.
The nearly axisymmetric final cores 
(for $n=1.5$ and $n=1.0$) will continue
to emit gravitational radiation as they evolve under the
secular instability; this has been calculated by
Lai and Shapiro \cite{LS-nascent}.

The actual values of the gravitational wave quantities
depend sensitively on the equatorial radius 
$R_{\rm eq}$ of the stellar
core when the dynamical instability takes
place.  This in turn depends on the 
astrophysical scenario in which the instability develops.
Consider, for example, the collapse of a rotating stellar
core of mass $M = 1.4 {\rm M}_{\odot}$ that has
$\beta > \beta_{\rm d}$ and is prevented from collapsing
further due to centrifugal forces.  The equatorial radius
$R_{\rm eq}$ of the core at which this centrifugal
hangup occurs determines the amplitude and frequency of 
the resulting gravitational radiation.
The simulations presented here use stiff equations of
state, which are appropriate only for stellar cores 
that have collapsed to near neutron star densities.
We therefore calculated the gravitational wave
amplitudes and frequencies from our models
 for two representative 
values of this parameter, 
$R_{\rm eq} = 10$ km and $R_{\rm eq} = 20$ km.
We remind the reader that since these simulations have been
done in the Newtonian limit, which breaks
down for $R_{\rm eq} \sim 10 - 20$ km,
these results must be viewed with appropriate
caution.

Table~\ref{table-h_scaled} shows the maximum amplitudes
$|h|$ of the gravitational waveforms and the characteristic
frequencies $f_{\rm grav}$ for these representative values
of $R_{\rm eq}$.  Wave amplitudes are given for sources
within the Milky Way ($r = 15$ kpc), the Local Group
($r = 1$ Mpc), and the Virgo Cluster ($r = 20$ Mpc).
 If the dynamical instability
occurs at $R_{\rm eq} \sim 20$ km, which is about twice
the typical neutron star radius, 
$f_{\rm grav}$ lies just outside the
frequency range of the broad-band interferometers
\cite{thorne96a}.
However, if such objects exist, they may potentially
be observed using specially designed narrow-band
interferometers \cite{narrow,KLT} or resonant detectors
\cite{resonant,TIGA}.  Of course, if hangup occurs 
at about the typical neutron star radius of
$R_{\rm eq} \sim 10$ km, the characteristic frequencies
become much larger. Since the star must be rotating 
differentially
to achieve $\beta > \beta_{\rm d}$, this last scenario
could only occur in a newly-formed neutron star
before its rotation becomes uniform (cf. \cite{ST}).

The maximum luminosity $L/L_0$, 
the energy emitted as gravitational radiation
$\Delta E/Mc^2$, and the angular momentum carried by the
waves $\Delta J/J_0$ is given for a core with 
mass $M = 1.4 {\rm M}_{\odot}$ and these same 
representative values of $R_{\rm eq}$ in 
Table~\ref{table-GW_scaled}.  Here, $L_0 = c^5/G$
and $J_0$ is the initial total angular momentum.
The largest integrated energy and angular momentum
losses are 
produced by the model with $n=0.5$.  Since this
model was still evolving when this run was stopped,
the final values will be larger.

There are several ways in which these calculations need
to be improved to provide greater understanding of
gravitational radiation from rotational instabilities.
In this paper, we have concentrated on models with 
stiff equations of state.  
As noted above, these models
are relevant for cores that have
already collapsed to radii near the typical neutron
star radius, $R_{\rm eq} \sim 10 - 20$ km.  However,
if centrifugal hangup occurs at $R_{\rm eq} \sim
100$ km the equation of state is expected to be
much
softer, with $n \gtrsim 3$.  Simulations 
of this
important case are currently in progress; these
are being done using Eulerian techniques since we
have found it easier to model the softer equations
of state in this manner \cite{CGS}.
Also, the very important and interesting question of
the final state of the objects following the
dynamical instability still remains to be fully
resolved.  In addition to longer runs with $n=0.5$,
this will involve a more detailed understanding of
the differences between our results and those of
Tohline, Durisen, and collaborators; we 
are making plans to 
pursue answers to these questions.  Finally, 
gravitational radiation reaction and other
general relativistic effects need to be included
in order to have good physical models for comparison
with future observations.
We intend to include these effects in our future work.

\acknowledgments
We are pleased to acknowledge interesting and
helpful conversations with Steven Cranmer,
Dong Lai, Stephen McMillan, and Scott Smith.
We are also grateful to  
Lars Hernquist for supplying a copy of TREESPH.
This work was supported in part by NSF grant PHY-9208914.
 The numerical simulations
were run at the Pittsburgh Supercomputing Center under
grant PHY910018P.

\newpage 

\begin{table}[p]
\begin{center}
\begin{tabular}{cccccccc}
Model  & $n$ & $N_{\varpi}$ & $N_z$ & $N_{it}$ &
$\beta$ &  $R_{\rm p}/R_{\rm eq}$ & $VR$  \\
\tableline
SCF1 & 1.5 & 225 & 225 & 20 &
0.30 & 0.20 &  $7.6 \times 10^{-5}$  \\
SCF2 & 1.0 & 129 &  70 & 16 &
0.30 & 0.23 &  $4.5 \times 10^{-5}$ \\
SCF3 & 0.5 & 129 &  70 & 22 &
0.30 & 0.25 &  $4.0 \times 10^{-5}$ \\
\end{tabular}
\end{center}
\caption{Properties of the initial axisymmetric equilibrium models
created using the SCF method.  $N_{\varpi}$ and $N_z$ are the
number of uniform grid zones in the $\varpi$ and $z$ directions, 
respectively. $N_{it}$ is the number of
iterations required for convergence to a solution with a tolerance of
$10^{-5}$. The axis ratio is $R_{\rm p}/R_{\rm eq}$.
The value of the virial parameter calculated on the
cylindrical grid is $VR$.
\label{table_SCF}}
\end{table}

\begin{table}
\begin{center}
\begin{tabular}{ccccccccc}
Model & $n$ & $N$ & $VR|_{\rm i}$ & $\beta_{\rm i}$ & time 
    & $ \left | \frac{E_{\rm i} - E_{\rm f}}{E_{\rm i}}\right| $
    & $\left | \frac{J_{\rm i} - J_{\rm f}}{J_{\rm i}} \right| $
    & CPU \\ 
    & & & & & [$t_D$] & & & (hr) \\
\tableline
Run 1 & 1.5 & 16096 & $4.5 \times 10^{-2}$ & 0.32 
& 35 & 0.020 & $\leq .001$ 
     	& 18.8 \\
Run 2 & 1.0 & 16619 & $3.7 \times 10^{-2}$ & 0.32
& 50 & 0.015 & .002 & 
	25.2 \\
Run 3 & 0.5 & 16526 & $1.5 \times 10^{-4}$ & 0.31
& 60 & 0.018 & .003 & 
	29.0 \\
\end{tabular}
\end{center}
\caption{Properties of the SPH models.
$N$ is the total number of particles 
in each model.  The fact that the
method used to generate the initial particle models 
does not allow strict control over the the number of
particles accepted into each model results in
somewhat unusual values of $N$. 
The subscripts ``i'' and ``f'' denote
the initial and final states of the model, respectively.
The  stability parameter of the particle model at the
initial time is $\beta_{\rm i}$.  
The duration of the run in units of the dynamical time
$t_{\rm D}$ is given in the column labled
``time''.
$E$ is the total energy, and $J$ is the
total angular momentum.  
All models were run on a Cray C90; the amount of
CPU time used is given for the  duration of the run.
\label{table_params}}
\end{table}

\begin{table}
\begin{center}
\begin{tabular}{ccccc}
Model 
& $M_{\rm core, f}$ & $J_{\rm core, f}$ & $\beta_{\rm core, f}$
     & $\beta_{\rm f}$ \\
 &  $[\%]$ & $[\%]$ &  &  \\
\tableline
Run 1 & 90 & 70 & 0.24 & 0.26 \\
Run 2 & 91 & 70 & 0.24 & 0.25 \\
Run 3 & 92 & 71 & 0.23 & 0.24 \\
\end{tabular}
\end{center}
\caption{Hydrodynamical results for the 
models.  
The core refers to matter
within cylindrical radius $\varpi = R_{\rm eq}$, 
where $R_{\rm eq}$ is the initial
equatorial radius, and the subscript ``f'' denotes the
final state of the model. 
\label{data-hydro}}
\end{table}

\begin{table}
\begin{center}
\begin{tabular}{ccccc}
Run  & $t$     & $M_{\rm shed}$ & $J_{\rm shed}$ 
& $\beta$ \\
   & [$t_D$] & $[\%]$     & $[\%]$     &   \\
\tableline
Run 1  &  0 & 0.0 & 0.0 & 0.32 \\
       & 17 & 7.3 & 24 & 0.27 \\
       & 23 & 8.6 & 28 & 0.26 \\
       & 35 & 9.5 & 30 & 0.26 \\
\tableline
Run 2  &  0 & 0.0 &  0.0 & 0.32 \\
       & 22 & 5.0 & 18 & 0.28 \\
       & 30 & 7.0 & 25 & 0.26 \\
       & 36 & 7.9 & 27 & 0.26  \\
       & 42 & 8.5 & 29 & 0.25 \\
       & 50 & 9.1 & 30 & 0.25 \\
\tableline
Run 3  &  0 & 0.0 &  0.0 & 0.31 \\
       & 25 & 4.1 & 12 & 0.29 \\
       & 35 & 6.2 & 19 & 0.28 \\
       & 43 & 6.7 & 22 & 0.26 \\
       & 50 & 7.1 & 24 & 0.26 \\
       & 58 & 7.4 & 26 & 0.25 \\
       & 60 & 8.4 & 29 & 0.24 \\
\end{tabular}
\end{center}
\caption{Properties of the models after each successive 
spiral arm ejection phase.
The mass $M_{\rm shed}$ and angular momentum $J_{\rm shed}$
shown are the {\em cumulative} mass and angular momentum lost after
each such episode. 
The core is defined as mass within $\varpi = R_{\rm eq}$;
see Fig.~\protect\ref{8.8}.  
The values of $\beta$
are obtained directly from the successive peaks corresponding 
to core recontraction in Fig.~\protect\ref{8.6}. The last
temporal point in each series corresponds to the end of the
run, and is not necessarily a time at which the 
core has reached maximum recontraction.
\label{table-mass-shed}}
\end{table} 

\begin{table}
\begin{center}
\begin{tabular}{ccc}
$n$  & $d \ln |A_2|/dt \mid_{\rm TVE}$ 
& $\sigma_2 \mid_{\rm TVE}$  \\
 & [$t_{\rm D}^{-1}$] & [$t_{\rm D}^{-1}$]  \\
\tableline
1.5  &  0.73 &  1.7   \\
1.0  &  0.56 &  1.5   \\
\end{tabular}
\end{center}
\caption{TVE bar mode
growth rates $d \ln |A_2|/dt$ and 
eigenfrequencies $\sigma_2$
for $n=1.5$ and $n=1.0$  These values are taken from
Ref.~\protect\cite{WT87}, where we have converted from 
their units.}
\label{tve-n}
\end{table}

\begin{table}
\begin{center}
\begin{tabular}{cccccc}
Model & max $|rh|$ & max $L/L_0$ &
 $(\Delta E/Mc^2)_{\rm f}$
 & max $dJ_z/dt$ &
    $(\Delta J_z/J_0)_{\rm f}$ \\
\tableline
Run 1 & 0.57 & 0.13 & 0.87 & 0.066 & 1.02 \\
Run 2 & 0.58 & 0.091 & 1.1 & 0.057 & 1.53  \\
Run 3 & 0.58 & 0.059 & 2.2 & 0.044 & 3.41  \\
\end{tabular}
\end{center}
\caption{Gravitational wave results for Runs 1, 2, and 3.
The peak values of $|rh|$, $L/L_0$, and
$dJ_z/dt$ throughout the run,
 and the final (cumulative) values of
$(\Delta E/Mc^2)$ and 
$(\Delta J_z/J_0)$ are given.
$L_0 = c^5/G$
and $J_0$ is the initial total angular momentum.
To obtain dimensional quantities, the scalings
given in the axis
labels of the corresponding 
Figs.~\protect\ref{8.14} - ~\protect\ref{8.17b}
must be applied; see 
Tables~\protect\ref{table-h_scaled} 
and~\protect\ref{table-GW_scaled}.
\label{data-GW}}
\end{table}

\begin{table}
\begin{center}
\begin{tabular}{ccc}
Model & $2f_{\rm bar}$ & $f_{\rm grav}$ \\
  & [$t_{\rm D}^{-1}$] &  [$t_{\rm D}^{-1}$] \\
\tableline
Run 1 & 0.30 & 0.32 \\
Run 2 & 0.24 & 0.30 \\
Run 3 & 0.19 & 0.23 \\
\end{tabular}
\end{center}
\caption{Frequencies for the models.
$f_{\rm bar}$ is the rotational frequency of the bar
and is calculated from the eigenfrequency
$\sigma_2$. $f_{\rm grav}$ is obtained from 
the gravitational wave energy spectrum
$dE/df$ shown in Fig.~\protect\ref{8.18}.
\label{table-freq}}
\end{table}

\begin{table}
\begin{center}
\begin{tabular}{ccccccc}
$R_{\rm eq}$ & max $|h|_{\rm MW}$ 
&  max $|h|_{\rm LG}$ 
& max $|h|_{\rm VC}$ & $f_{\rm grav}$ 
& $f_{\rm grav}$ & $f_{\rm grav}$ \\
 & ($r=15$ kpc) & ($r = 1$ Mpc) 
& ($r = 20$ Mpc) & ($n=1.5$) & ($n=1.0$) 
& ($n=0.5$) \\
\tableline
10 km & $5 \times 10^{-19}$ 
& $8 \times 10^{-21}$ & $4 \times 10^{-22}$
& 4900 Hz & 4100 Hz & 3100 Hz \\
20 km & $3 \times 10^{-19}$ 
& $4 \times 10^{-21}$ & $2 \times 10^{-22}$
& 1700 Hz & 1400 Hz & 1100 Hz \\
\end{tabular}
\end{center}
\caption{The maximum amplitudes of the gravitational
waveform $|h|$ and the characteristic frequencies
$f_{\rm grav}$ are given for two representative 
values of the equatorial radius $R_{\rm eq}$.
The core is taken to have mass 
$M = 1.4 {\rm M}_{\odot}$.  
The waveform amplitudes $|h|$ are given for 
sources located within the Milky Way
($r=15$ kpc), the Local Group ($r = 1$ Mpc),
and the Virgo Cluster ($r = 20$ Mpc).
Notice that $|h|$ is
essentially independent of the polytropic index $n$.
These values were obtained by applying the 
appropriate scalings to the data given in
Tables~\protect\ref{data-GW} 
and~\protect\ref{table-freq}.
\label{table-h_scaled}}
\end{table}

\begin{table}
\begin{center}
\begin{tabular}{cccc}
$R_{\rm eq}$ & max $L/L_0$ & $\Delta E/Mc^2$ 
& $\Delta J/J_0$ \\
\tableline
10 km & 
$2 - 5 \times 10^{-5}$ & $4 - 9 \times 10^{-3}$
& $2 - 7 \times 10^{-2}$ \\
20 km & 
$6 - 20 \times 10^{-7}$ & $4 - 8 \times 10^{-4}$
& $4 - 10 \times 10^{-3}$ \\
\end{tabular}
\end{center}
\caption{The maximum luminosity $L/L_0$,
the energy emitted as gravitational radiation
$\Delta E/Mc^2$, and the angular momentum carried by
the waves $\Delta J/J_0$ are given for two representative
values of the equatorial radius $R_{\rm eq}$.
$L_0 = c^5/G$
and $J_0$ is the initial total angular momentum.
The core is taken to have mass 
$M = 1.4 {\rm M}_{\odot}$.  The lower and upper limits for
$L/L_0$ are produced for $n=0.5$ and $n=1.5$,
respectively; the values for $n=1.0$ are between these
two limits.  However, the lower values of 
$\Delta E/Mc^2$ and $\Delta J/J_0$ correspond to the
case $n=1.5$. The larger values are produced by 
the model with $n=0.5$; since this model was still evolving
when it was stopped, these values will be larger
once the model reaches its final state.
These values were obtained by applying the 
appropriate scalings to the data given in
Table~\protect\ref{data-GW}.
\label{table-GW_scaled}}
\end{table}

\clearpage

\begin{figure}[p]
\caption{The normalized equatorial plane  
density is shown
for the iterated contraction 
initial models Run 1 ($n=1.5$), 
Run 2 ($n=1.0$), and Run 3 ($n=0.5$).   
Here, the
equatorial plane is taken to include all particles 
within $z=\pm 0.01 R_{eq}$.
The solid curve in each frame
represents the SCF equatorial plane density, and
$\rho_c$ is the the central SCF
density. 
\label{6.20}}
\end{figure}

\begin{figure}[p]
\caption{The normalized angular
velocity is shown for the initial models of Runs
1, 2 and 3. All particles are plotted 
in this figure.  In each frame, 
the solid curve gives the angular
velocity for the corresponding SCF initial
model and $\Omega_c$ is the SCF central angular
velocity.
\label{6.21}}
\end{figure}

\begin{figure}[p]
\caption{Particle positions are shown projected onto the
equatorial plane for various times during the evolution of
Run 1 with $n=1.5$.  All particles are plotted.  The vertical
axis is $y/R_{\rm eq}$ and the horizontal axis is
$x/R_{\rm eq}$. The system rotates in the counterclockwise
direction.
\label{7.5}}
\end{figure}

\begin{figure}[p]
\caption{Density contours in the equatorial plane are shown
for Run 1 with $n = 1.5$.  The frames are taken at the same
times as the corresponding particle plots in 
Fig.~\protect\ref{7.5}.  The contour levels are the same
in all frames, and are spaced a factor of 10 apart, going 
down 4 decades below the maximum (central) SCF initial
density.  
The contours were calculated using kernel 
interpolation on a $100\times100$
Cartesian grid covering the same spatial area in the 
equatorial plane as the
frames in Figure~\protect\ref{7.5}.
\label{7.6}}
\end{figure}

\begin{figure}[p]
\caption{Same as Fig.~\protect\ref{7.5} for Run 2 with
$n=1.0$.
\label{8.1}}
\end{figure}

\begin{figure}[p]
\caption{Same as Fig.~\protect\ref{7.6} for Run 2 with
$n=1$.  These frames correspond to the particle plots 
shown in Fig.~\protect\ref{8.1}.
\label{8.2}}
\end{figure}

\begin{figure}[p]
\caption{Same as Fig.~\protect\ref{7.5} for Run 3 with
$n=0.5$.
\label{8.3}}
\end{figure}

\begin{figure}[p]
\caption{Same as Fig.~\protect\ref{7.6} for Run 3 with
$n=0.5$.  These frames correspond to the particle plots 
shown in Fig.~\protect\ref{8.3}.
\label{8.4}}
\end{figure}

\begin{figure}[p]
\caption{The distributions of mass $m(\varpi)/M$
and angular momentum $J(\varpi)/J_0$
are shown for Run 1 (solid line), Run 2 (dashed line), 
and Run 3 (dot-dashed line).  Frames (a) and (b) show the
initial models, and frames (c) and (d) show the final
states. Here, $M$ is the total mass and $J_0$ is the
initial total angular momentum.
\label{8.8}}
\end{figure}

\begin{figure}[p]
\caption{The mass $\Delta m/M$ and angular momentum
$\Delta J/J_0$ distributions are shown for Run 1.
Here, $\Delta m$ is the mass within a cylindrical shell of
thickness $d\varpi$ at radius $\varpi$, and 
similarly for $\Delta J$; $J_0$ is the
total
initial angular momentum. The solid lines show the values
at time $t = 13.4 t_{\rm D}$ and the dashed lines at
$t = 18.5 t_{\rm D}$.  Frames (b) and (d) show enlargements
of (a) and (c), respectively.
\label{7.7}}
\end{figure}

\begin{figure}[p]
\caption{The normalized angular velocity is shown for the
final states of Runs 1, 2, and 3.
Here, $\Omega_c$ is the central angular velocity for the
initial SCF model. All particles are plotted.
\label{final_omega}}
\end{figure}

\begin{figure}[p]
\caption{The behavior of the stability parameter
$\beta = T_{\rm rot}/|W|$ (solid line) and 
$T_{\rm tot}/|W|$ (dashed line) is shown as a function 
of time for Runs 1, 2, and 3.  Here,   
$T_{\rm rot}$ is the rotational kinetic energy,
$T_{\rm tot}$ is the total kinetic energy, and
$W$ is the gravitational potential energy.
\label{8.6}}
\end{figure}

\begin{figure}[p]
\caption{The growth of the Fourier components 
$m=1, 2, 3,$ and $4$
for Run 1 with $n=1.5$.  These values were
obtained in the density ring at $\varpi = 0.36 R_{\rm eq}$
in the equatorial plane $z=0$.
\label{modes-n=1.5}}
\end{figure}

\begin{figure}[p]
\caption{Same as Fig.~\protect\ref{modes-n=1.5}
for Run 2 with $n=1.0$.
\label{modes-n=1.0}}
\end{figure}

\begin{figure}[p]
\caption{Same as Fig.~\protect\ref{modes-n=1.5}
for Run 3 with $n=0.5$.  
\label{modes-n=0.5}}
\end{figure}

\begin{figure}[p]
\caption{The gravitational waveform $rh_+$ for an observer
located at $\theta = \phi = 0$ at distance $r$ from the
source for Runs 1, 2, and 3.
\label{8.14}}
\end{figure}

\begin{figure}[p]
\caption{The gravitational wave luminosity $L/L_0$ for 
Runs 1, 2, and 3.  Here, $L_0 = c^5/G$.
\label{8.15}}
\end{figure}

\begin{figure}[p]
\caption{The energy $\Delta E/Mc^2$ emitted as gravitational
radiation for Runs 1, 2, and 3.
\label{8.16}}
\end{figure}

\begin{figure}[p]
\caption{The rate $dJ_z/dt$ at which angular momentum is carried
away by gravitational radiation for Runs 1, 2, and 3.
\label{8.17a}}
\end{figure}

\begin{figure}[p]
\caption{The angular momentum $\Delta J/J_0$ carried by the 
gravitational waves for Runs 1, 2, and 3.  
Here, $J_0$ is the total initial 
angular momentum. 
\label{8.17b}}
\end{figure}

\begin{figure}[p]
\caption{The gravitational wave energy spectrum $dE/df$ is
shown as a function of frequency $f$
for Runs 1, 2, and 3.
\label{8.18}}
\end{figure}


\begin{references}

\bibitem{LIGO92}A. Abramovici, {\it et al.}, 
Science {\bf 256}, 325 
(1992).

\bibitem{VIRGO} C. Bradaschia, {\it et al.}, 
Nucl. Instrum. Methods
A {\bf 289}, 518 (1990).

\bibitem{GEO} K. Danzmann, {\it et al.}, in
 {\it Relativistic Gravity
Research}, Proceedings of the 81WE-Heraus-Seminar,
 Bad Hannef, Germany,
edited by J. Ehlers and G. Sch\"{a}fer
 (Springer-Verlag, Berlin, 1992).

\bibitem{resonant}
W. Johnson and S. Merkowitz, Phys. Rev. Lett. {\bf 70}, 2367
(1993)

\bibitem{TIGA}
G. Harry, T. Stevenson, and H. Paik, Phys. Rev. D
{\bf 54}, 2409 (1996).

\bibitem{schutz86}B. Schutz, in {\it Dynamical Spacetimes and Numerical
Relativity}, edited by J. Centrella (Cambridge University Press,
New York, 1986).

\bibitem{thorne96a}K. Thorne, in
{\it  Compact Stars in Binaries}, Proceedings of IAU Symposium 165,
edited by J. van Paradijs, E. van den Heuvel, and E. Kuulkers
(Kluwer, Dordrecht, 1996).  

\bibitem{thorne96b} K. Thorne, in 
{\it  Proceedings of the
Snowmass 95 Summer Study on Particle and Nuclear Astrophysics and
Cosmology}, eds. E. W. Kolb and R. Peccei (World Scientific,
Singapore, 1995); also published in  {\it Particle
Physics, Astrophysics \& Cosmology},  Proceedings of the
SLAC Summer Institute on Particle Physics, 
eds.\ Jennifer Chan \& Lilian
DePorcel (SLAC--Report--484, Stanford Linear Accelerator Center,
Stanford, CA, 1996). 

\bibitem{wagoner}
R. Wagoner, Astrophys. J. {\bf 278}, 345 (1984).

\bibitem{schutz89}
B. Schutz, Class. Quantum Gravity {\bf 6}, 1761 (1989).

\bibitem{tassoul}J. Tassoul, {\it Theory of Rotating Stars}
(Princeton University Press, Princeton, 1978).

\bibitem{ST}S. Shapiro and S. Teukolsky, {\it Black Holes, White
Dwarfs, and Neutron Stars} (Wiley, New York, 1983).

\bibitem{DT85}R. Durisen and J. Tohline, in {\it Protostars and
Planets II},
edited by D. Black and M. Matthews (Univ. of Arizona Press, Tucson,
1985).

\bibitem{managan85}
R. Managan, R. Astrophys. J. {\bf 294}, 463 (1985).

\bibitem{IFD}
J. Imamura, J. Friedman, and R. Durisen,
Astrophys. J. {\bf 294}, 474 (1985).

\bibitem{ITDPY}
J. Imamura, J. Toman, R. Durisen, B. Pickett, and S. Yang,
Astrophys. J. {\bf 444}, 363 (1995).

\bibitem{PDD}
B. Pickett, R. Durisen, and G. Davis, Astrophys. J.
{\bf 458}, 714 (1996).

\bibitem{LRS-APJS}D. Lai, F. Rasio, and S. Shapiro, 
 Astrophys. J. Suppl. {\bf 88}, 205 (1993).

\bibitem{TDM}J. Tohline, R. Durisen, and M. McCollough, Astrophys.
 J. {\bf 298}, 234 (1985).

\bibitem{DGTB}R. Durisen, R. Gingold, J. Tohline, and A. Boss,
Astrophys. J. {\bf 305}, 281 (1986).

\bibitem{WT87}H. Williams and J. Tohline, Astrophys. J. {\bf 315},
 594 (1987).

\bibitem{PRL}J. Houser, J. Centrella, and S. Smith,
Phys. Rev. Lett. {\bf 72}, 1314 (1994).

\bibitem{new_phd} K. New, Ph.D. Thesis, Louisiana State
University (1996).

\bibitem{comparison} S. Smith, J. Houser, and J. Centrella,
Astrophys. J. {\bf 458}, 236 (1996).

\bibitem{jlh_phd}J. Houser, Ph.D. Thesis, Drexel University (1996).

\bibitem{WT88}H. Williams and J. Tohline, Astrophys. J. 
{\bf 334}, 449 (1988).

\bibitem{CM}J. Centrella and S. McMillan, Astrophys. J.
{\bf 416}, 719 (1993).

\bibitem{SPH}L. Lucy, Astron. J. {\bf 82},
 1013 (1977); R. Gingold 
and J. Monaghan, Mon. Not. R. Astron. Soc. 
{\bf 181}, 375 (1977); see
J. Monaghan, Ann. Rev. Astron. Astrophys. 
{\bf 30}, 543 (1992) for a
review.

\bibitem{HK}L. Hernquist and N. Katz, Astrophys. 
J. Suppl. {\bf 70}, 
419 (1989).

\bibitem{tree}J. Barnes and P. Hut, Nature {\bf324}, 446 (1986);
L. Hernquist, Astrophys. J. Suppl. {\bf 64}, 715 (1987).

\bibitem{MTW}C. Misner, K. Thorne, and J. Wheeler {\it Gravitation} 
(Freeman, New York, 1973).

\bibitem{thorne87}K. Thorne, 
in {\it 300 Years of Gravitation}, edited by S.\ Hawking and W.\
Israel (Cambridge University Press, New York, 1987).

\bibitem{FE}L. S. Finn and C. Evans, Astrophys. J. {\bf 351},
588 (1990).

\bibitem{SC}S. Smith and J. Centrella, in {\it Approaches to
Numerical Relativity}, edited by R.\ d'Inverno (Cambridge
University Press, New York, 1992).

\bibitem{OM}J. Ostriker and J. Mark, Astrophys. J. {\bf 151},
 1075 (1968).

\bibitem{BO}P. Bodenheimer and J. Ostriker, Astrophys. J.
{\bf 180}, 159 (1973).

\bibitem{hachisu86}I. Hachisu, Astrophys. J. Suppl. {\bf 61},
479 (1986).

\bibitem{NumRec}
W. H. Press, B. P. Flannery, S. A. Teukolsky, and
W. T. Vetterling,  {\it Numerical Recipes}
(Cambridge University Press, New York, 1992)

\bibitem{ZCM1} X. Zhuge, J. Centrella, and S. McMillan,
Phys. Rev. D {\bf 50}, 6247 (1994).

\bibitem{ZCM2} X. Zhuge, J. Centrella, and S. McMillan,
Phys. Rev. D, submitted.

\bibitem{Powers} D. L. Powers, {\it Boundary Value Problems},
3rd Edition, (Harcourt, Brace, and Jovanovich, Florida, 1987).

\bibitem{chandra69} S. Chandrasekhar, {\it Ellipsoidal 
Figures of Equilibrium}, (Yale Univ. Press, New Haven, 1969).

\bibitem{Bonnell}I. Bonnell, Mon. Not. R. Astr. Soc. 
{\bf 269}, 837 (1994); I. Bonnell and M. Bate,
Mon. Not. R. Astr. Soc. {\bf 271}, 999 (1994).

\bibitem{LS-nascent}D. Lai and S. Shapiro,
Astrophys. J. {\bf 442}, 259 (1995).

\bibitem{narrow}B. Meers, Phys. Rev. D {\bf 38}, 2317 (1988);
K. Strain and B. Meers, Phys. Rev. Lett. {\bf 66}, 1391 (1991);
A. Krolak, J. Lobo, and B. Meers, 
Phys. Rev. D {\bf 43}, 2470 (1991);
{\bf 47}, 2184 (1993).

\bibitem{KLT}D. Kennefick, D. Laurence, and K. Thorne,
Phys. Rev. D, in preparation.

\bibitem{CGS}J. Centrella, J. Gao, and S. Smith, work in
progress.



\end{references}
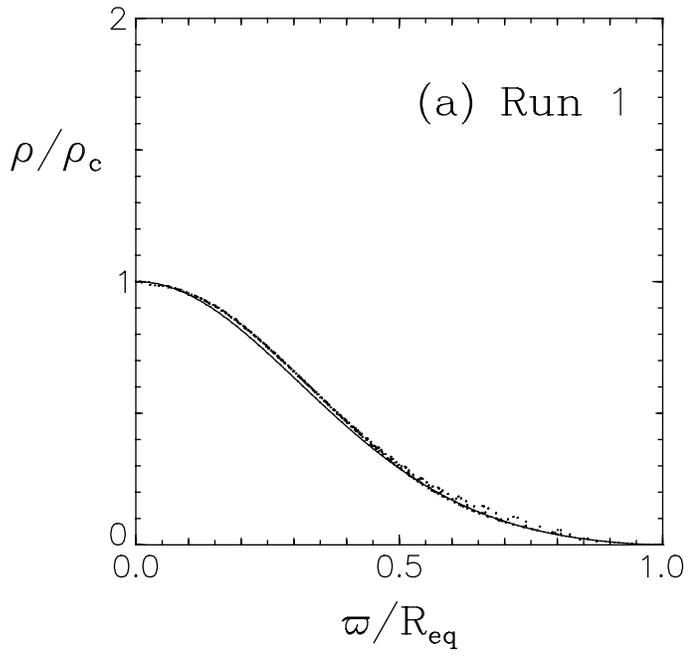
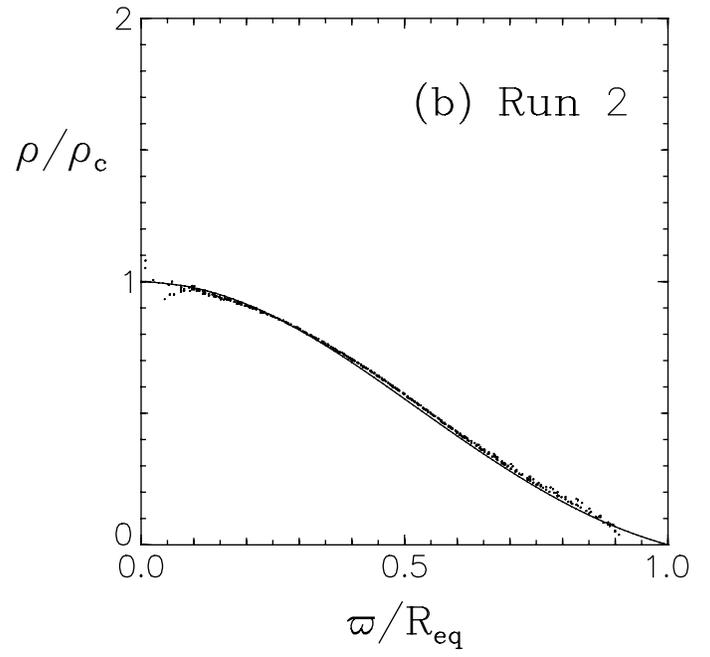
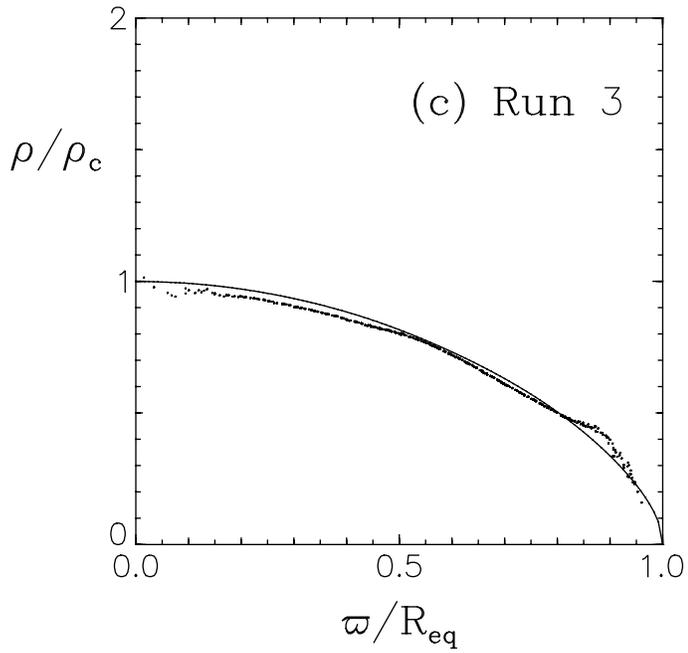
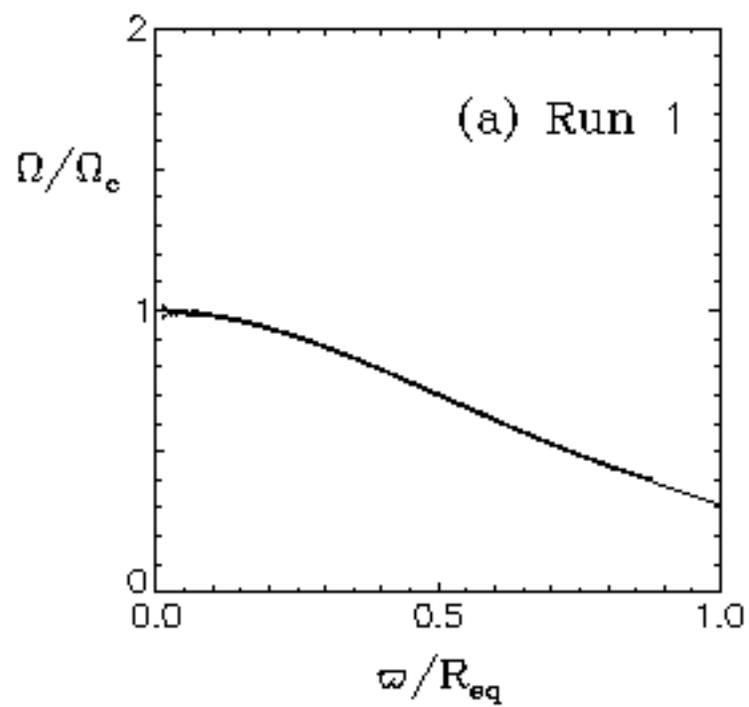
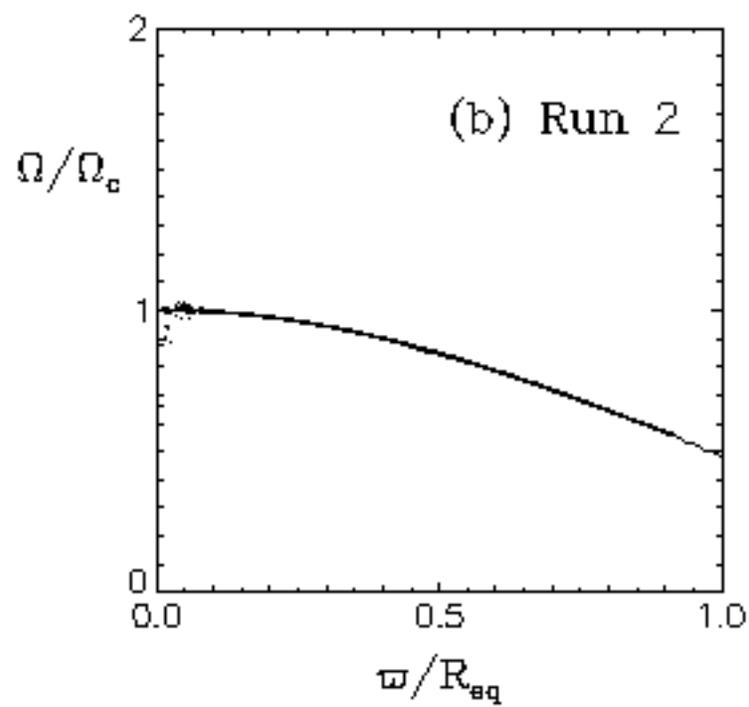
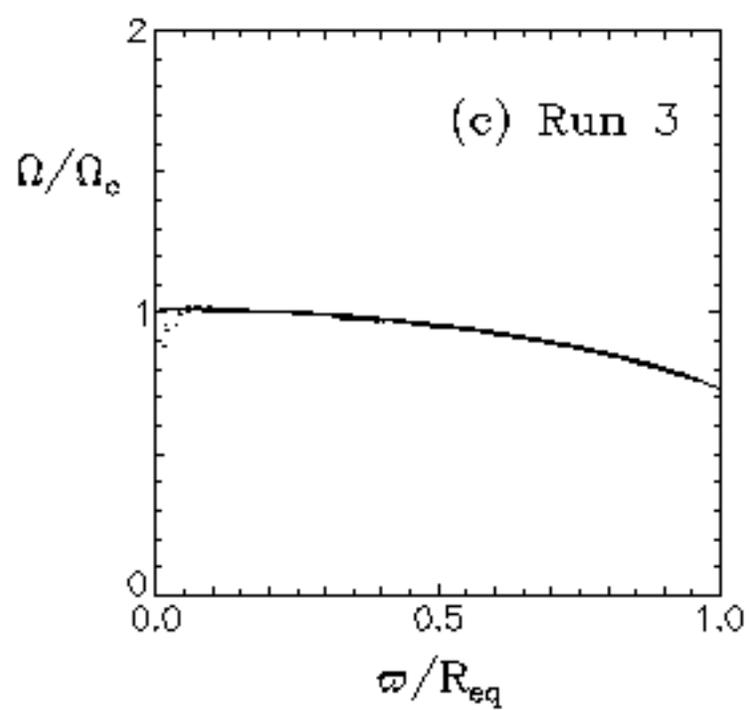
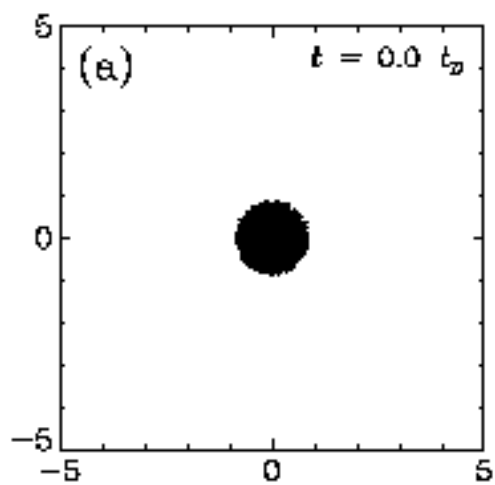
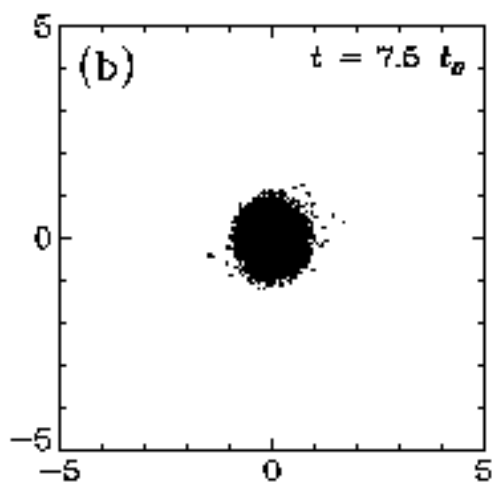
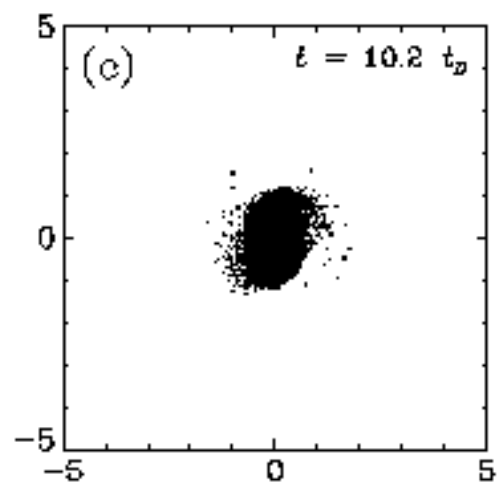
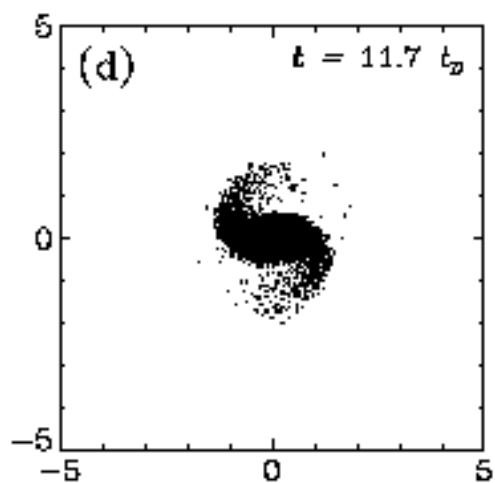
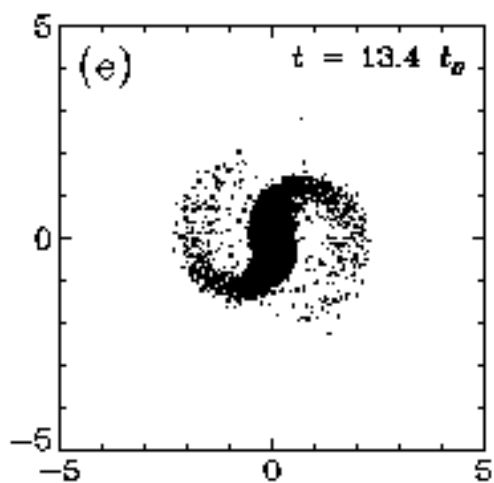
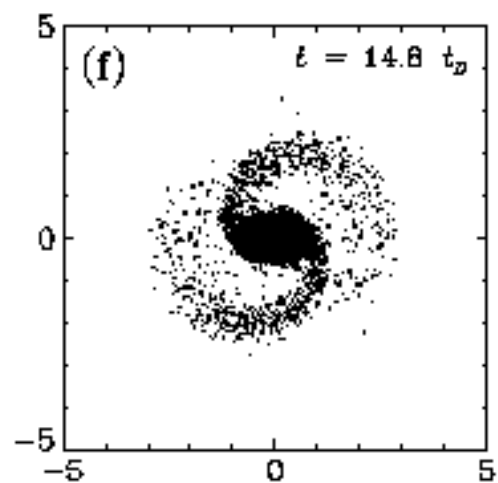
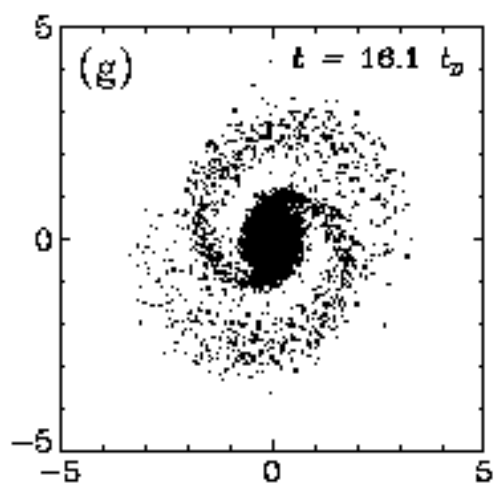
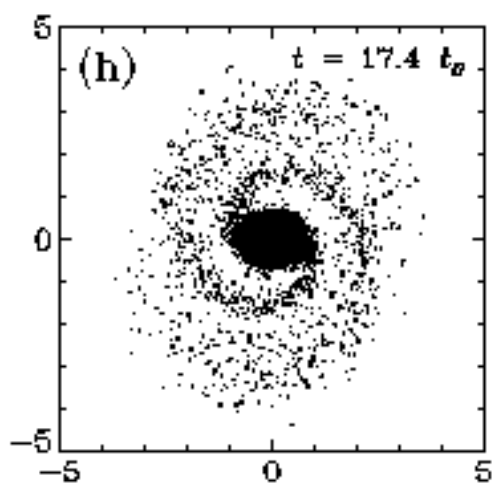
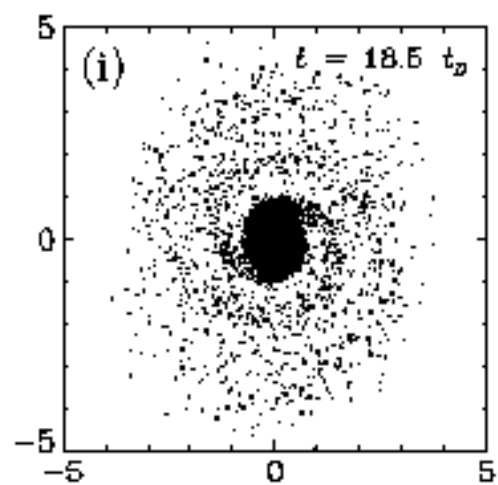
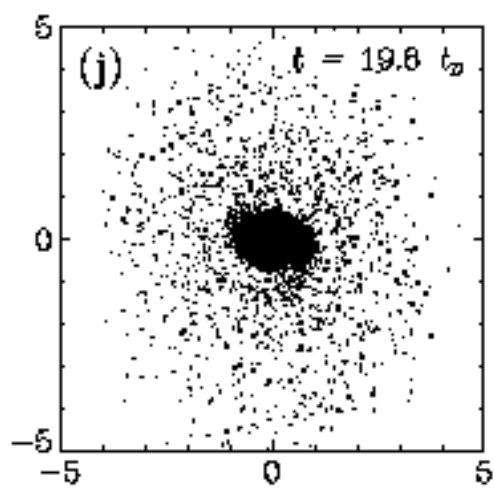
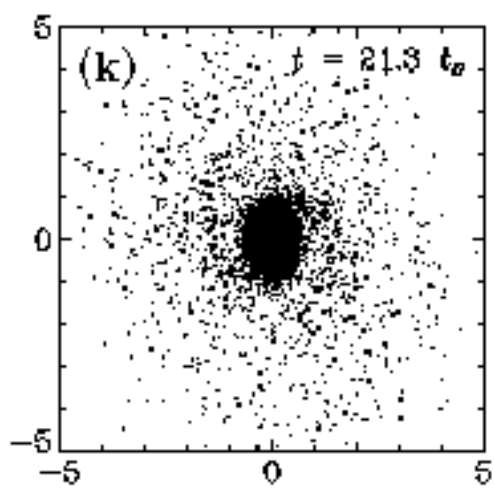
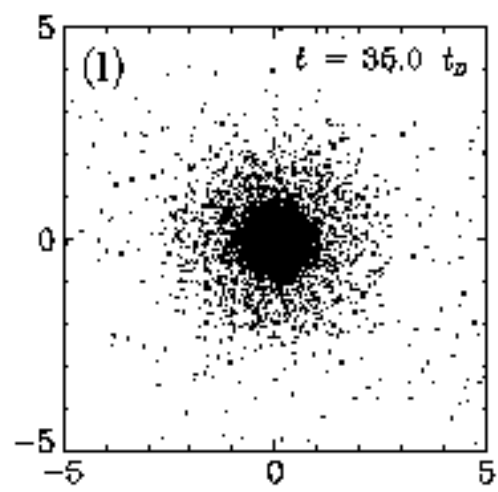
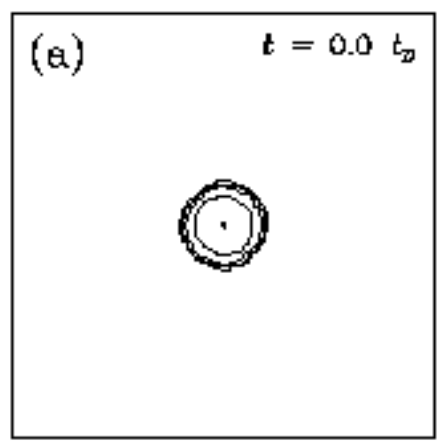
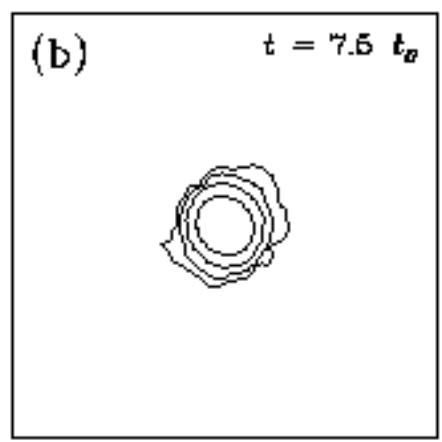
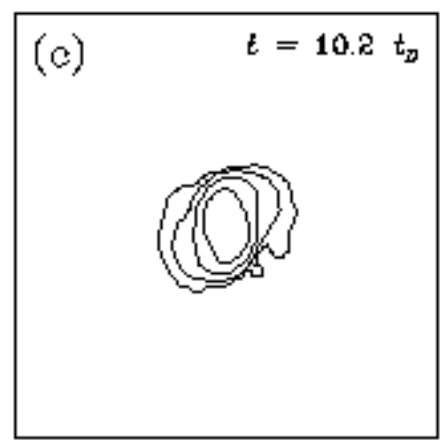
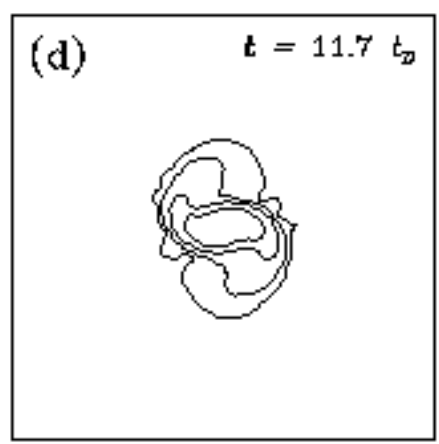
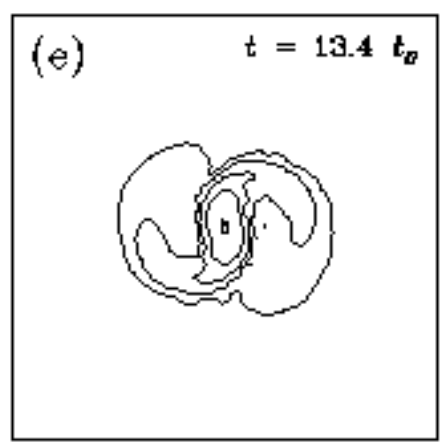
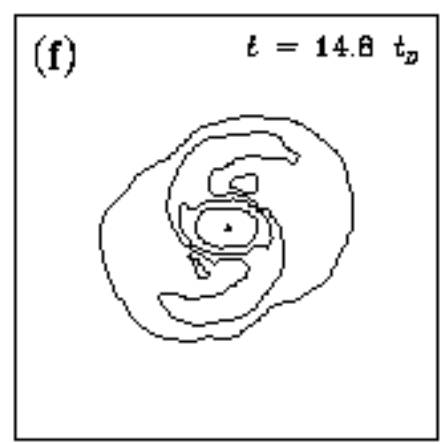
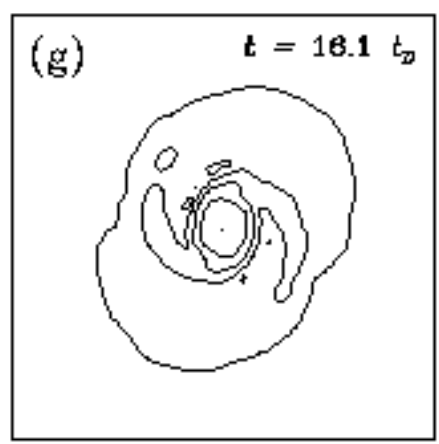
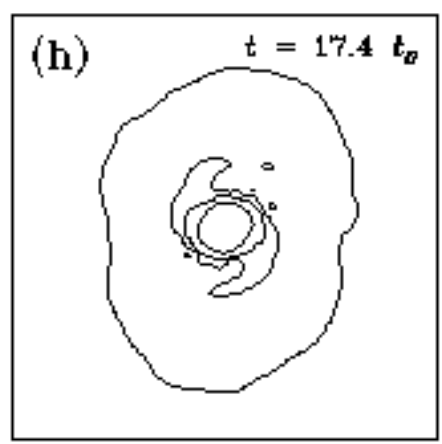
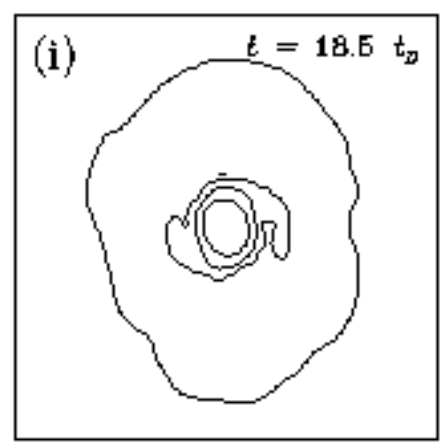
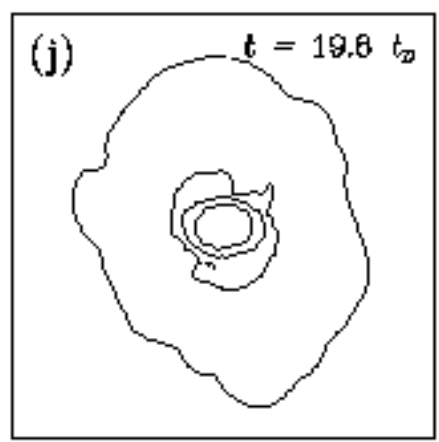
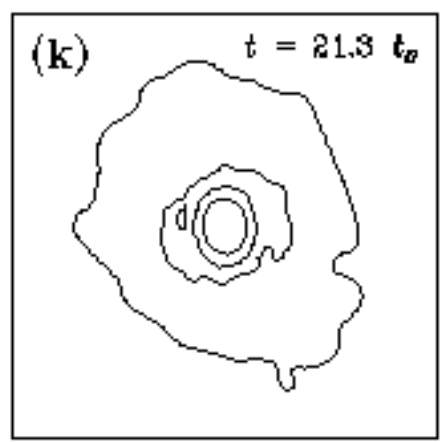
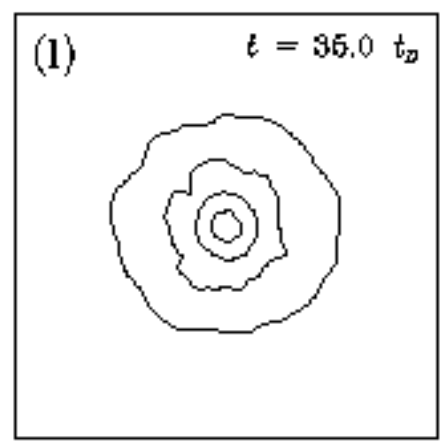
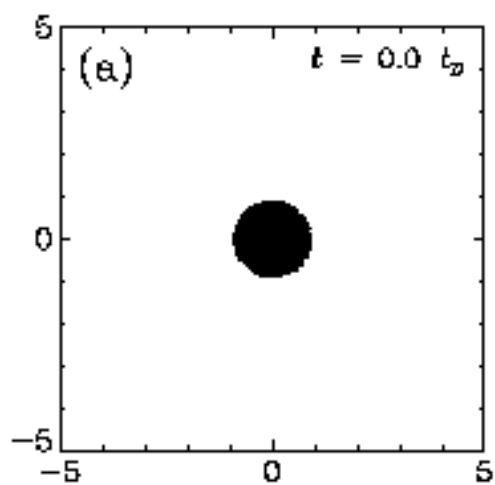
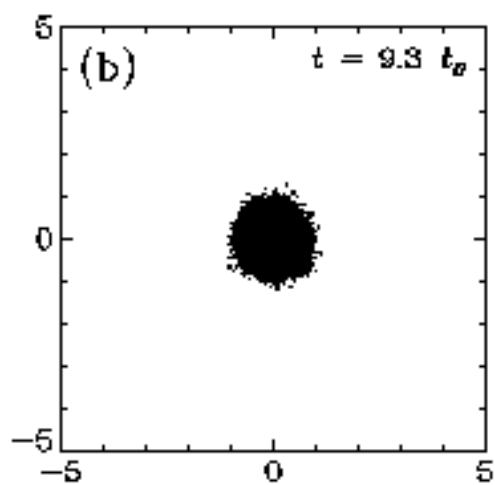
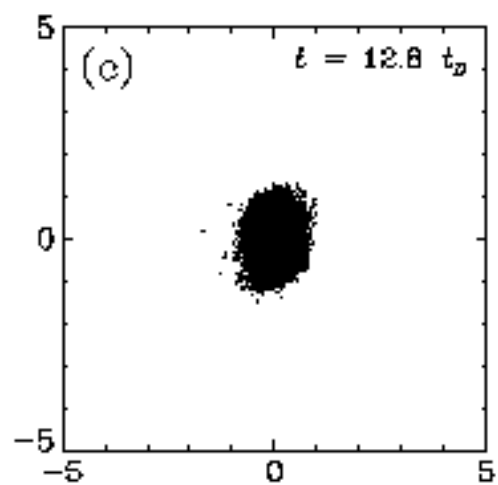
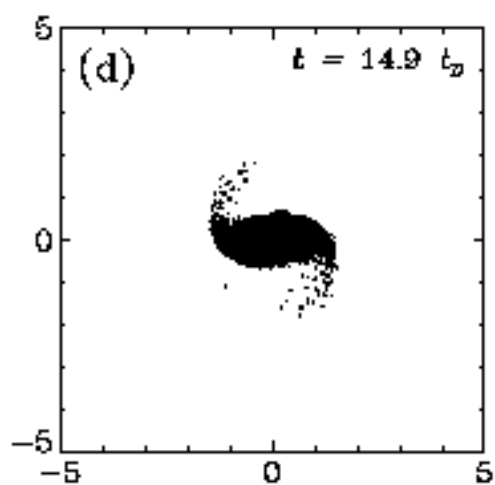
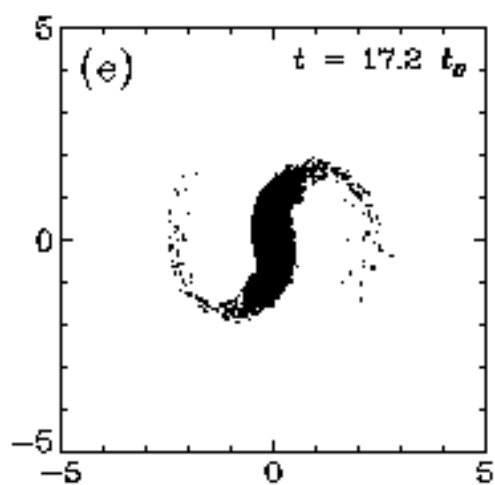
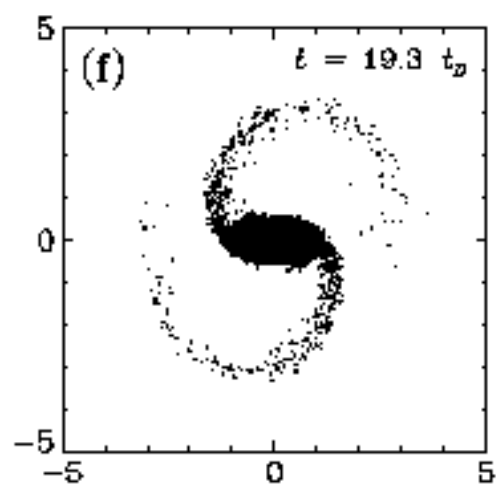
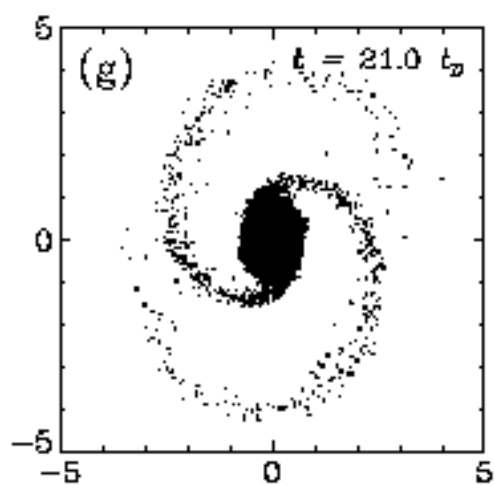
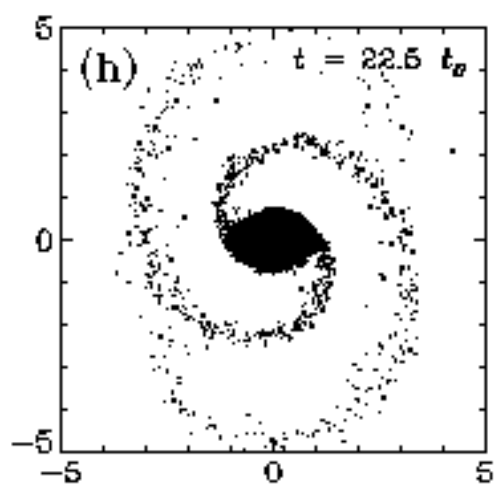
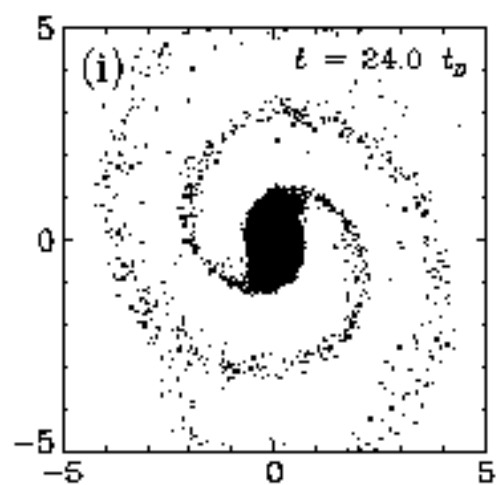
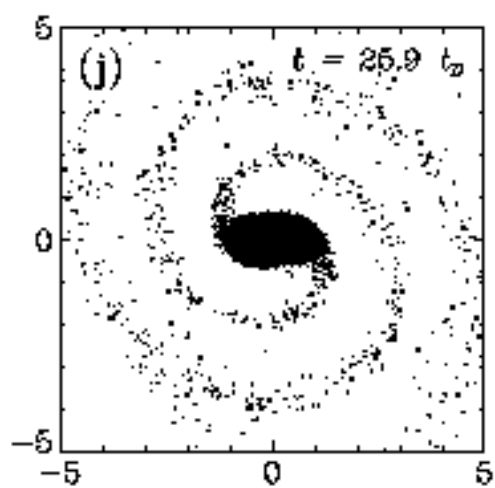
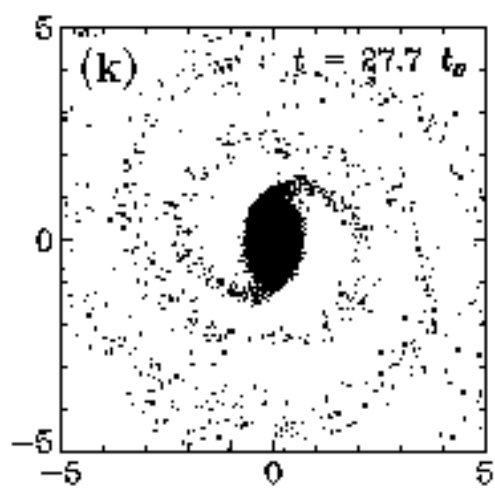
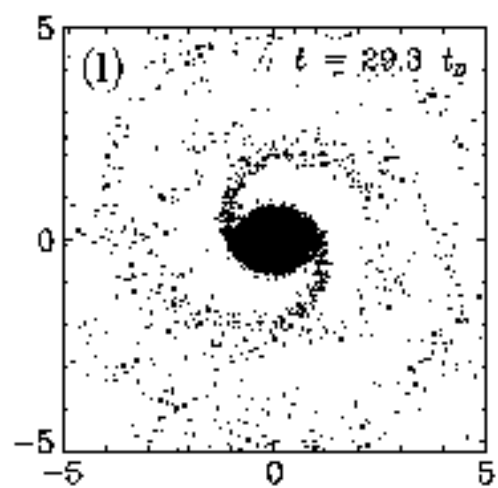
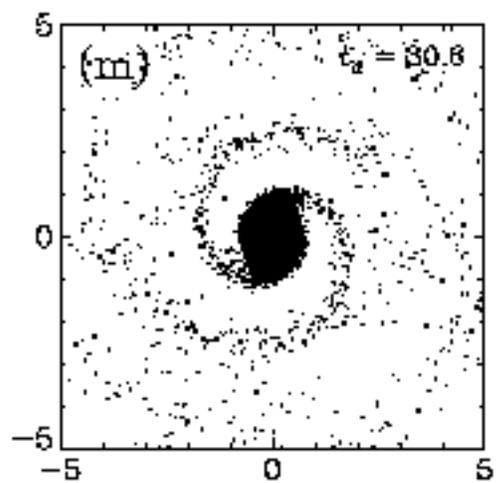
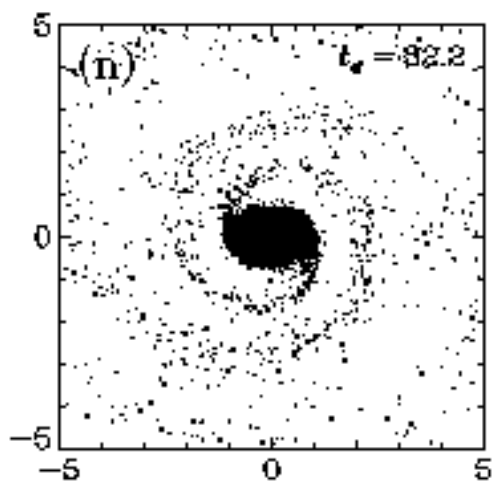
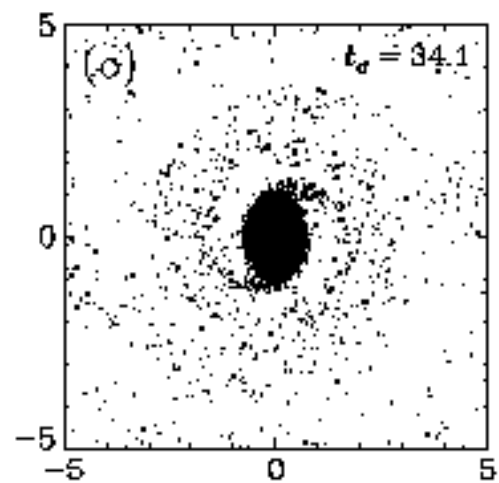
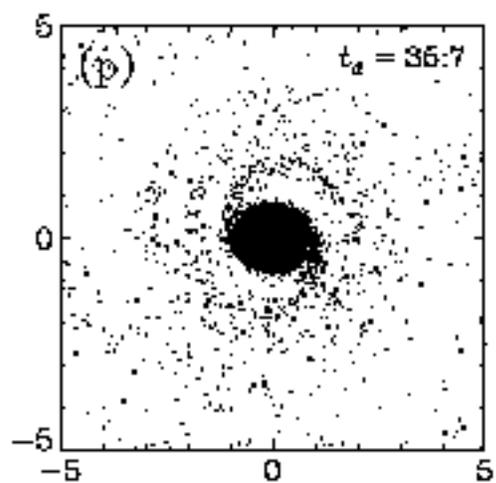
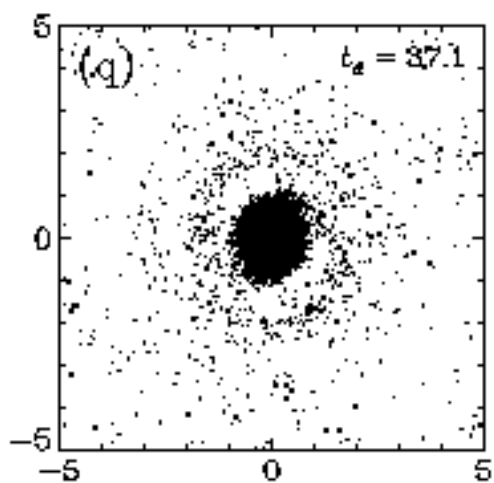
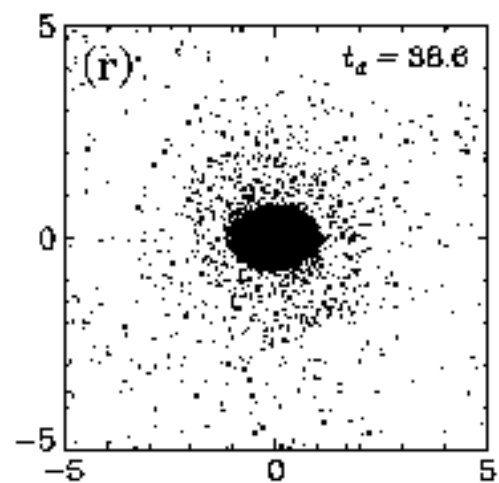
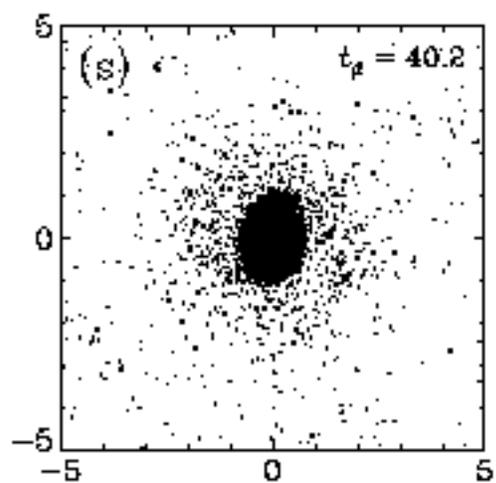
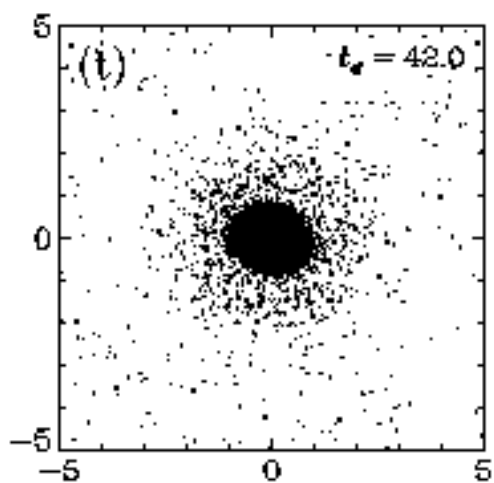
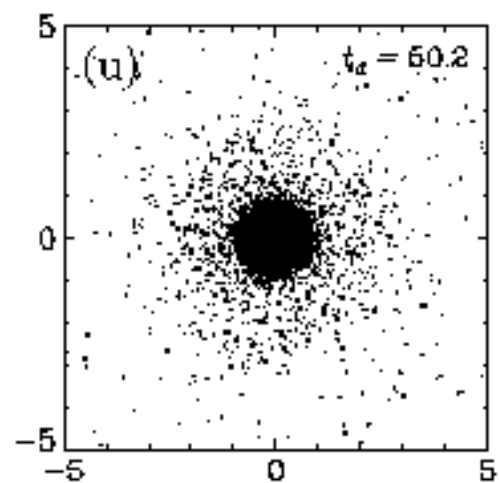
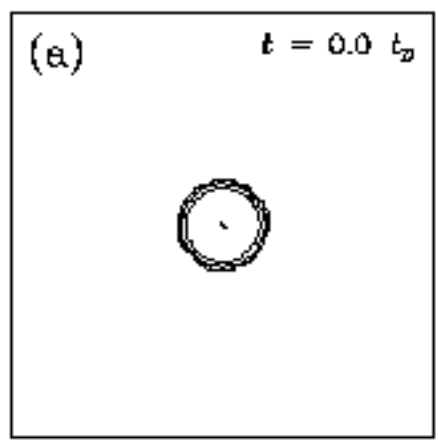
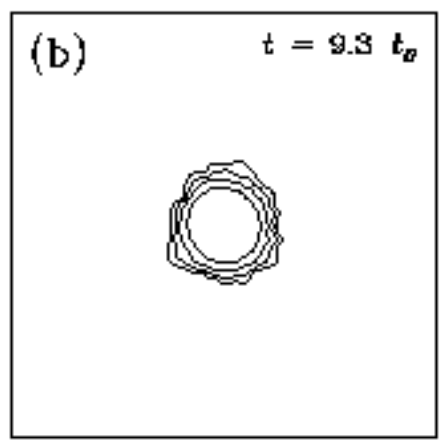
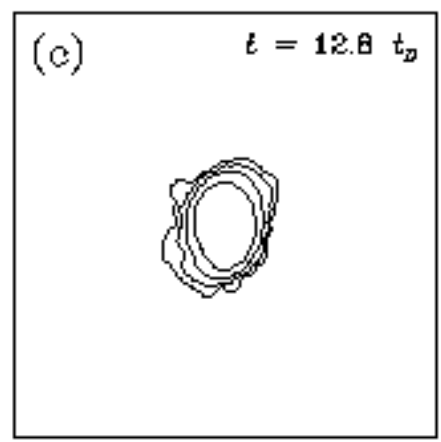
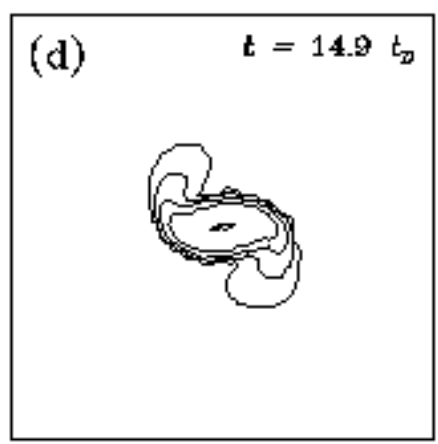
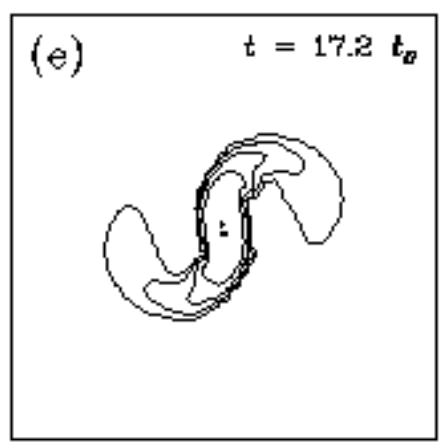
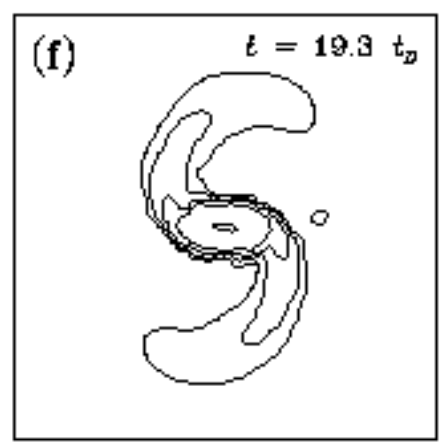
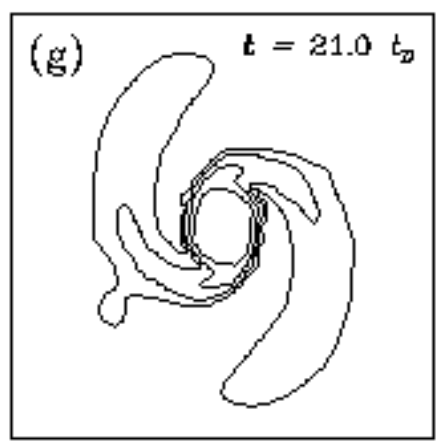
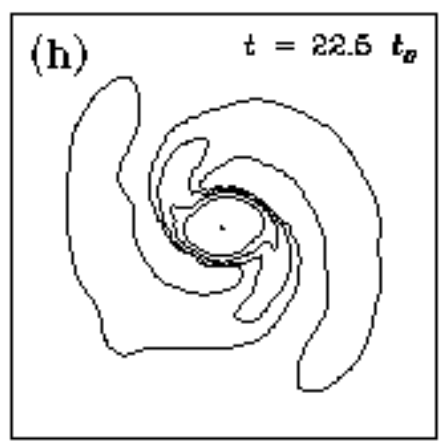
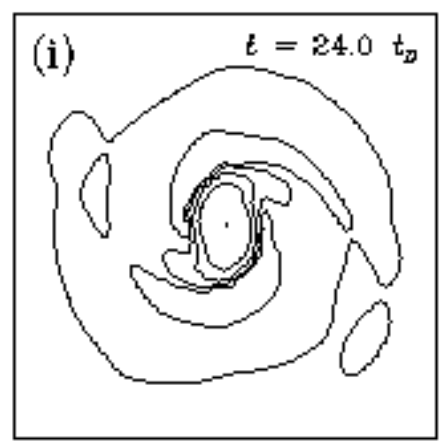
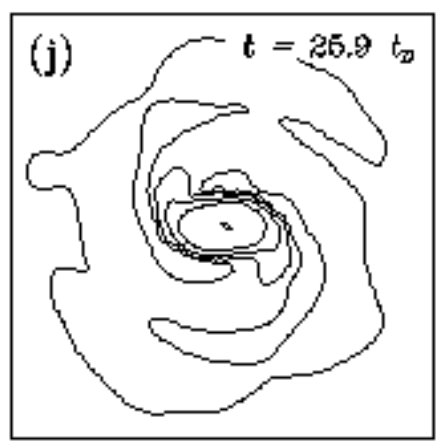
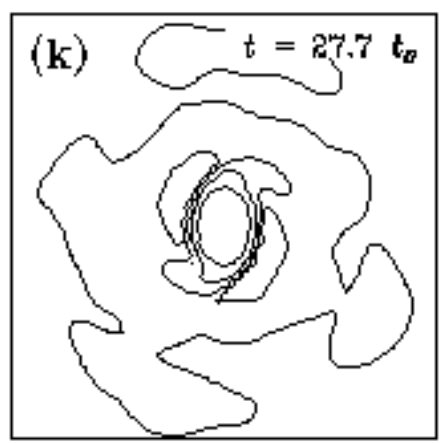
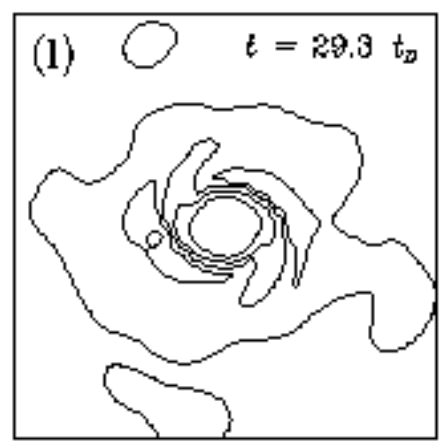
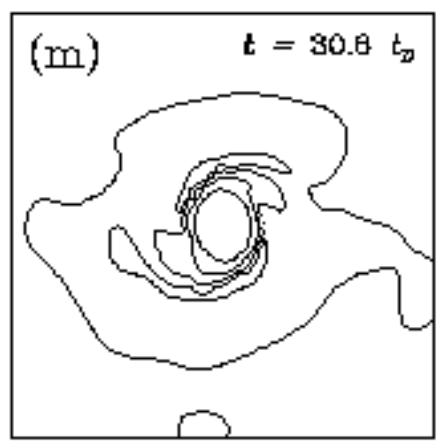
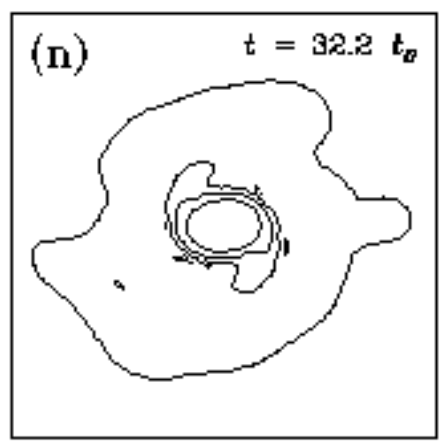
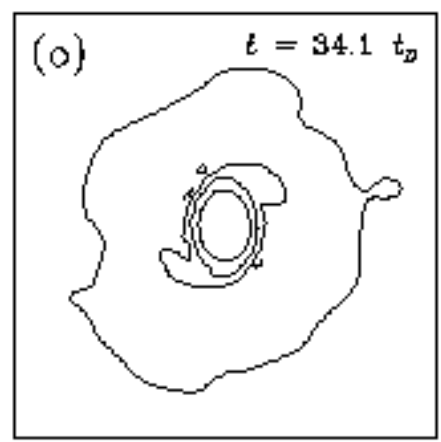
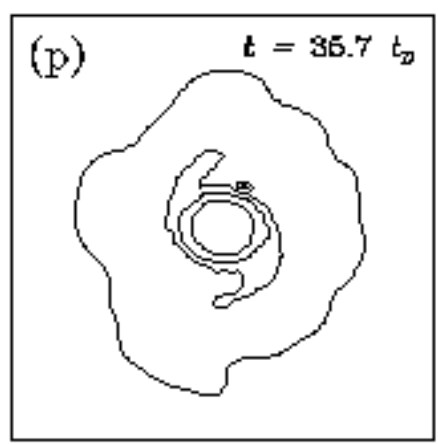
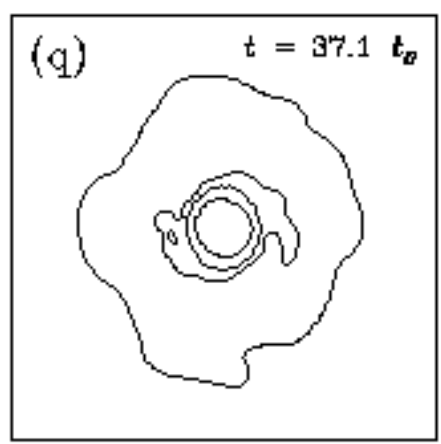
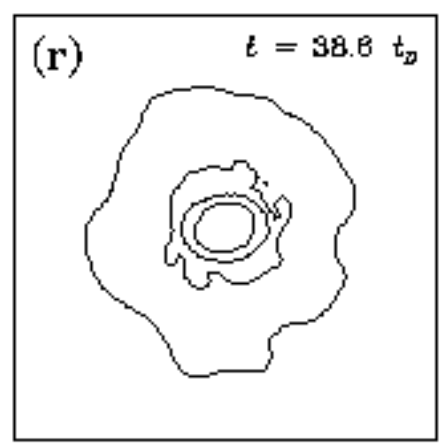
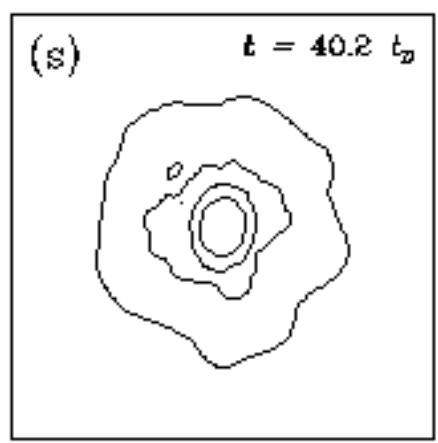
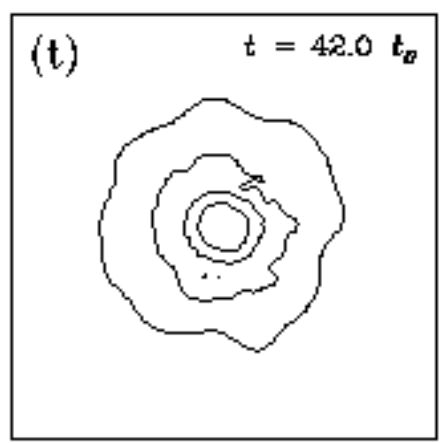
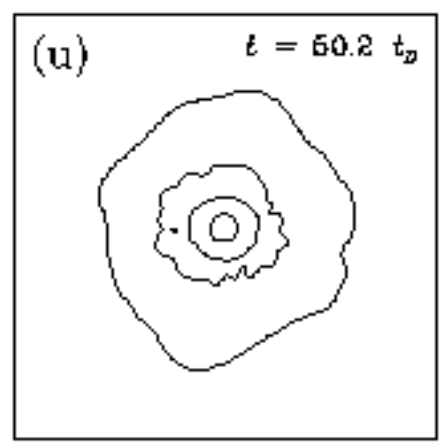
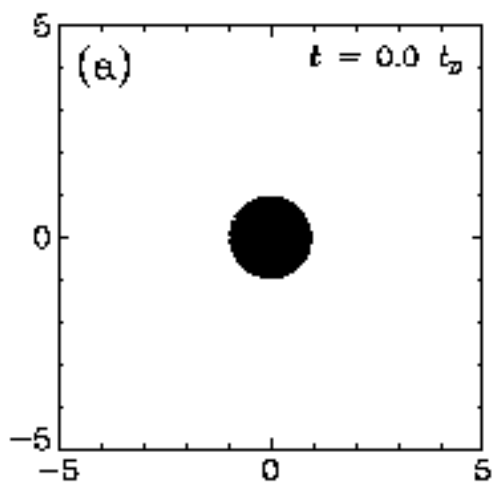
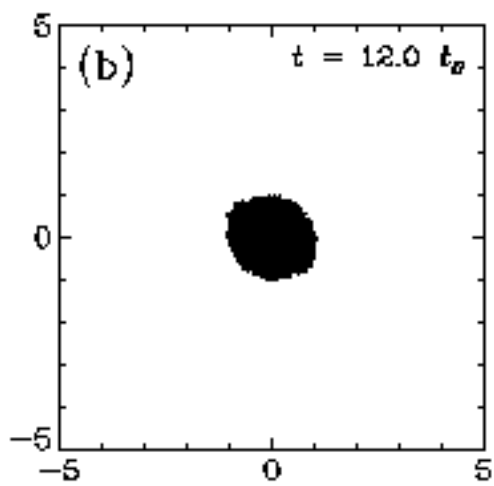
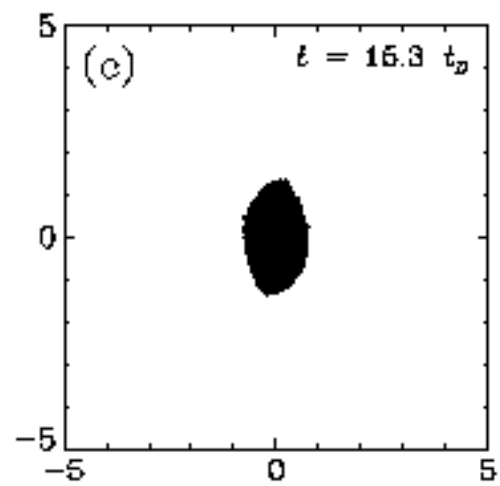
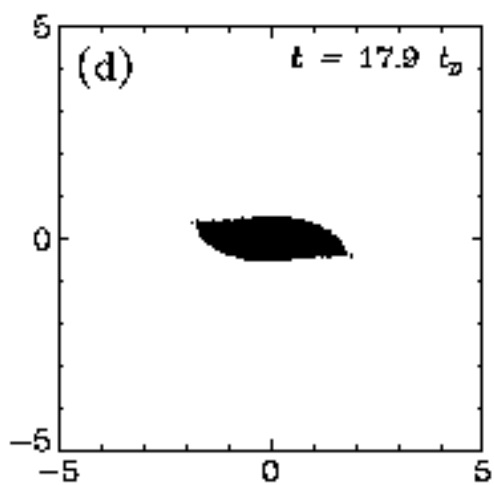
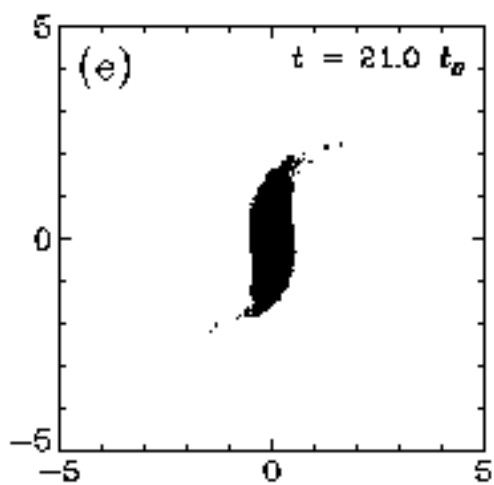
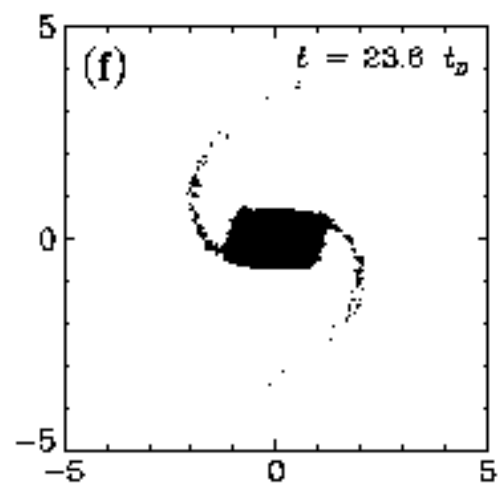
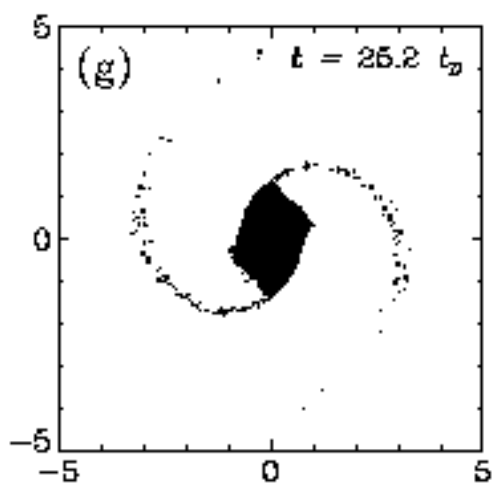
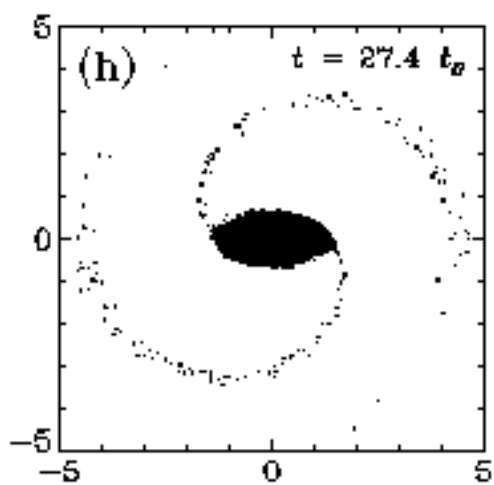
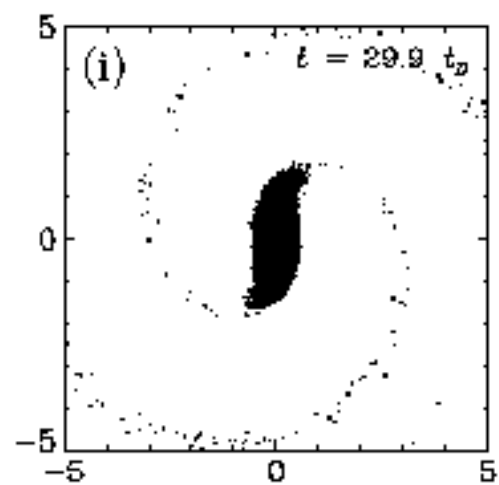
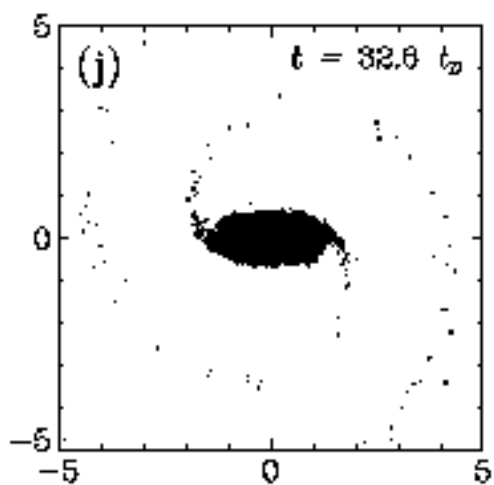
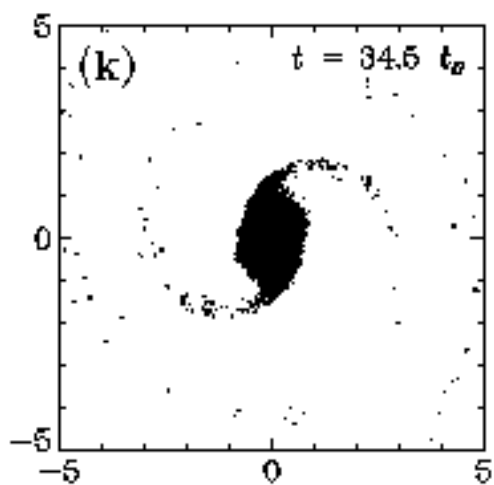
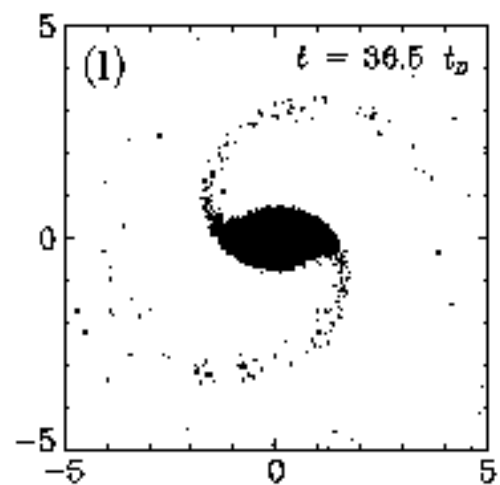
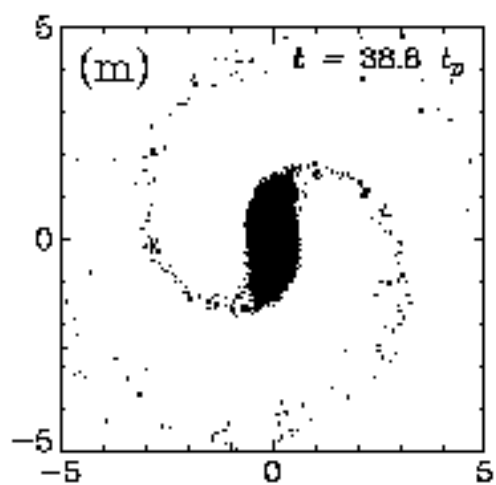
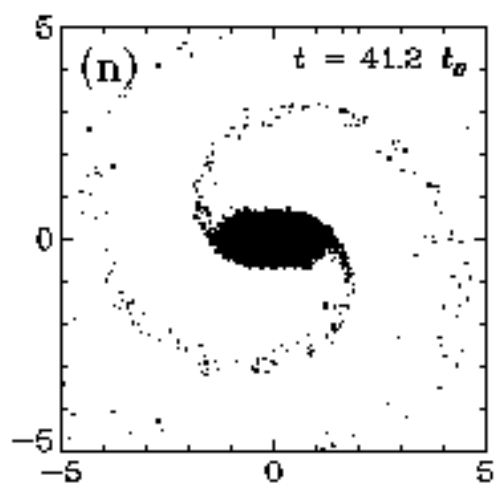
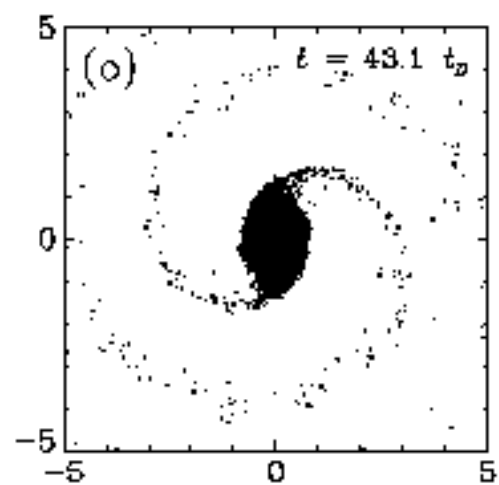
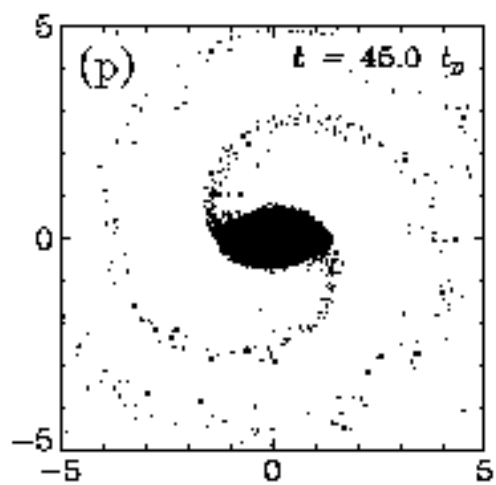
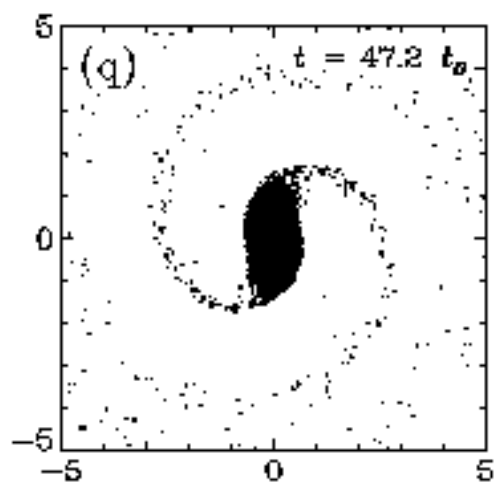
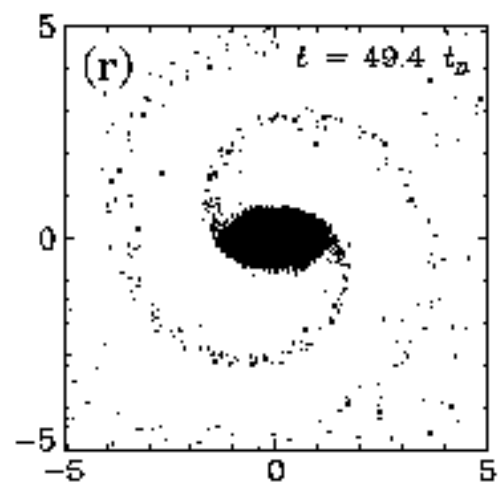
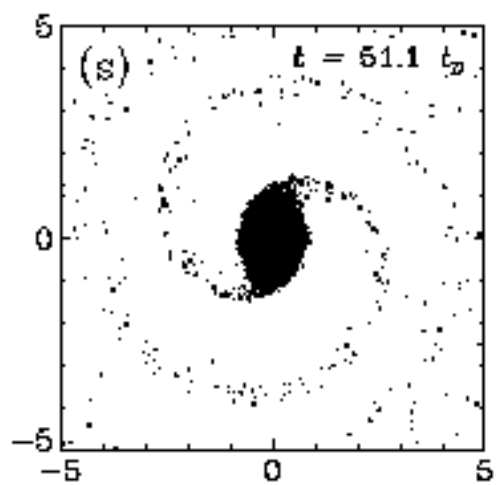
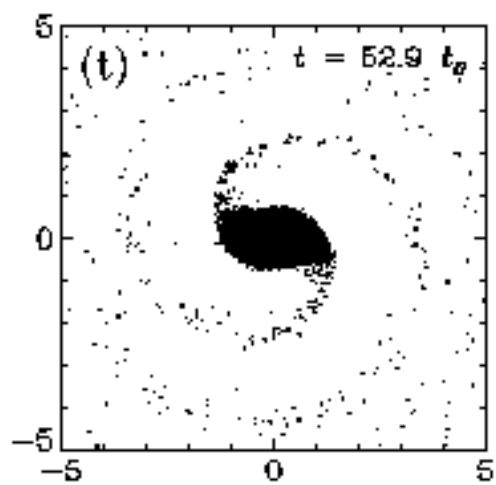
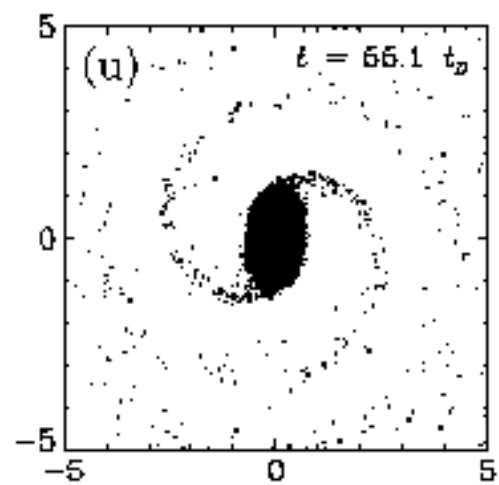
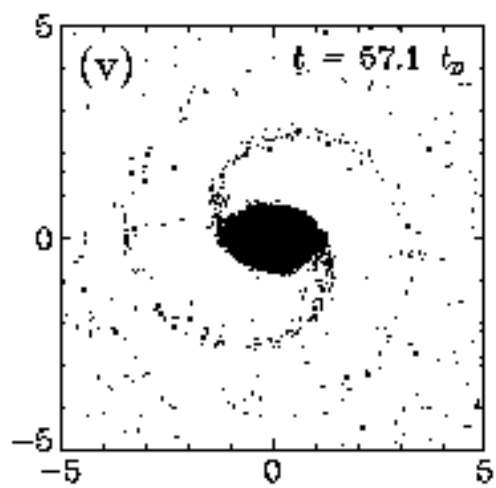
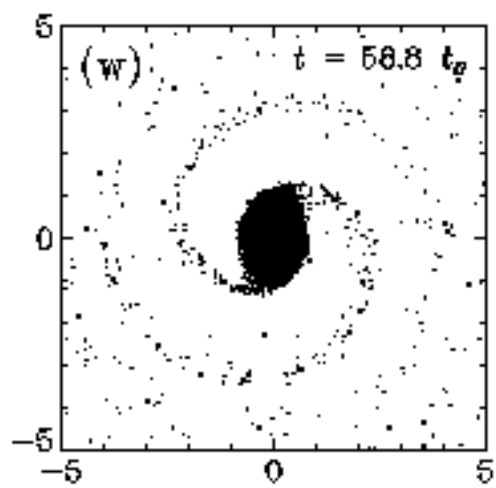
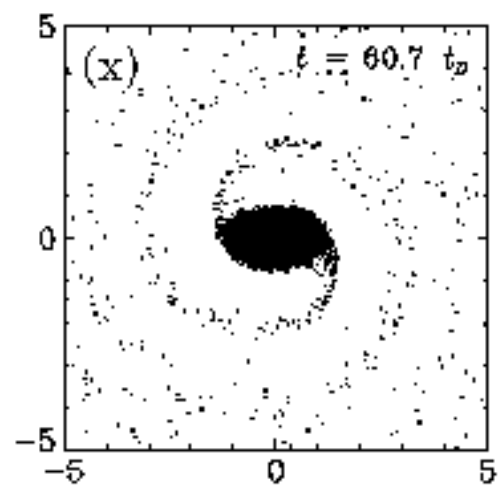
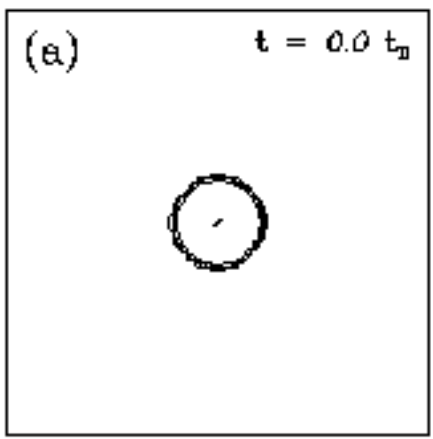
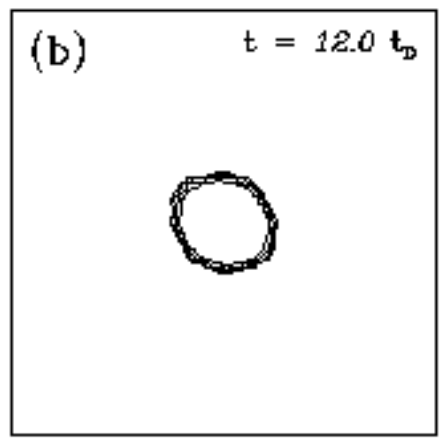
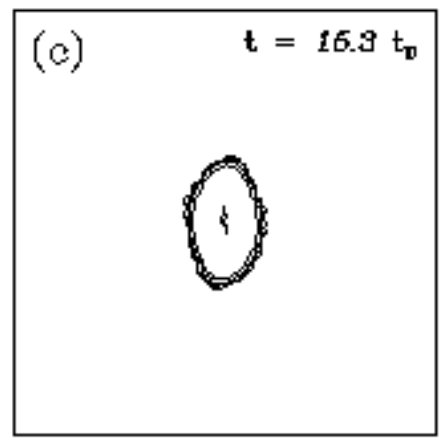
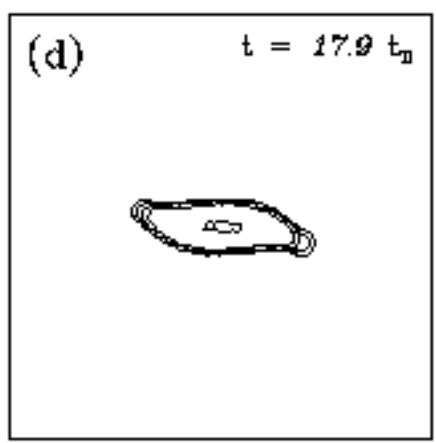
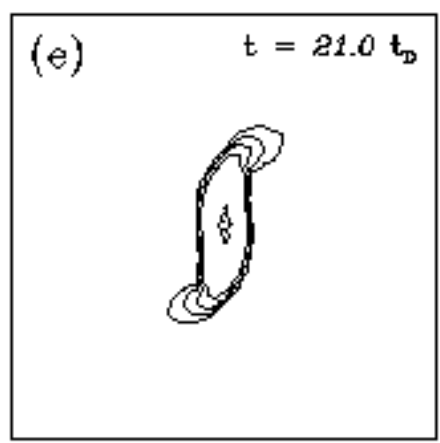
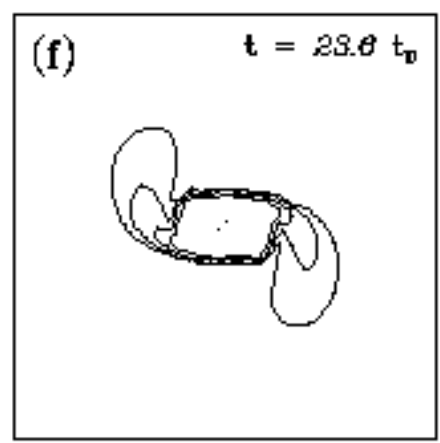
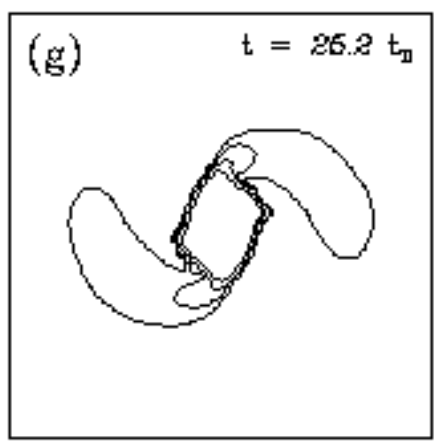
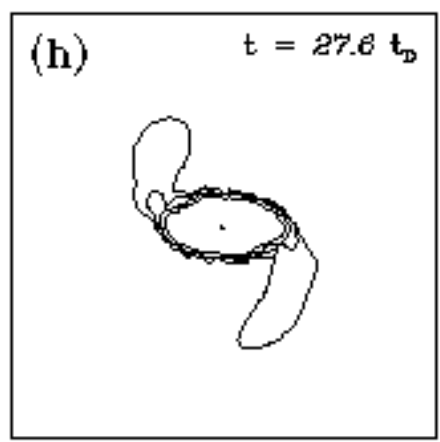
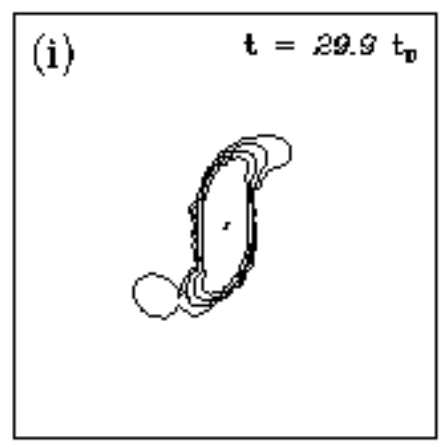
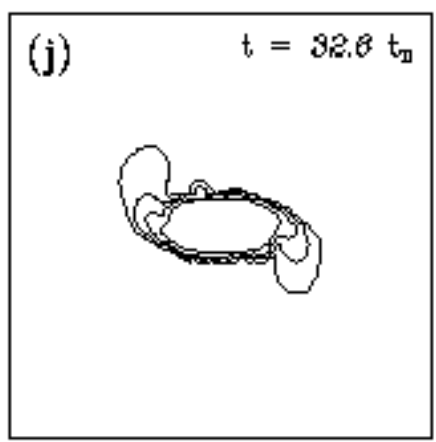
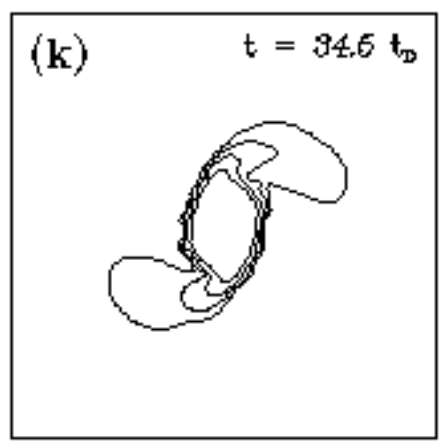
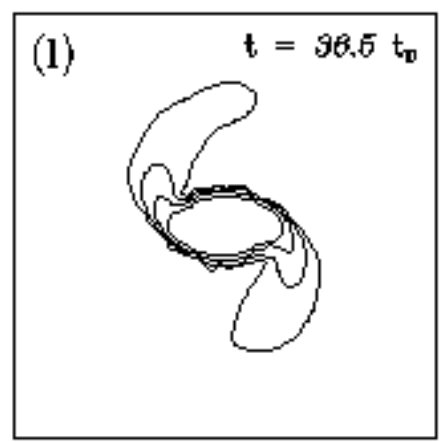
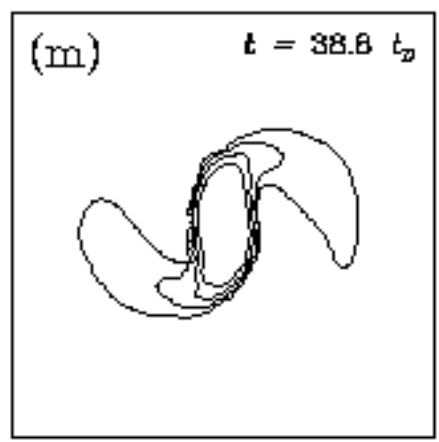
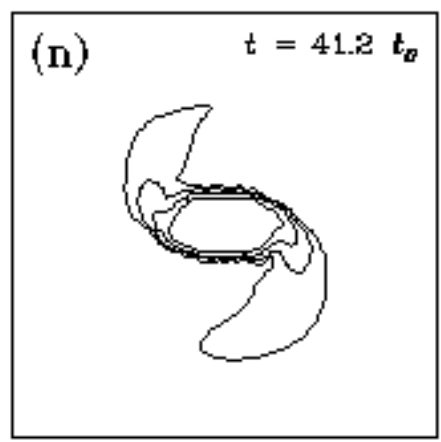
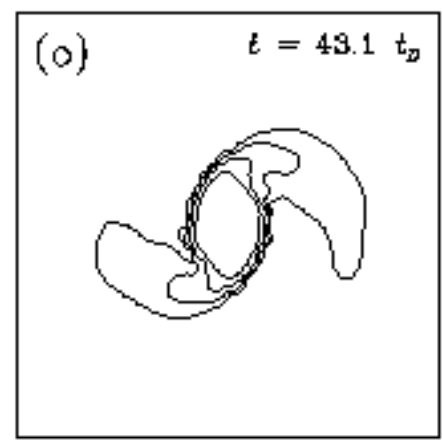
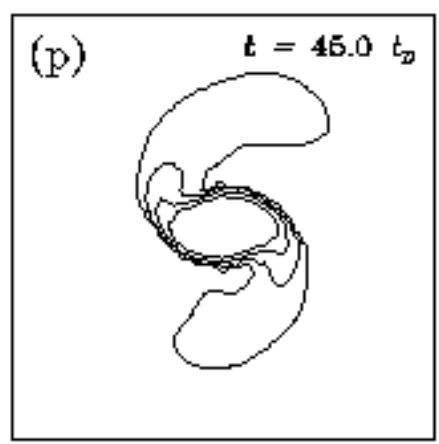
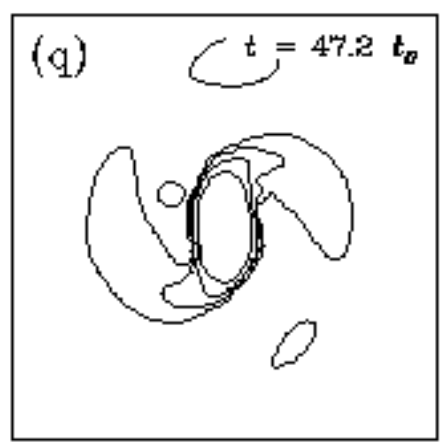
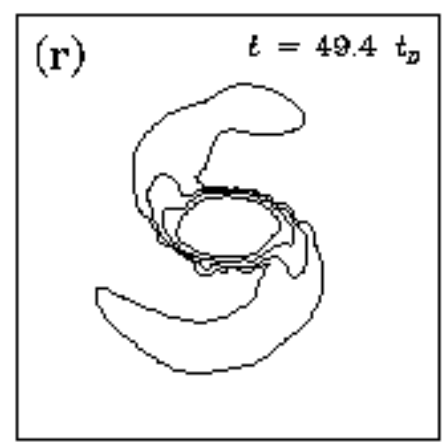
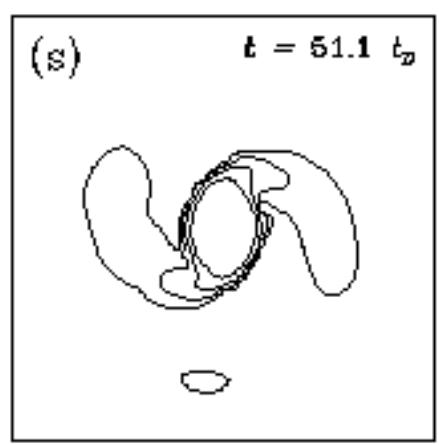
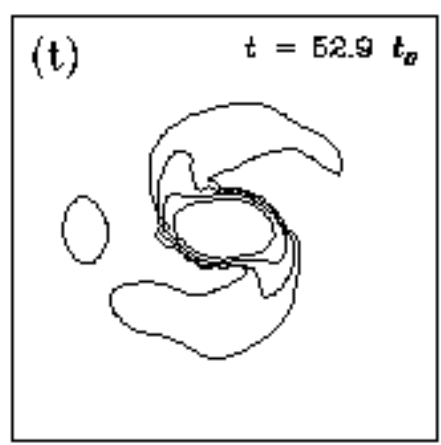
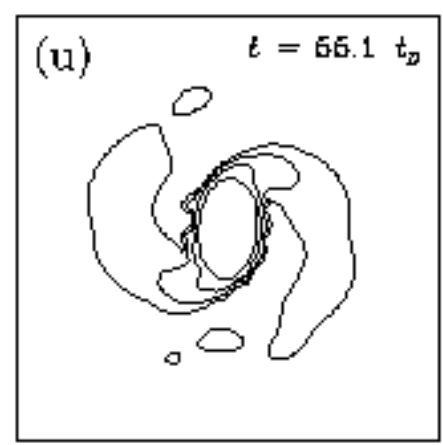
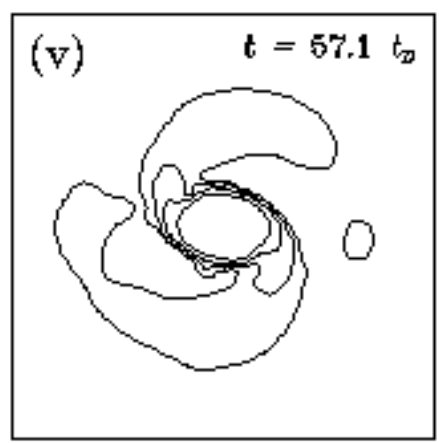
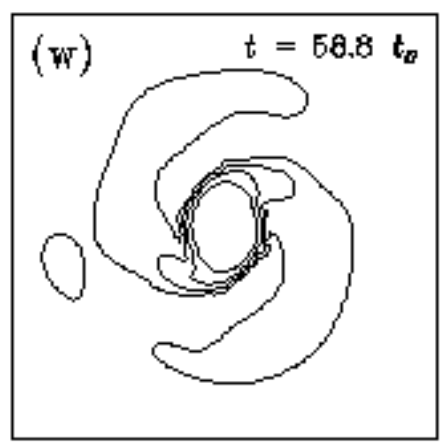
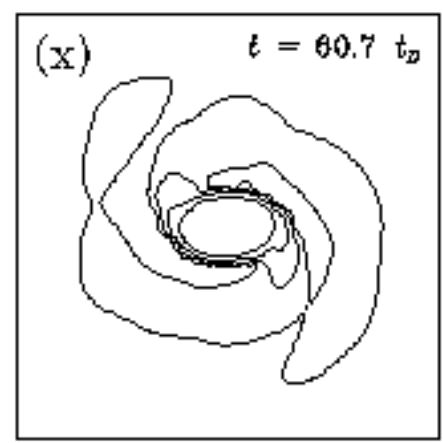
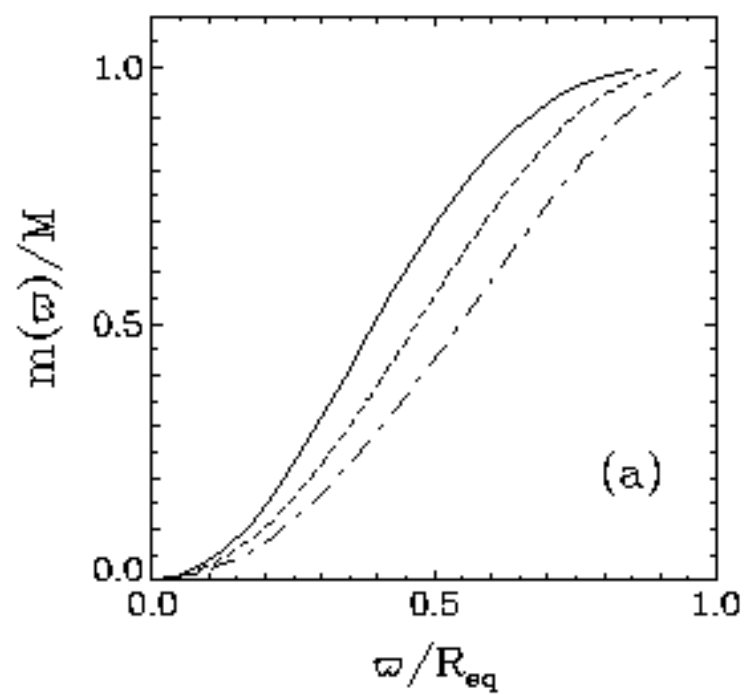
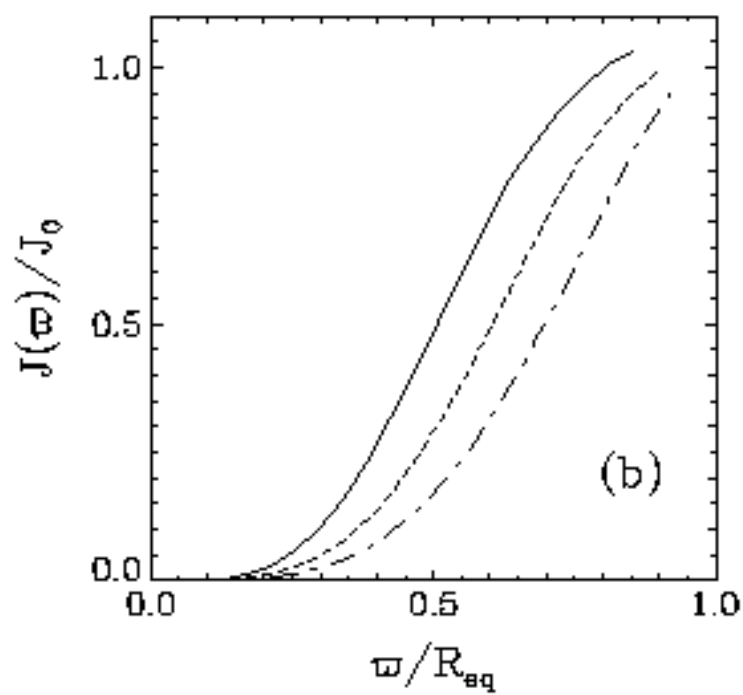
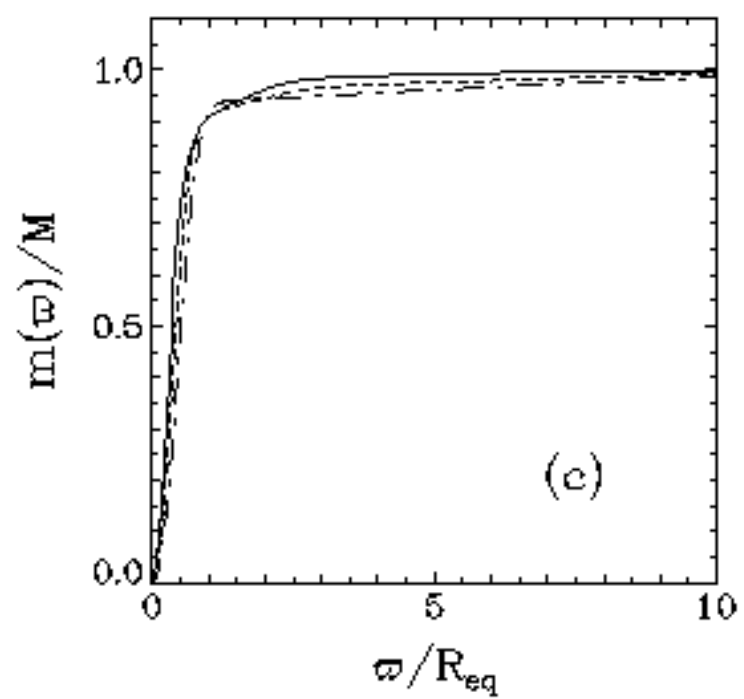
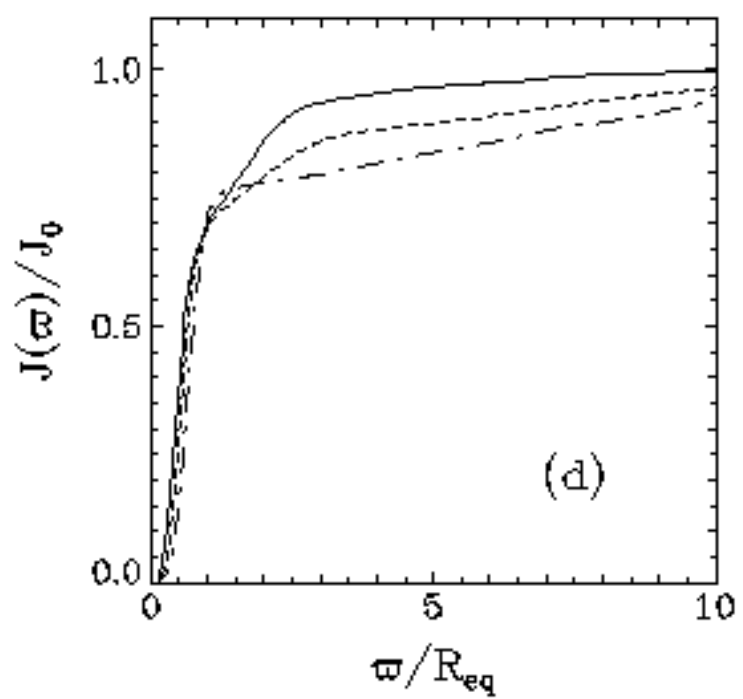
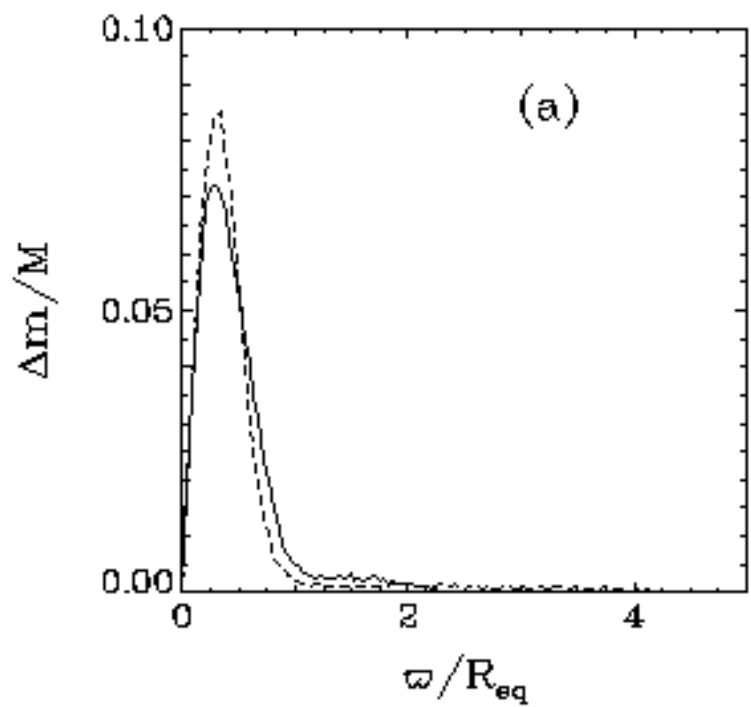
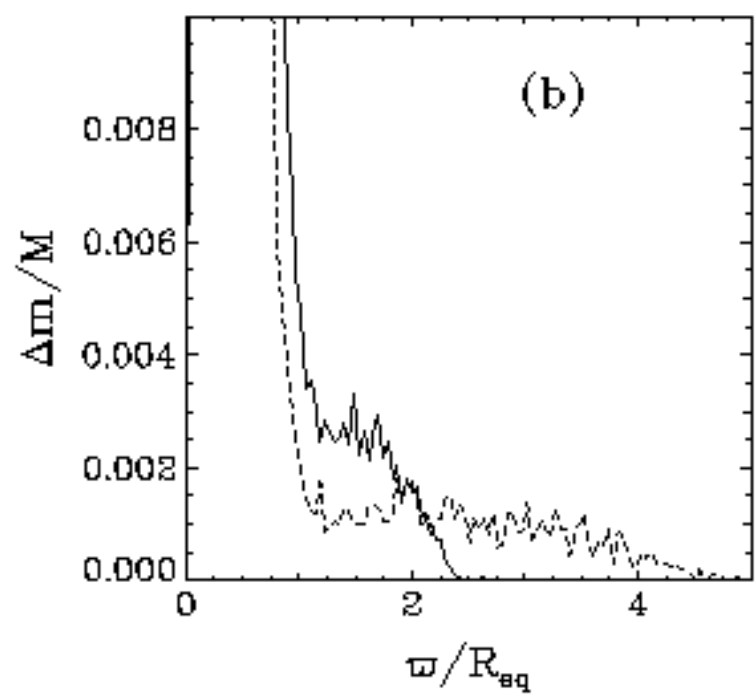
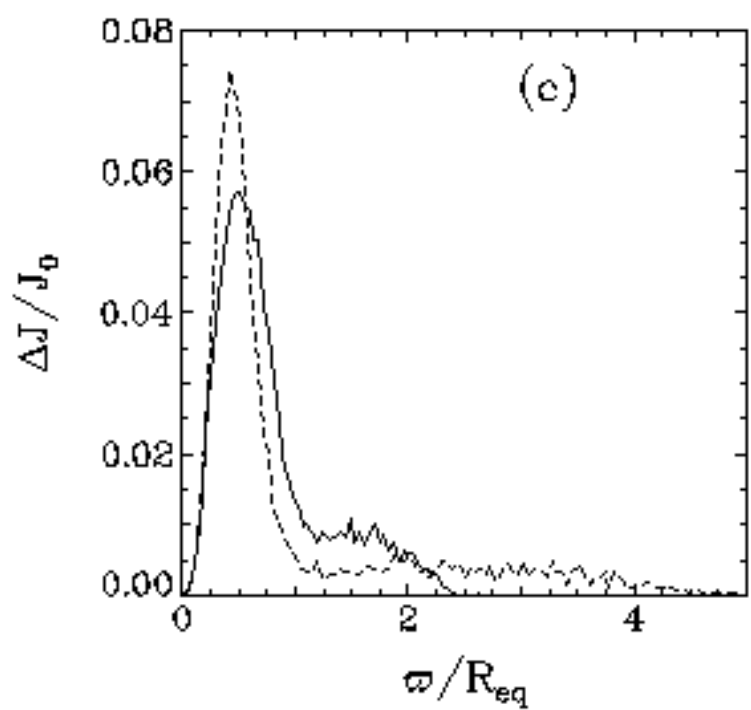
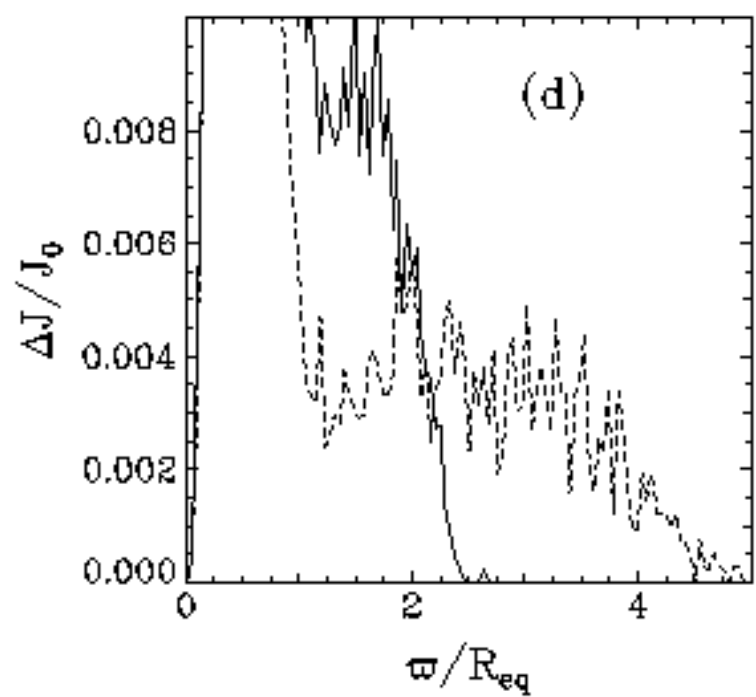
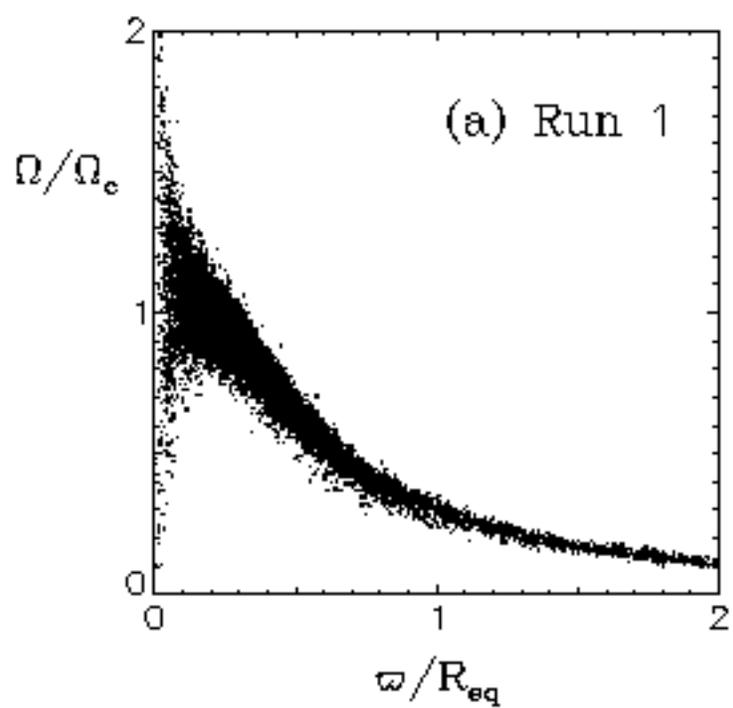
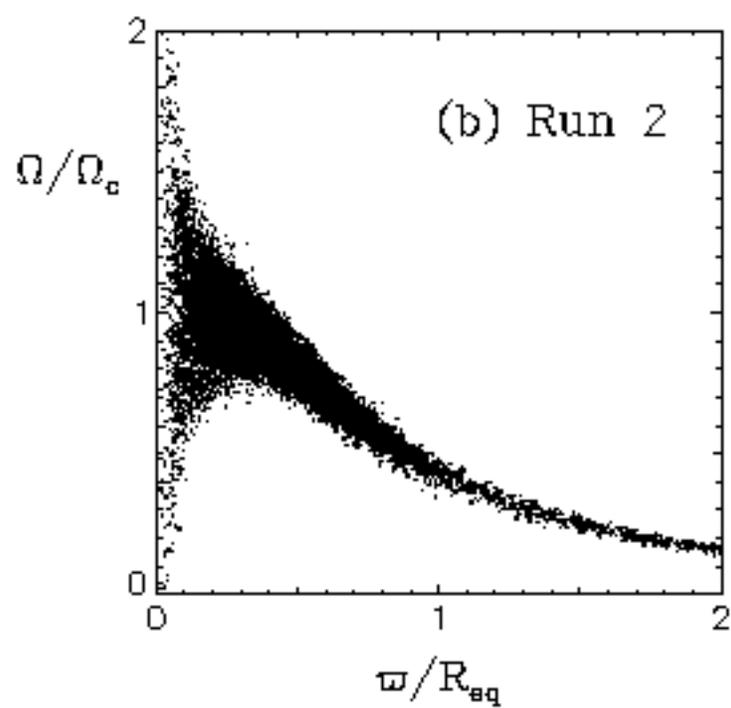
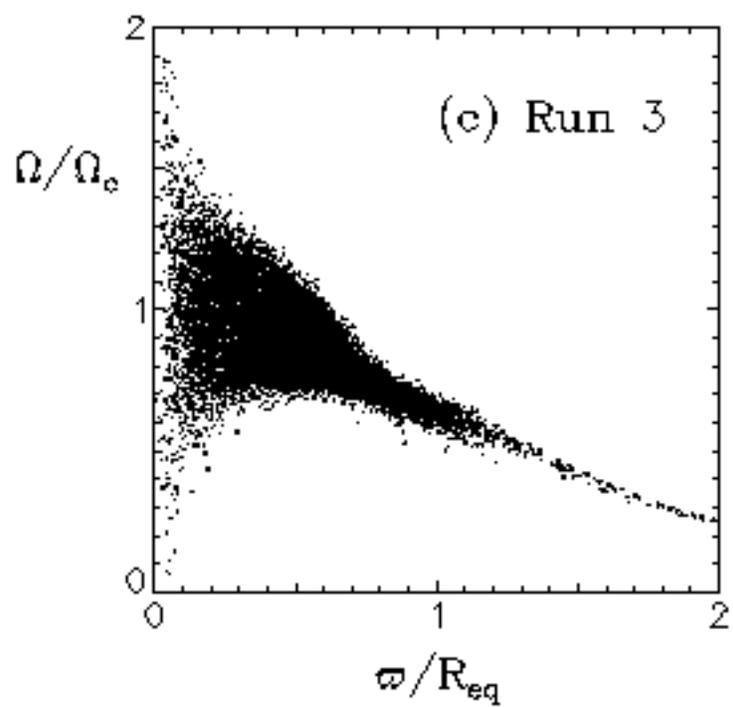
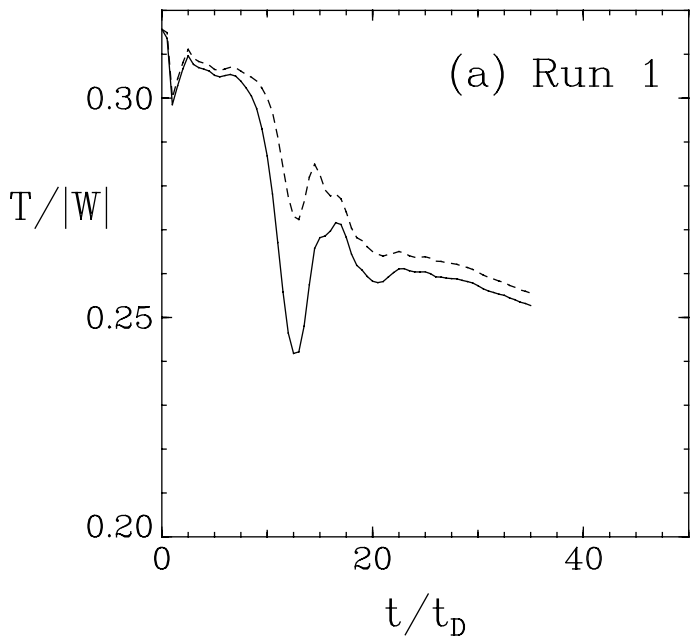
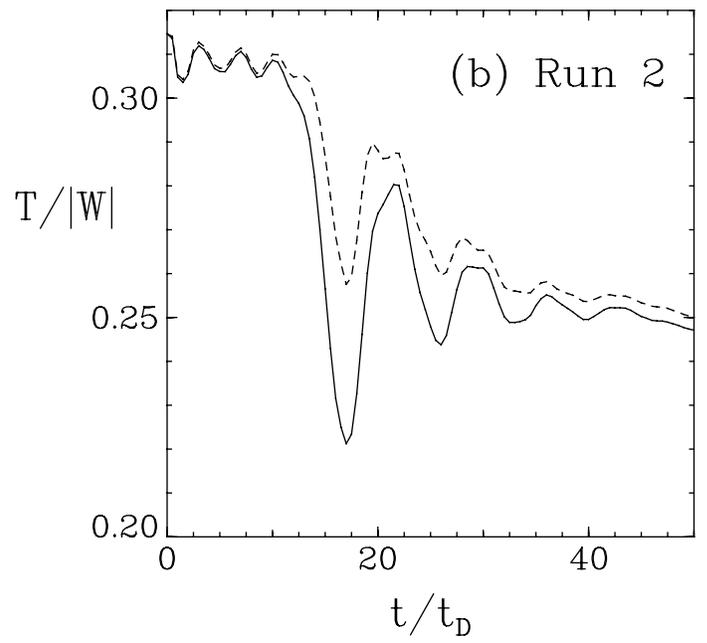
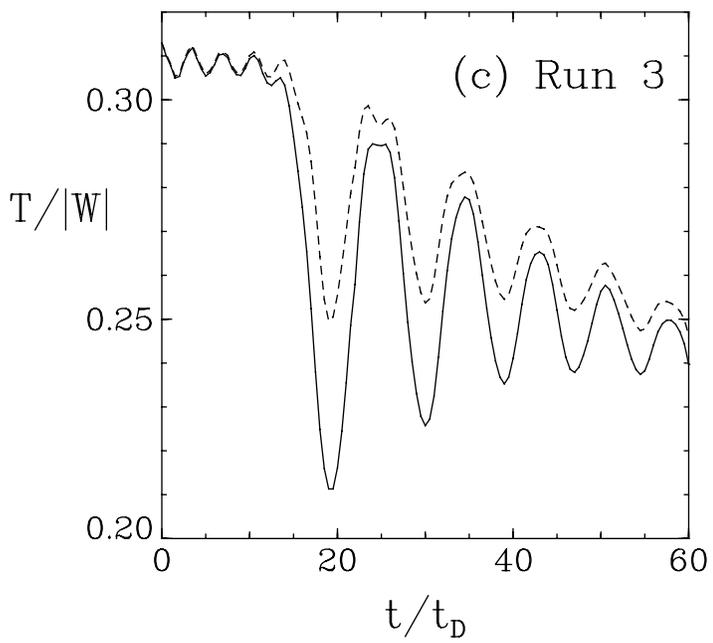
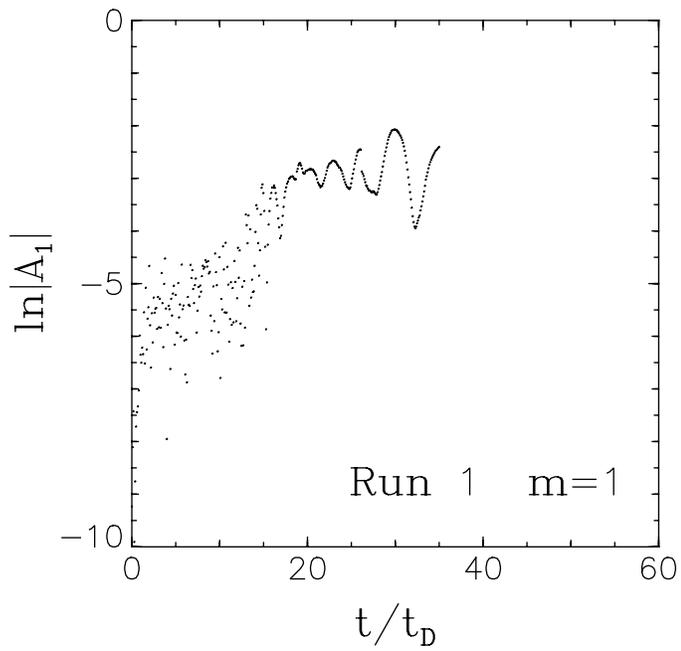
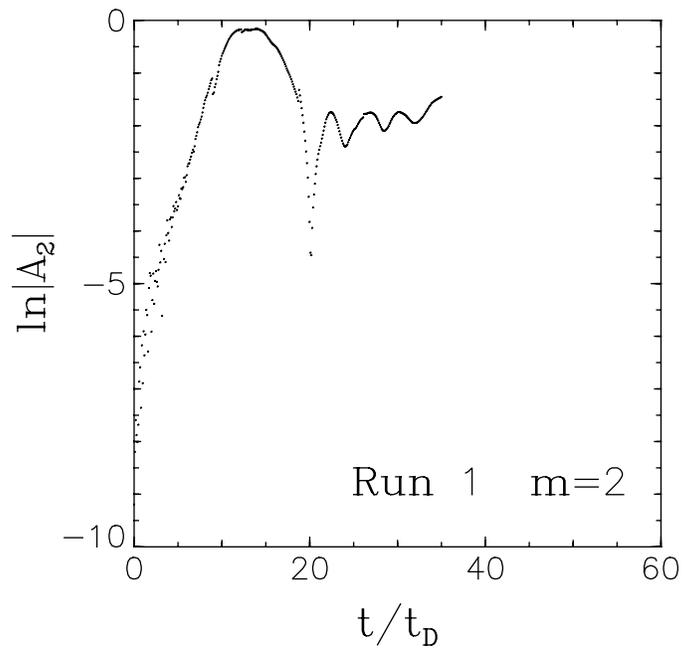
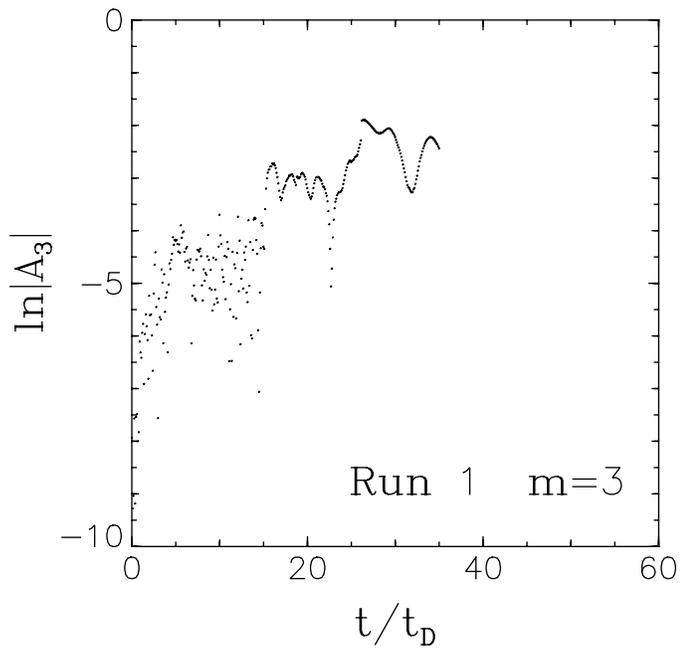
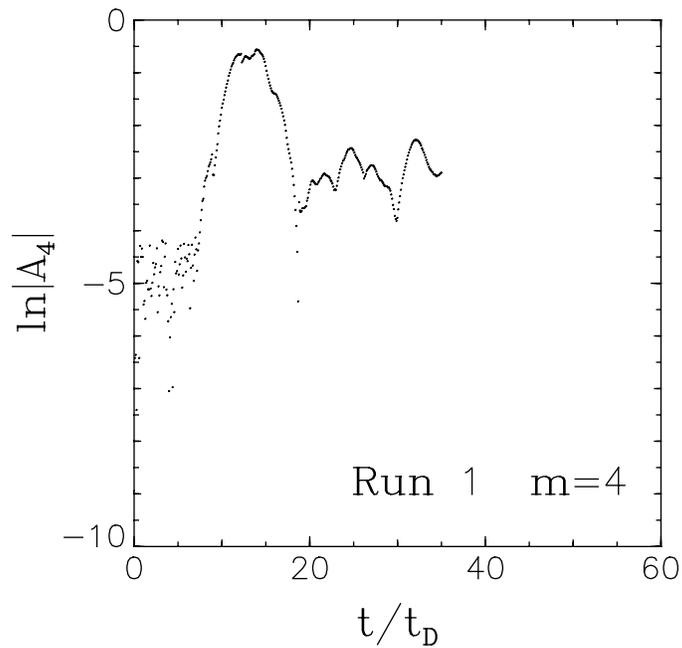
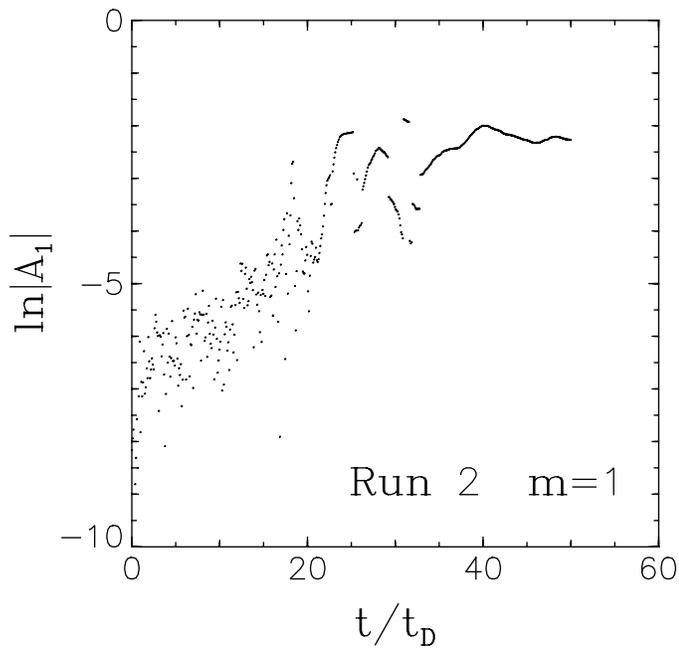
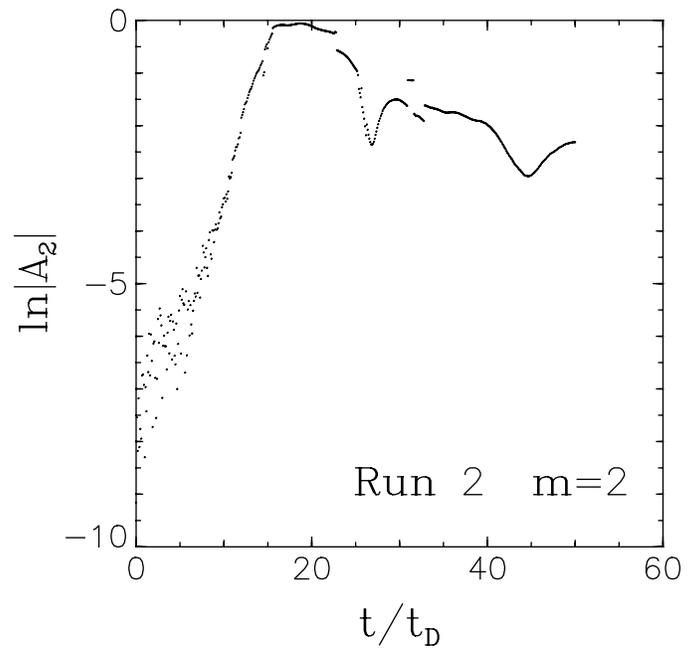
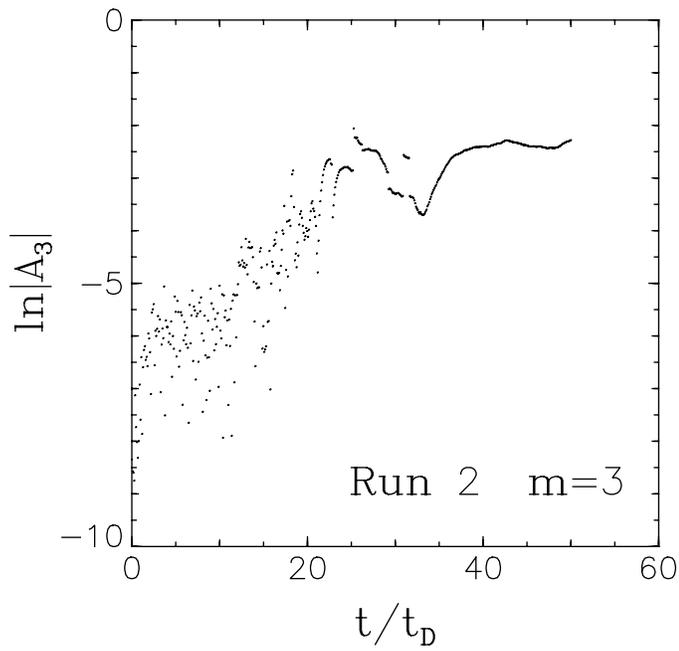
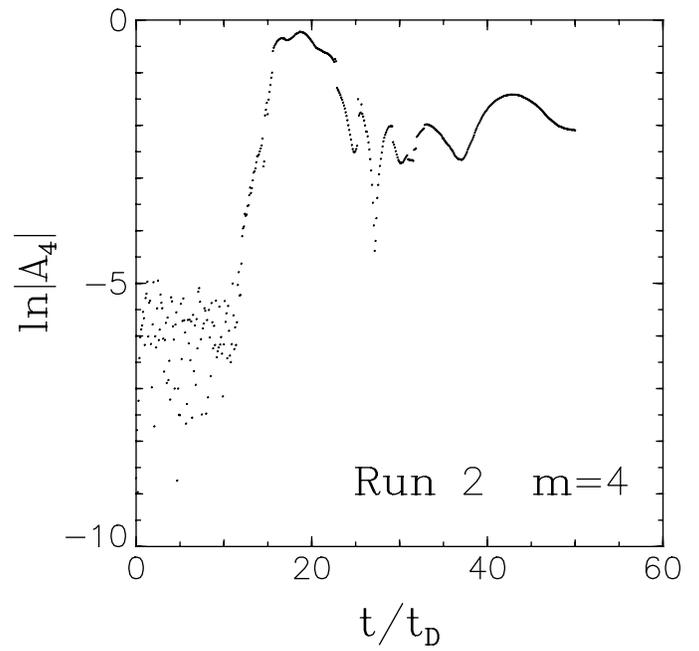
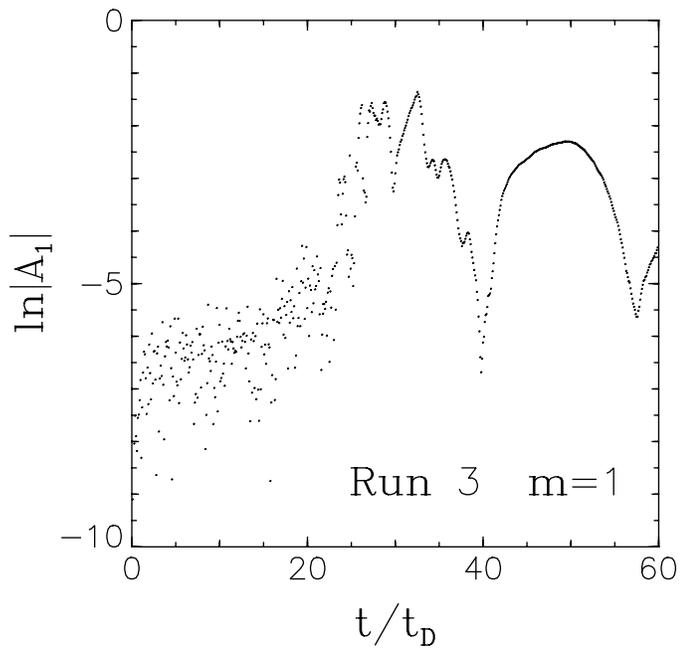
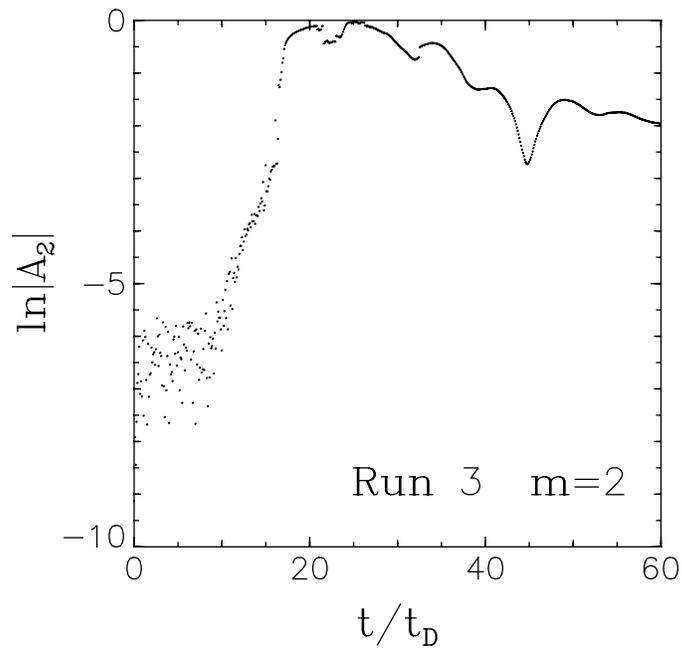
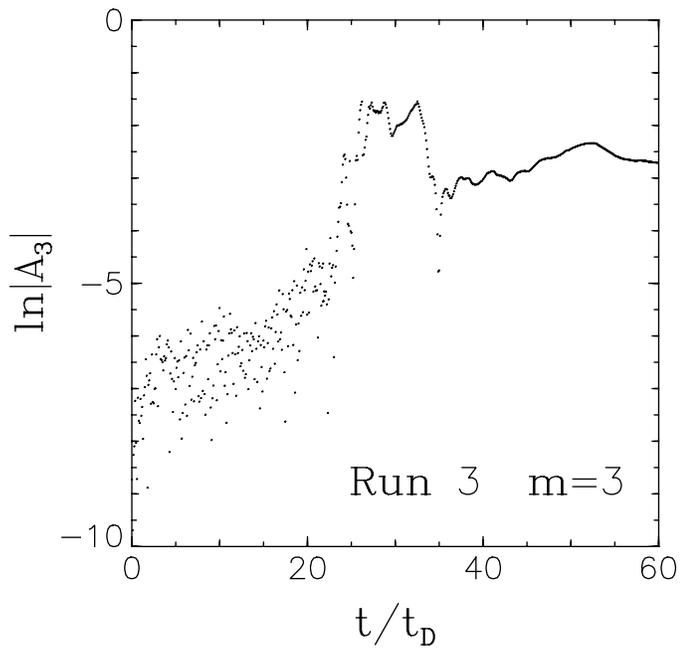
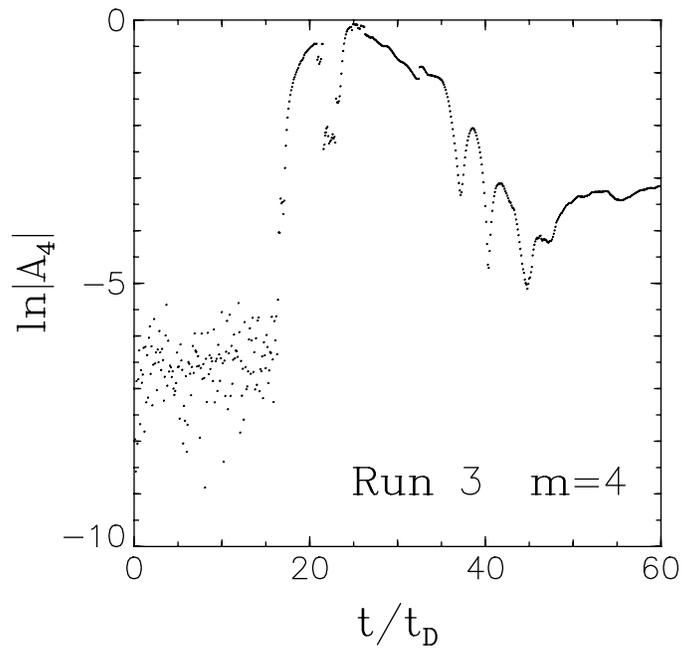
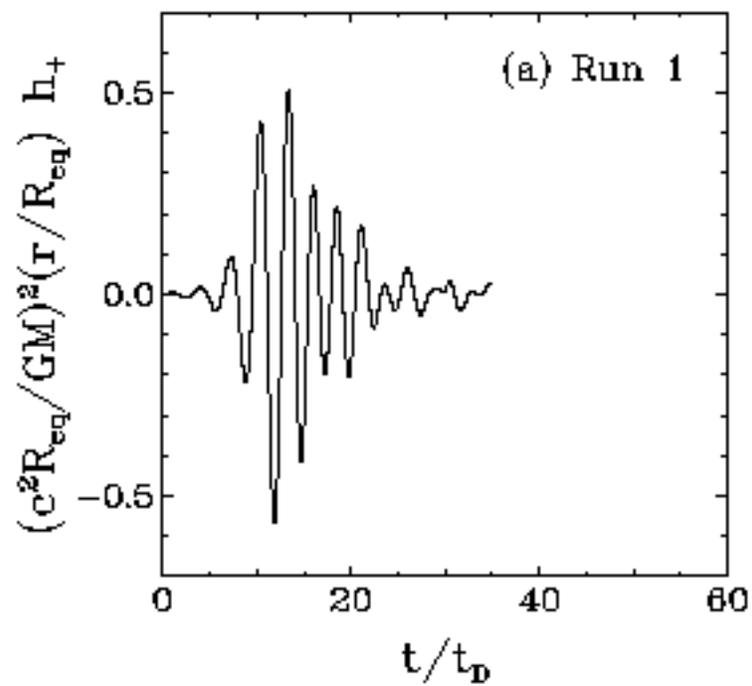
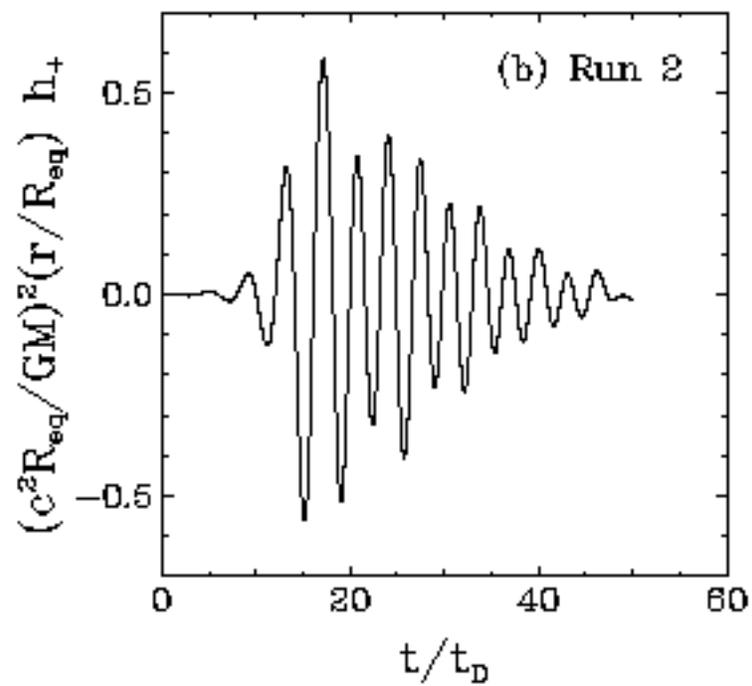
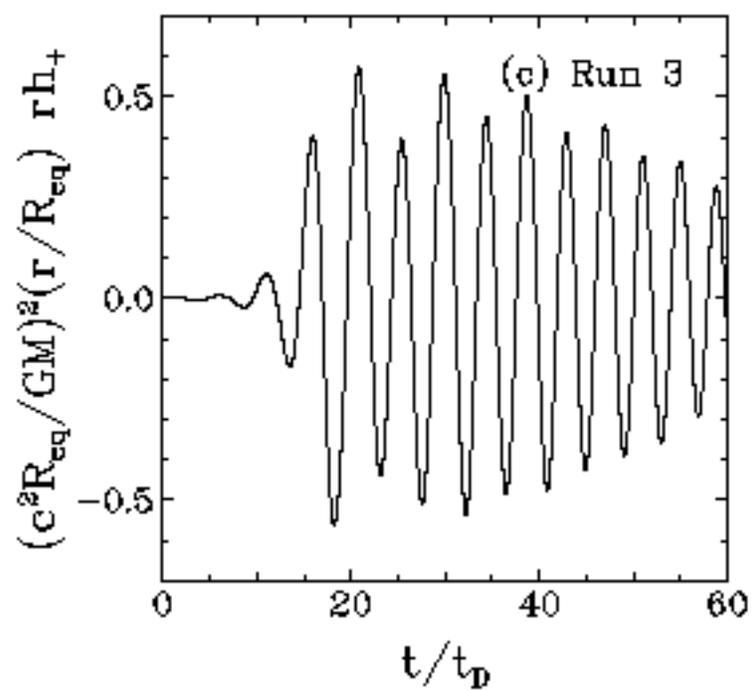
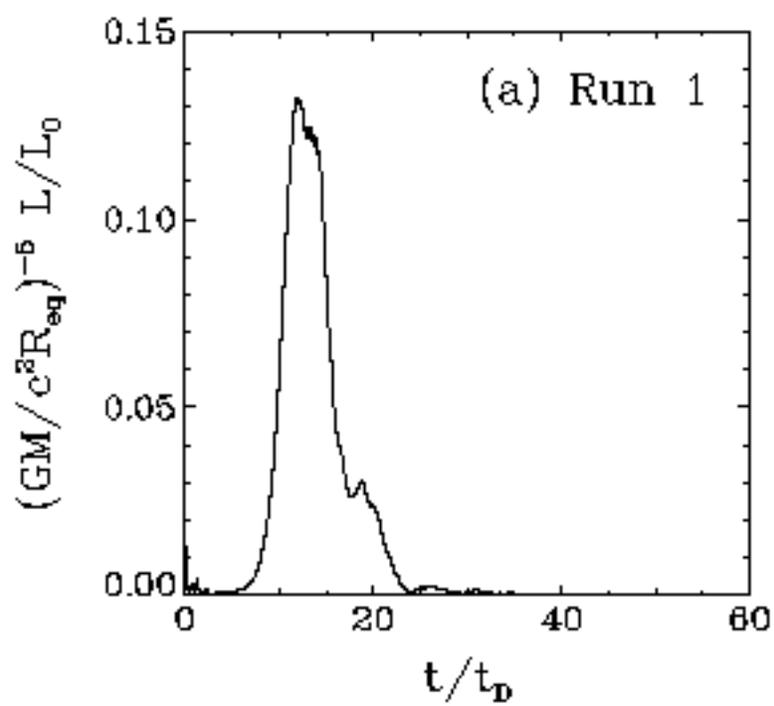
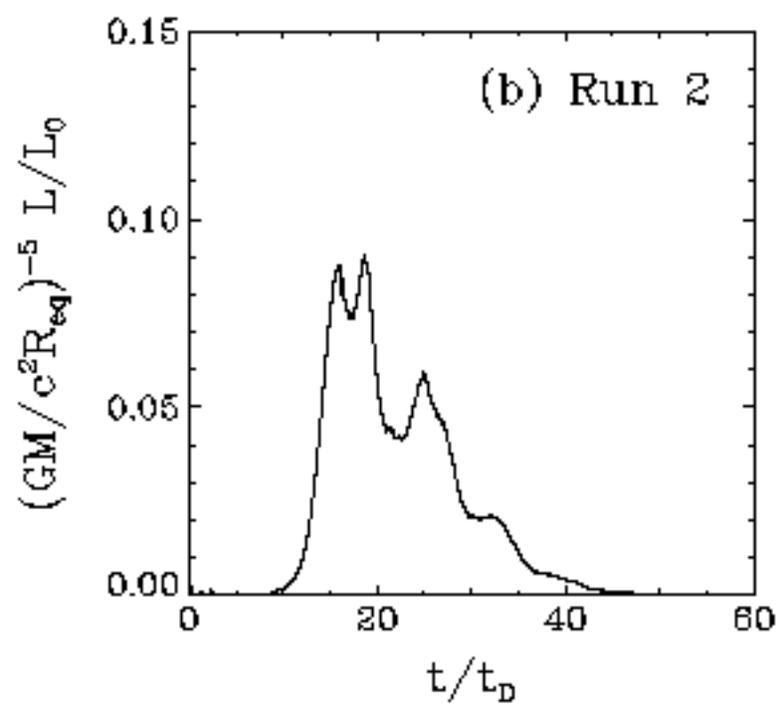
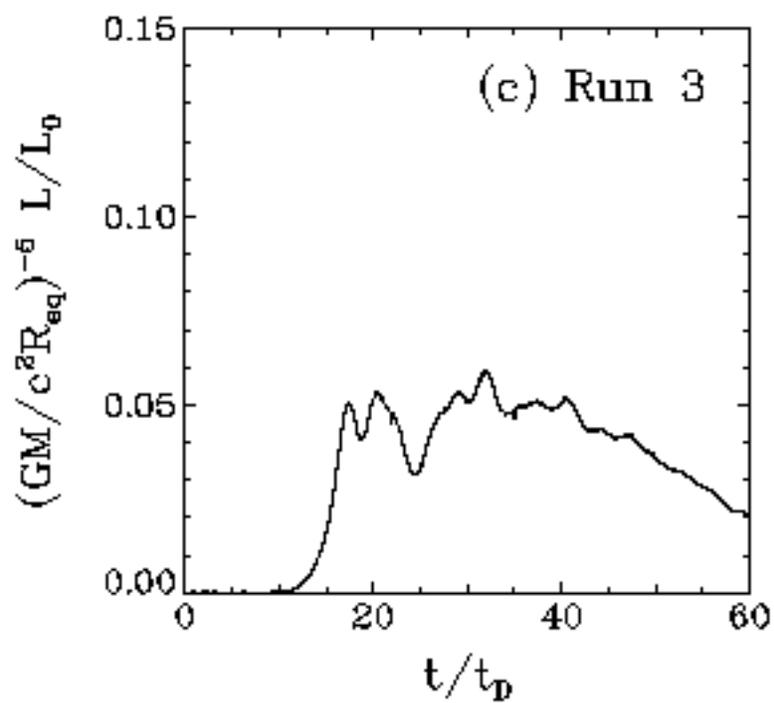
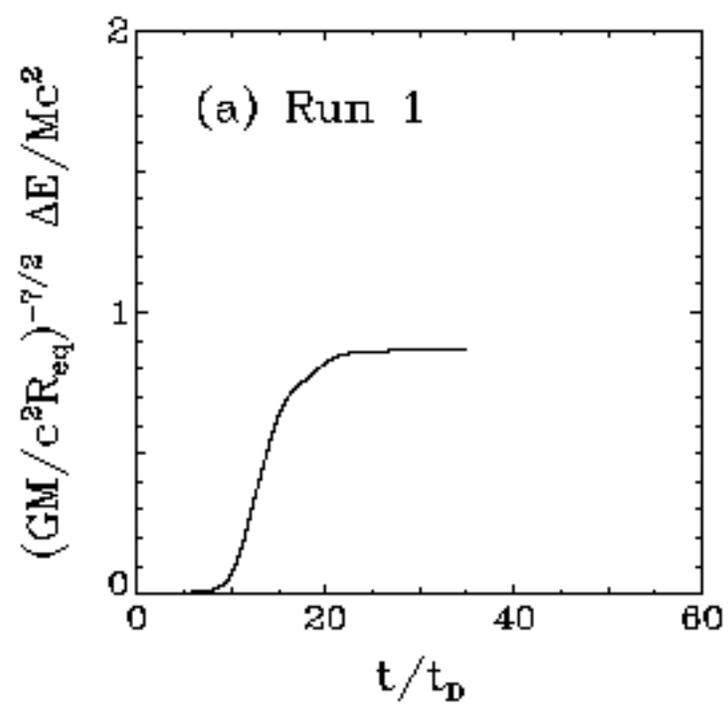
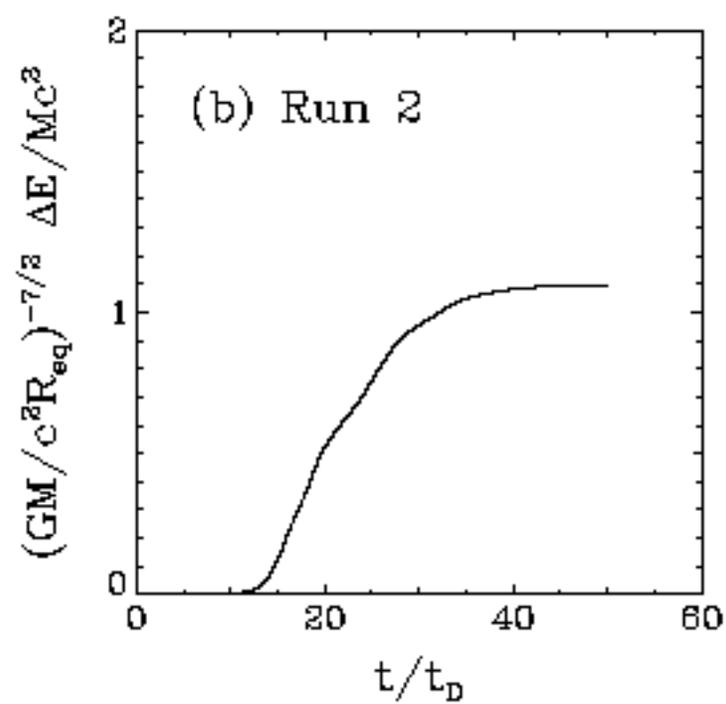
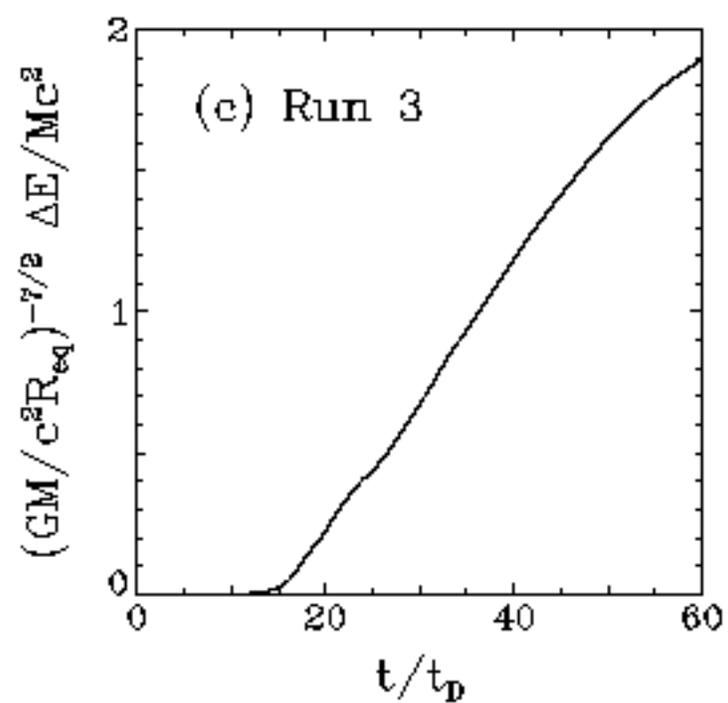
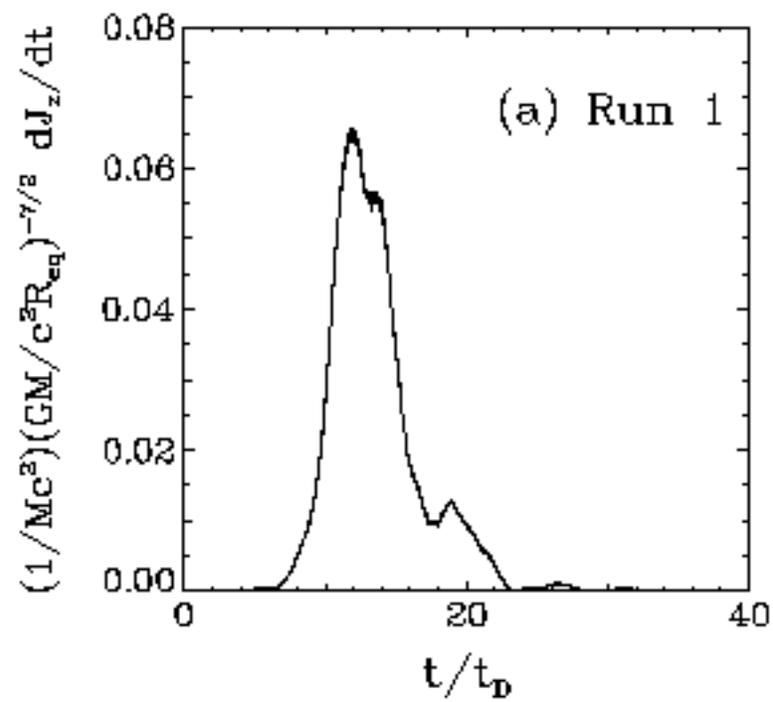
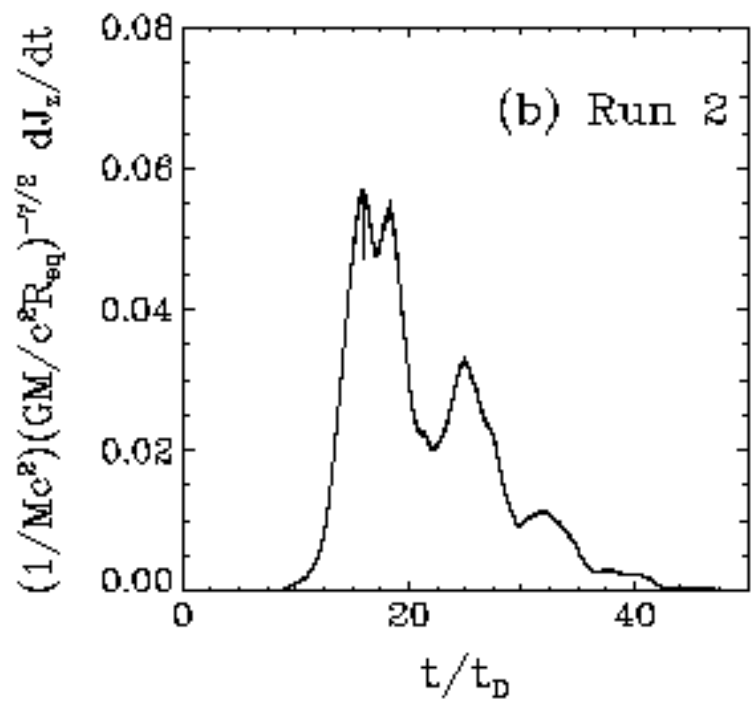
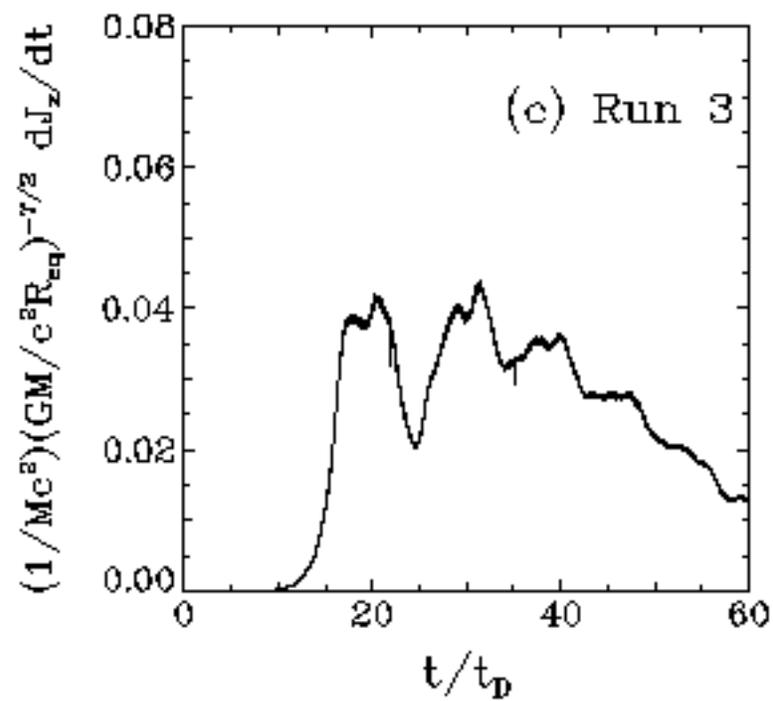
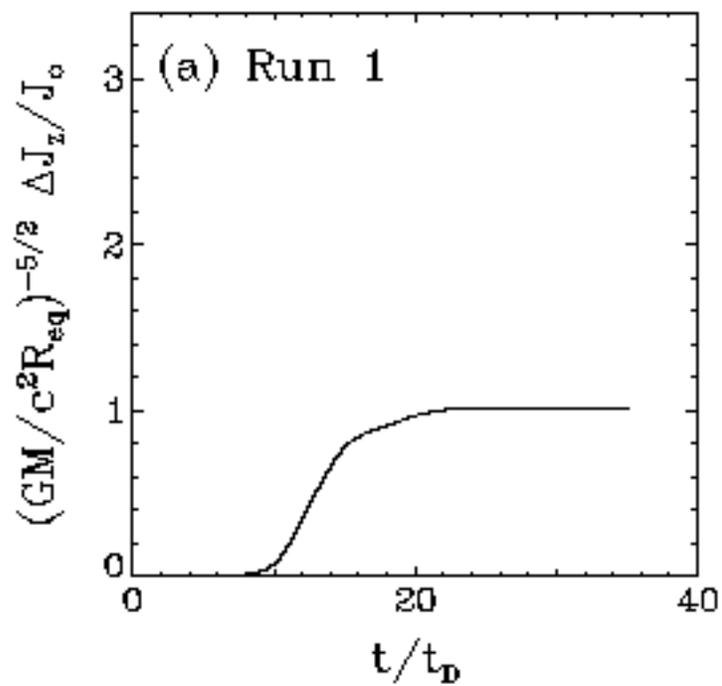
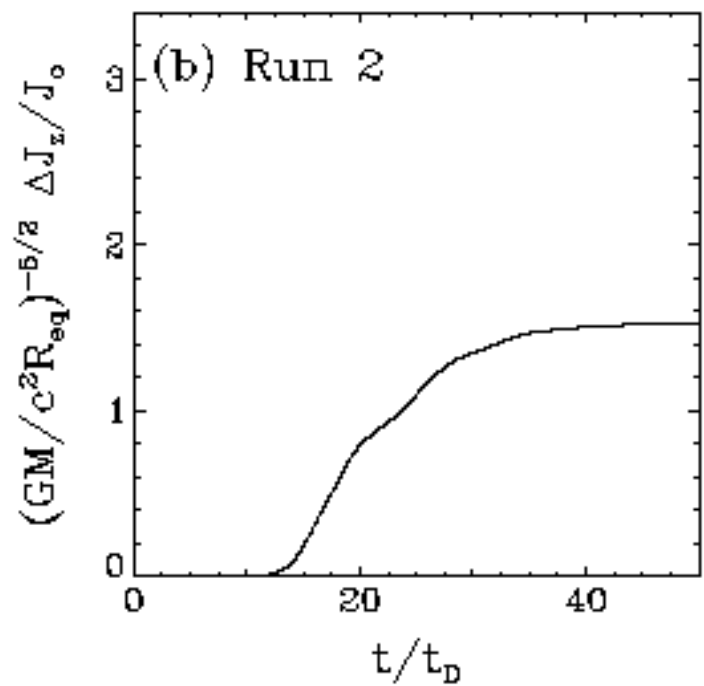
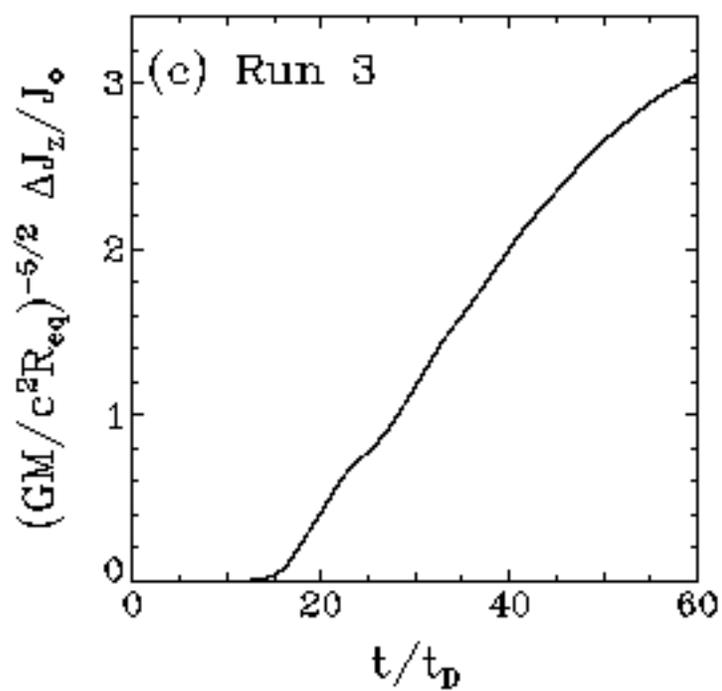
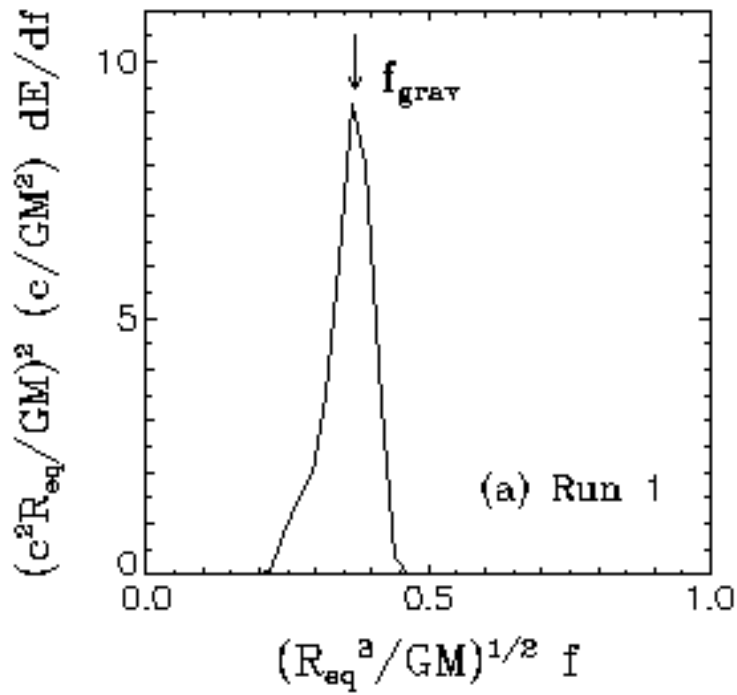
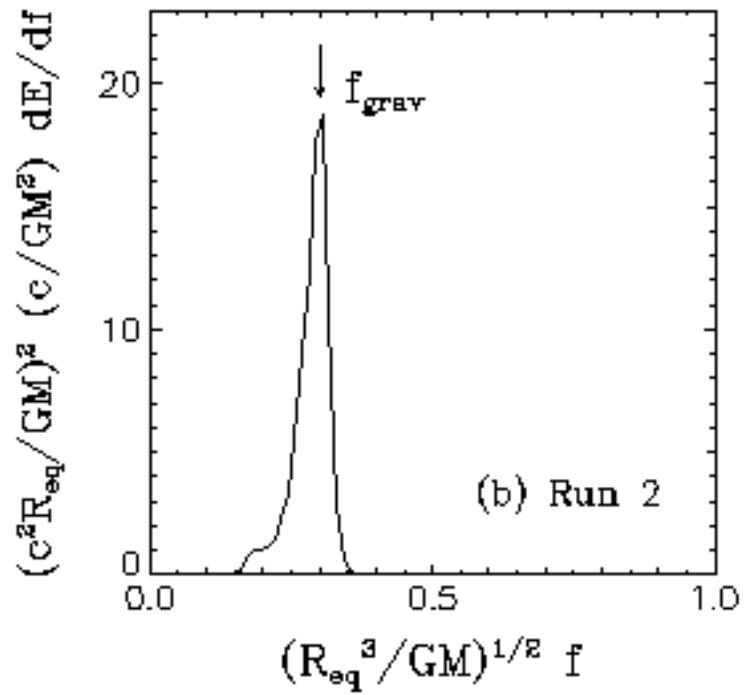
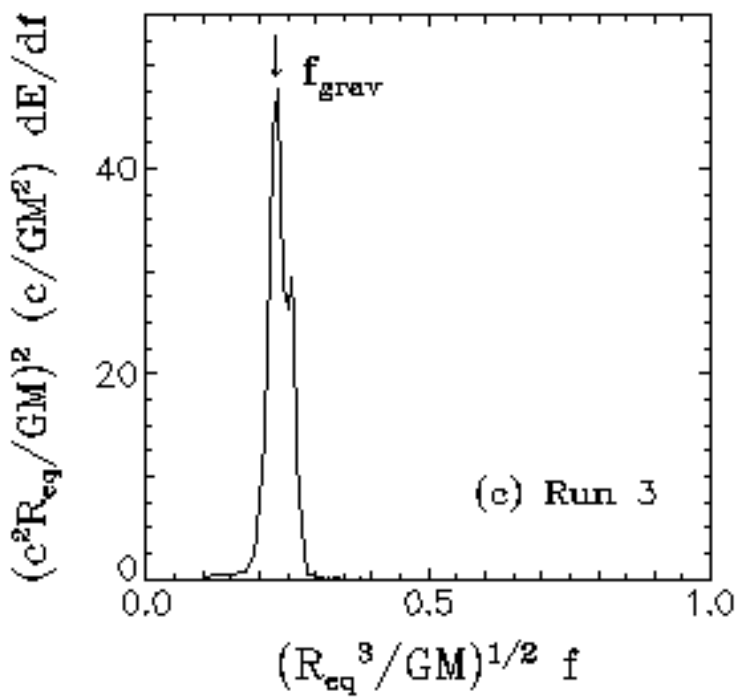
\end{document}